\documentclass[aps,preprintnumbers,showpacs,twocolumn,superscriptaddress]{revtex4}

\usepackage{amsmath}
\usepackage{amssymb}
\usepackage{color}
\usepackage{dcolumn}
\usepackage{graphicx}
\usepackage{epstopdf}           
\usepackage{latexsym}
\usepackage{times}

\input{epsf.sty}

\newcommand{\tg}{\tilde\gamma}
\newcommand{\tG}{\tilde\Gamma}
\newcommand{\tA}{\tilde A}

\newcommand{\dt}{(\partial_t - {\cal L}_\beta)\;}
\DeclareMathOperator{\tr}{\ensuremath{\mathrm{tr}}}

\newcommand{\bea}{\begin{eqnarray}}
\newcommand{\eea}{\end{eqnarray}}
\newcommand{\beq}{\begin{equation}}
\newcommand{\eeq}{\end{equation}}
\newcommand{\lm}{\ell m}

\newcommand{\mbf}[1]{\ensuremath{{\boldsymbol #1}}}

\begin{document}

\title{Recoil velocities from equal-mass binary black-hole mergers:
a systematic investigation of spin-orbit aligned configurations}

\author{Denis Pollney}
\affiliation{
  Max-Planck-Institut f\"ur Gravitationsphysik,
  Albert-Einstein-Institut,
  Potsdam-Golm, Germany
}

\author{Christian Reisswig}
\affiliation{
  Max-Planck-Institut f\"ur Gravitationsphysik,
  Albert-Einstein-Institut,
  Potsdam-Golm, Germany
}

\author{Luciano Rezzolla}
\affiliation{
  Max-Planck-Institut f\"ur Gravitationsphysik,
  Albert-Einstein-Institut,
  Potsdam-Golm, Germany
}
\affiliation{
  Department of Physics and Astronomy,
  Louisiana State University,
  Baton Rouge, LA, USA
}

\author{B\'{e}la Szil\'{a}gyi}
\affiliation{
  Max-Planck-Institut f\"ur Gravitationsphysik,
  Albert-Einstein-Institut,
  Potsdam-Golm, Germany
}

\author{Marcus Ansorg}
\affiliation{
  Max-Planck-Institut f\"ur Gravitationsphysik,
  Albert-Einstein-Institut,
  Potsdam-Golm, Germany
}

\author{Barrett Deris}
\affiliation{
  Center for Computation \& Technology,
  Louisiana State University,
  Baton Rouge, LA, USA
}
\affiliation{
  Department of Physics,
  University of California at San Diego,
  La Jolla, CA, USA
}

\author{Peter Diener}
\affiliation{
  Center for Computation \& Technology,
  Louisiana State University,
  Baton Rouge, LA, USA
}
\affiliation{
  Department of Physics and Astronomy,
  Louisiana State University,
  Baton Rouge, LA, USA
}

\author{Ernst Nils Dorband}
\affiliation{
  Max-Planck-Institut f\"ur Gravitationsphysik,
  Albert-Einstein-Institut,
  Potsdam-Golm, Germany
}

\author{Michael Koppitz}
\affiliation{
  Max-Planck-Institut f\"ur Gravitationsphysik,
  Albert-Einstein-Institut,
  Potsdam-Golm, Germany
}

\author{Alessandro Nagar}
\affiliation{
Dipartimento di Fisica, Politecnico di Torino
and INFN, sez.\ di Torino, Italy.
}

\author{Erik Schnetter}
\affiliation{
  Center for Computation \& Technology,
  Louisiana State University,
  Baton Rouge, LA, USA
}
\affiliation{
  Department of Physics and Astronomy,
  Louisiana State University,
  Baton Rouge, LA, USA
}

\date{July 16, 2007}


\begin{abstract}
  Binary black-hole systems with spins aligned with the orbital
  angular momentum are of special interest, as studies indicate that
  this configuration is preferred in nature due to non-vacuum
  environmental interactions, as well as post-Newtonian (PN)
  spin-orbit couplings. If the spins of the two bodies differ, there
  can be a prominent beaming of the gravitational radiation during the
  late plunge, causing a recoil of the final merged black hole. In
  this paper we perform an accurate and systematic study of recoil
  velocities from a sequence of equal-mass black holes whose spins are
  aligned with the orbital angular momentum, and whose individual
  spins range from $a = +0.584$ to $-0.584$.  In this way we extend
  and refine the results of a previous study which concentrated on the
  anti-aligned portion of this sequence, to arrive at a consistent
  maximum recoil of $448\pm 5\,\mathrm{km/s}$ for anti-aligned models
  as well as to a phenomenological expression for the recoil velocity
  as a function of spin ratio. Quite surprisingly, this relation
  highlights a nonlinear behavior, not predicted by the PN estimates,
  and can be readily employed in astrophysical studies on the
  evolution of binary black holes in massive galaxies.  An essential
  result of our analysis, without which no systematic behavior can be
  found, is the identification of different stages in the waveform,
  including a transient due to lack of an initial linear momentum in
  the initial data.  Furthermore, by decomposing the recoil
  computation into coupled modes, we are able to identify a pair of
  terms which are largely responsible for the kick, indicating that an
  accurate computation can be obtained from modes up to
  $\ell=3$. Finally, we provide accurate measures of the radiated
  energy and angular momentum, finding these to increase linearly with
  the spin ratio, and derive simple expressions for the final spin and
  the radiated angular momentum which can be easily implemented in
  $N$-body simulations of compact stellar systems. Our code is
  calibrated with strict convergence tests and we verify the
  correctness of our measurements by using multiple independent
  methods whenever possible.
\end{abstract}

\maketitle

\section{Introduction} 
\label{sec:intro}

Recent developments in numerical relativity have solved the problem of
stably evolving black hole initial data for useful timescales, and
opened the door to studies of physical phenomena resulting from
strong-field gravitational interactions. A result of particular
interest to astrophysics is an accurate calculation of the recoil
velocity which is generated during an asymmetric collision of a black-hole
binary. It is well known that a binary with unequal masses or spins of
the individual bodies will radiate gravitational energy asymmetrically.
This results in an uneven flux, which gives a net linear momentum to
the final black hole, often called a ``kick''~\cite{peres:1962,
Bekenstein:1973mi}. While estimations of kick velocities have been
available for some time~\cite{fitchett:1983, 1984MNRAS.211..933F,
Favata:2004wz}, the largest part of the system's acceleration is
generated in the final orbits of the binary system, and as such
requires fully relativistic calculations to be determined accurately.

Over the past year, a number of numerical relativity simulations have
been carried out to determine recoil velocities in various sections of
the parameter space of binary black-hole systems. The first systems to
be studied were unequal mass systems with moderate mass ratios, where
the first calculations were performed by the Penn
State~\cite{Herrmann:2006ks} and Goddard~\cite{Baker:2006vn} groups,
with simulations at mass ratios near the estimated peak of the
Fitchett formula~\cite{1984MNRAS.211..933F}.  A more extensive study,
exploring a large number of models between mass ratios 0.25 to 1.0,
was carried out by the Jena group~\cite{Gonzalez:2006md}, providing
for the first time a mapping of the unequal mass parameter space with
fully relativistic simulations.  The recoils from systems in which the
bodies had spin were first considered by a number of studies in the
first half of this year. The Penn State group examined a sequence of
equal mass binaries with spins equal and anti-aligned, determining
that the largest recoil possible from such an evolution is of the
order of $475\,\mathrm{km/s}$~\cite{Herrmann:2007ac}. At the same
time, in Ref.~\cite{Koppitz-etal-2007aa} we studied a sequence of
models in which the spins are anti-aligned, but of different
magnitude, and arrived at a similar estimate of $450\,\mathrm{km/s}$.
The Jena and Brownsville (now Rochester) groups showed that extremely
large kicks are possible from particular configurations of misaligned
spins, measuring recoils as high as
$2500\,\mathrm{km/s}$~\cite{Gonzalez:2007hi}, and extrapolating to
$4000\,\mathrm{km/s}$ for the maximally spinning
case~\cite{Campanelli:2007ew, Campanelli:2007cg}. Such spin
configurations have recently been studied in more detail
in~\cite{Bruegmann:2007zj}. Velocities of this magnitude have a number
of astrophysical implications for models of galaxy mergers.

In this paper, we expand on the work performed
in~\cite{Koppitz-etal-2007aa}, extending it in a number of different
ways. First, we consider a larger sequence of aligned but unequal
spins with spin-ratios ranging from $-1$ to $+1$, where the spins are
aligned (or anti-aligned) with the orbital angular momentum.  Our
interest in this set of models is motivated by the fact that there are
strong indications that binary black-hole systems having spins aligned
with the orbital angular momentum are preferred in nature.
Post-Newtonian (PN) studies in vacuum have in fact shown that in
vacuum the gravitational spin-orbit coupling has a tendency to bring
about such an alignment from generic initial
conditions~\cite{Schnittman:2004vq}.  Furthermore, in astrophysical
situations where there is likely to be at least some component of
interstellar matter inducing a dissipative dynamics, there is also a
tendency to align~\cite{Bogdanovic:2007hp}.

We describe the influence of the initial dynamics on the radiated
waveforms and the importance of suitable vector integration
constants to remove these effects when determining the final recoil
velocity. These vectors, in fact, capture the information about the
net linear momentum that the spacetime has built-up during its past
evolution and prior to the actual numerical evolution and can result
into a significant correction. We discuss how to use the results
obtained to derive a phenomenological expression for the recoil
velocity as a function of the spin ratio. Finally, we also compute how
the angular momentum of the system is redistributed between radiation
and the spin of the final black hole, providing useful expressions as
functions of the spin ratio.

The paper is organized as follows: Section~\ref{sec:NumericalMethods}
describes the code, as well as the initial data construction, and
calibration tests. In Section~\ref{sec:kick_calculation} we discuss
the calculation of the recoil velocity from gravitational-wave data on
a large sphere. We introduce and compare two methods, one based on the
Newman-Penrose $\Psi_4$ scalar which is the usual method that has been
adopted in recent numerical studies, and another which is based on
perturbations of Schwarzschild black holes modeled by a
gauge-invariant formalism. Though the two methods are based on quite
different underlying assumptions, they agree very well in their
estimation of physical quantities, and in particular the recoil
velocity. Section~\ref{sec:results} describes evolutions of the
aligned-spin sequence and the dependence of the recoil velocity on
the spin-ratio. We find that the data show an almost linear behavior
at large negative spin-ratios, as predicted by PN calculations.
However taking into account also results from positive spin-ratios,
the data suggest a nonlinear (quadratic) dependence and we give a
phenomenological expression for the recoil velocity as a function of
the spin ratio.  Extrapolating our results to the case of maximally
rotating black holes, we find that the maximum recoil velocity
attainable by spin-orbit aligned configurations is $448\pm
5\,\mathrm{km/s}$. Finally, we discuss the radiation of mass and
angular momentum for these evolutions, determining the parameters of
the isolated final black holes and show the excellent conservation of
mass and angular momentum recorded in our simulations.  Again we
provide phenomenological expressions for the relative amount of
radiated mass and spin as functions of the initial spin ratio.

In the following equations we use Greek indices (running from 0 to 3)
to denote components of four-dimensional objects and Latin indices
(running from 1 to 3) for three-dimensional ones that are defined on
space-like foliations of the space-time.

\section{Mathematical and Numerical Setup}
\label{sec:NumericalMethods}

The data presented in this paper were produced using the CCATIE code,
a three-dimensional finite differencing code based on the Cactus
Computational Toolkit~\cite{Goodale02a, cactusweb1}. The current code
is an evolution of previous versions which implemented an excision
method and co-rotating coordinates~\cite{Alcubierre99d, Alcubierre02a,
Alcubierre2003:pre-ISCO-coalescence-times}. The main features of the
code, in particular the evolution equations, remain the same. However,
some modifications have been introduced in the gauge evolution to
accommodate ``moving punctures'' which has proven to be an effective
way to evolve black hole spacetimes~\cite{Baker:2006yw,
Campanelli:2005dd}. This method simply removes any restrictions on
movement of the punctures from their initial locations, allowing them
to be advected on the grid.

\subsection{Evolution system}

We evolve a conformal-traceless ``$3+1$'' formulation of the Einstein
equations~\cite{Nakamura87, Shibata95, Baumgarte99, Alcubierre99d}, in
which the spacetime is decomposed into three-dimensional spacelike
slices, described by a metric $\gamma_{ij}$, its embedding in the full
spacetime, specified by the extrinsic curvature $K_{ij}$, and the gauge
functions $\alpha$ (lapse) and $\beta^i$ (shift) that specify
a coordinate frame (see Sect.~\ref{sec:Gauges} for details on how we
treat gauges and~\cite{York79} for a general description of the $3+1$
split). The particular
system which we evolve transforms the standard ADM variables as
follows. The 3-metric $\gamma_{ij}$ is conformally transformed via
\begin{equation}
  \label{eq:def_g}
  \phi = \frac{1}{12}\ln \det \gamma_{ij}, \qquad
  \tilde{\gamma}_{ij} = e^{-4\phi} \gamma_{ij},
\end{equation}
and the conformal factor $\phi$ evolved as an independent variable,
whereas $\tilde{\gamma}_{ij}$ is subject to the constraint
$\det \tilde{\gamma}_{ij} = 1$. The extrinsic curvature is
subjected to the same conformal transformation, and its trace
$\tr K_{ij}$ evolved as an independent variable. That is, in place of
$K_{ij}$ we evolve:
\begin{equation}
  \label{eq:def_K}
  K \equiv \tr K_{ij} = g^{ij} K_{ij}, \qquad
  \tilde{A}_{ij} = e^{-4\phi} (K_{ij} - \frac{1}{3}\gamma_{ij} K),
\end{equation}
with $\tr\tilde{A}_{ij}=0$. Finally, new evolution variables
\begin{equation}
  \label{eq:def_Gamma}
  \tilde{\Gamma}^i = \tilde{\gamma}^{jk}\tilde{\Gamma}^i_{jk}
\end{equation}
are introduced, defined in terms of the Christoffel symbols of
the conformal 3-metric.

The Einstein equations specify a well known set of evolution equations
for the listed variables and are given by 
\begin{align}
  \label{eq:evolution}
  \dt \tg_{ij} & = -2 \alpha \tA_{ij},  \\
  \dt \phi & = - \frac{1}{6} \alpha K, \\
  \dt \tA_{ij} & = e^{-4\phi} [ - D_i D_j \alpha 
   + \alpha R_{ij}]^{TF} \nonumber\\
   & + \alpha (K \tA_{ij} - 2 \tA_{ik} \tA^k{}_j), \\
  \dt K & = - D^i D_i \alpha
   + \alpha (\tA_{ij} \tA^{ij} + \frac{1}{3} K^2), \\
  \partial_t \tG^i & = \tilde\gamma^{jk} \partial_j\partial_k \beta^i
    + \frac{1}{3} \tilde\gamma^{ij}  \partial_j\partial_k\beta^k
    \nonumber\\
   & + \beta^j\partial_j \tilde\Gamma^i
   - \Tilde\Gamma^j \partial_j \beta^i 
   + \frac{2}{3} \tilde\Gamma^i \partial_j\beta^j \nonumber \\
   & - 2 \tilde{A}^{ij} \partial_j\alpha
   + 2 \alpha ( 
   \tilde{\Gamma}^i{}_{jk} \tilde{A}^{jk} + 6 \tilde{A}^{ij}
   \partial_j \phi \nonumber\\
   & - \frac{2}{3} \tg^{ij} \partial_j K ),
\end{align}
where $R_{ij}$ is the three-dimensional Ricci tensor, $D_i$
the covariant derivative associated with the three metric $\gamma_{ij}$
and ``TF'' indicates the trace-free part of tensor objects. The
Einstein equations also lead to a set of physical constraint equations
that are satisfied within each spacelike slice,
\begin{align}
  \label{eq:einstein_ham_constraint}
  \mathcal{H} &\equiv R^{(3)} + K^2 - K_{ij} K^{ij} = 0, \\
  \label{eq:einstein_mom_constraints}
  \mathcal{M}^i &\equiv D_j(K^{ij} - \gamma^{ij}K) = 0,
\end{align}
which are usually referred to as Hamiltonian and momentum constraints.
Here $R^{(3)}=R_{ij} \gamma^{ij}$ is the Ricci scalar on a three-dimensional
time slice.  Our specific choice of evolution variables introduces
five additional constraints,
\begin{align}
  \det \tilde{\gamma}_{ij} & = 1, 
    \label{eq:gamma_one}\\
  \tr \tilde{A}_{ij} & = 0,
    \label{eq:trace_free_A}\\
  \tilde{\Gamma}^i & = \tilde{\gamma}^{jk}\tilde{\Gamma}^i_{jk}.
   \label{eq:Gamma_def}
\end{align}
Our code actively enforces the algebraic
constraints~(\ref{eq:gamma_one}) and~(\ref{eq:trace_free_A}).  The
remaining constraints, $\mathcal{H}$, $\mathcal{M}^i$,
and~(\ref{eq:Gamma_def}), are not actively enforced, and can be used
as monitors of the accuracy of our numerical solution.
See~\cite{Alcubierre02a} for a more comprehensive discussion of the
these points.

\subsection{Gauges}
\label{sec:Gauges}

We specify the gauge in terms of the standard ADM lapse
function, $\alpha$, and shift vector, $\beta^a$~\cite{misner73}.
We evolve the lapse according to the ``$1+\log$'' slicing
condition:
\begin{equation}
  \partial_t \alpha - \beta^i\partial_i\alpha 
    = -2 \alpha (K - K_0),
  \label{eq:one_plus_log}
\end{equation}
where $K_0$ is the initial value of the trace of the extrinsic
curvature, and
equals zero for the maximally sliced initial data we consider here.
The shift is evolved using the hyperbolic $\tilde{\Gamma}$-driver
condition~\cite{Alcubierre02a},
\begin{eqnarray}
  \partial_t \beta^i - \beta^j \partial_j  \beta^i & = & \frac{3}{4} \alpha B^i\,,
  \\
  \partial_t B^i - \beta^j \partial_j B^i & = & \partial_t \tilde\Gamma^i 
    - \beta^j \partial_j \tilde\Gamma^i - \eta B^i\,,
\end{eqnarray}
where $\eta$ is a parameter which acts as a damping coefficient.  The
advection terms on the right-hand-sides of these equations were not
present in the original definitions of~\cite{Alcubierre02a}, where
co-moving coordinates were used, but have been added following the
experience of~\cite{Baker05a, Baker:2006mp}, and are required for
correct advection of the puncture in ``moving-puncture'' evolutions.

\subsection{Numerical methods}  
\label{sec:NumericalSpecifications}

Spatial differentiation of the evolution variables is performed via
straightforward finite-differencing using fourth-order accurate
centered stencils for all but the advection terms for each variable,
which are upwinded in the direction of the shift. Vertex-centered
adaptive mesh-refinement (AMR) is employed using nested
grids~\cite{Schnetter-etal-03b, carpetweb} with a $2:1$ refinement
for successive grid levels, and the highest resolution concentrated
in the neighborhood of the individual horizons. Individual apparent
horizons are located every few time steps during the
evolution~\cite{Thornburg95, Thornburg2003:AH-finding_nourl}.

The time steps on each grid are set by the Courant condition and thus
the spatial grid resolution for that level, with the time evolution
being carried out using fourth-order accurate Runge-Kutta integration
steps. Boundary data for finer grids are calculated with spatial
prolongation operators employing 5th-order polynomials, and
prolongation in time employing 2nd-order polynomials. The latter
allows a significant memory saving, requiring only three time levels
to be stored, with little loss of accuracy due to the long dynamical
timescale relative to the typical grid time step.

In the results presented below we have used 8 levels of mesh
refinement with finest grid resolutions of $h/M=0.030$, $0.024$, and
$0.018$; we will refer to these resolutions as ``low'', ``medium'' and
``high'' respectively. We find that the medium (\textit{i.e.},
$h=0.024\,M$) fine-grid resolution is typically good enough to
accurately represent the dynamics which we are studying here and will
be used hereafter as our fiducial resolution. In this case, the
wave-zone grid has a resolution of $h=1.536\,M$. In addition, when
measuring the convergence order (see discussion in
Sect.~\ref{sec:ct}), we have also used a ``very-high'' resolution of
$h/M=0.012$ which therefore gives a factor of 2 refinement with
respect to the ``medium'' resolution; this should be contrasted with
similar convergence tests recently discussed in the literature and in
which the refinement factor is much smaller.

The finest grids are centered on each black hole, with a radius about
$50\%$ larger than the apparent horizon. A single grid resolution
covers the region between $r=20\,M$ and $r=80\,M$, in which our wave
extraction is carried out. The outer (coarsest) grid extends to a
spatial position which is large compared with the evolution time of
the system. In particular, it ranges from $256\,M$ in each coordinate
direction for the binaries which merge rapidly, up to $768\,M$ for the
binaries which inspiral more slowly because of the spin-orbit
interaction. In all cases, artificial wave-like boundary conditions
are used, and although these are not explicitly constraint-preserving,
they do not introduce major violations of the constraints as long as
they are placed sufficiently far away from the central black holes
(\textit{i.e.}, with a light-crossing time which is large as compared
to the time for the merger).  Furthermore, for the models considered
here, in which all spins are directed along the $z$-axis of our
Cartesian grids, it is possible to use a reflection symmetry condition
across the $z=0$ plane.  Tests against the runs on a full grid show
that this symmetry is preserved to a high degree in our simulations
(\textit{i.e.}, with differences below $10^{-14}$) so that this
symmetry boundary has no influence on the dynamics.

\subsection{Initial data}
\label{sec:initial_data}

\begin{table*}[t]
\caption{The puncture initial data parameters defining the binaries:
  location $\pm x/M$, linear momenta $\pm p/M$, mass parameters
  $m_i/M$, spins $S_i/M^2$, dimensionless spins $a_i$, ADM mass
  $M_{_{\rm ADM}}$ measured at infinity, and ADM angular momentum
  $J_{_{\mathrm{ADM}}}$ computed from Eq.~(\ref{eq:InitialSpin}). Note
  that we set $M_1=M_2=1/2$ [\textit{cf.}, Eq.~(\ref{eq:AHMass})].}
\begin{ruledtabular}
\begin{tabular}{l|ccccccrccl}
Model &
\multicolumn{1}{c}{$\pm x/M$} &
\multicolumn{1}{c}{$\pm p/M$} & 
\multicolumn{1}{c}{$m_1/M$} &
\multicolumn{1}{c}{$m_2/M$} & 
\multicolumn{1}{c}{$S_1/M^2$} &
\multicolumn{1}{c}{$S_2/M^2$} & 
\multicolumn{1}{c}{$a_1$} &
\multicolumn{1}{c}{$a_2$} &
\multicolumn{1}{c}{$M_{_{\rm ADM}}/M$} &
\multicolumn{1}{c}{$J_{_{\mathrm{ADM}}}/M^2$}\\
\hline
$r0$  & 3.0205 & 0.1366 & 0.4011 & 0.4009 & -0.1460 & 0.1460 & -0.5840 & 0.5840 &  0.9856 & 0.8252\\
$r1$  & 3.1264 & 0.1319 & 0.4380 & 0.4016 & -0.1095 & 0.1460 & -0.4380 & 0.5840 &  0.9855 & 0.8612\\
$r2$  & 3.2198 & 0.1281 & 0.4615 & 0.4022 & -0.0730 & 0.1460 & -0.2920 & 0.5840 &  0.9856 & 0.8979\\
$r3$  & 3.3190 & 0.1243 & 0.4749 & 0.4028 & -0.0365 & 0.1460 & -0.1460 & 0.5840 &  0.9857 & 0.9346\\
$r4$  & 3.4100 & 0.1210 & 0.4796 & 0.4034 &  0.0000 & 0.1460 &  0.0000 & 0.5840 &  0.9859 & 0.9712\\
$r5$  & 3.5063 & 0.1176 & 0.4761 & 0.4040 &  0.0365 & 0.1460 &  0.1460 & 0.5840 &  0.9862 & 1.007\\
$r6$  & 3.5988 & 0.1146 & 0.4638 & 0.4044 &  0.0730 & 0.1460 &  0.2920 & 0.5840 &  0.9864 & 1.044\\
$r7$  & 3.6841 & 0.1120 & 0.4412 & 0.4048 &  0.1095 & 0.1460 &  0.4380 & 0.5840 &  0.9867 & 1.081\\
$r8$  & 3.7705 & 0.1094 & 0.4052 & 0.4052 &  0.1460 & 0.1460 &  0.5840 & 0.5840 &  0.9872 & 1.117\\ 
\hline
$r0l$ & 4.1924 & 0.1073 & 0.4066 & 0.4065 & -0.1460 & 0.1460 & -0.5840 & 0.5840 &  0.9889 & 0.8997 \\
$r0s$ & 2.8186 & 0.1441 & 0.3997 & 0.3994 & -0.1460 & 0.1460 & -0.5840 & 0.5840 &  0.9849 & 0.8123
\end{tabular}
\end{ruledtabular}
\vskip -0.25cm
\label{tbl:parameters}
\end{table*}

The initial data are constructed applying the ``puncture''
method~\cite{Brandt97b}, which uses Bowen-York extrinsic curvature and
solves the Hamiltonian constraint equation numerically as
in~\cite{Ansorg:2004ds}.

We have considered a sequence of binaries for which the initial spin
of one of the black holes is held fixed at
$\mbf{S}_2/M^2=0.146\,\mbf{e}_z$, and the spin of the other black hole
is $\mbf{S}_1/M^2 = (a_1/a_2) \mbf{S}_2/M^2$, where the spin ratio
$a_1/a_2$ takes the values $-1,\,-3/4,\ldots\,,3/4,1$, and $M$ is the
sum of the black hole masses, $M = M_1 + M_2$.  Thus the black hole
spins are anti-aligned when $a_1/a_2$ is negative and aligned when it
is positive. In all cases the initial data parameters are chosen such
that the black hole masses are
\begin{equation}
M_i = \sqrt{\frac{A_i}{16 \pi} + \frac{4\pi S_i^2}{A_i}} = \frac{1}{2},
\label{eq:AHMass}
\end{equation}
\cite{Smarr73a, Christodoulou70} where $A_i$ is the area of the $i$-th
apparent horizon.

For the orbital initial data parameters we use the effective potential
method introduced in~\cite{Cook94} and extended to spinning
configurations in~\cite{Pfeiffer:2000um}. The effective potential
method is a way of choosing the initial data parameters such that the
required physical parameters (e.g.\ masses and spins) are
obtained to describe a binary black-hole system on a quasi-circular
orbit.

The free parameters to be chosen for the puncture initial data are:
the puncture coordinate locations $\mbf{C}_i$, the puncture mass
parameters $m_i$, the linear momenta $\mbf{p}_i$, and the individual
spins $\mbf{S}_i$. Since we are interested in quasi-circular orbits we
work in the zero momentum frame and choose $\mbf{p}_1 = -\mbf{p}_2$ to
be orthogonal to $\mbf{C}_2 - \mbf{C}_1$. The physical parameters we
want to control are: the black hole mass ratio $M_1/M_2$, the orbital
angular momentum $\mbf{L}=\mbf{C}_1 \times \mbf{p}_1 + \mbf{C}_2
\times \mbf{p}_2$ 
(see
for example~\cite{Cook94, Cook00a, Pfeiffer:2000um})
and the dimensionless spin parameters
$a_i=\mbf{S}_i/M_i^2$. In order to choose the input parameters that
correspond to the desired physical parameters we have to use a
non-linear root finding procedure, since the physical
parameters depend non-linearly on the input parameters and it is not
possible to invert the problem analytically.

As detailed in~\cite{Pfeiffer:2000um}, when the black-hole spins are
taken as parameters, it is possible to reduce the number of
independent input variables, so that at a given separation
${\boldsymbol {\bar C}} \equiv |\mbf{C}_2 - \mbf{C}_1|/m_1$, the
independent input parameters are: ${\bar q} \equiv m_1/m_2$ and the
dimensionless magnitude of the linear momentum $p/m_1$. Using a
Newton-Raphson method, we solve for ${\bar q}$ and $p/m_1$ so that
$M_1/M_2 = 1$ and the system has a given dimensionless orbital angular
momentum, $L/(\mu M)$ where $\mu = m_1 m_2 / M^2$ is the reduced mass.
For such a configuration the initial data solver~\cite{Ansorg:2004ds}
returns a very accurate value for $M_{_{\rm ADM}}$, which together
with the accurate irreducible mass calculated by the apparent horizon
finder~\cite{Thornburg95,Thornburg2003:AH-finding} makes it possible
to calculate an accurate value of the dimensionless binding energy
\begin{equation}
E_b/\mu = (M_{_{\rm ADM}}-M_1-M_2)/\mu.
\end{equation}
The quasi-circular initial data parameters are then obtained by
finding the minimum in $E_b/\mu$ for varying values of ${\boldsymbol
{\bar C}}$ while keeping the required orbital angular momentum $L/(\mu
M)$ constant.

We chose a fixed orbital angular momentum $L/(\mu M) = 3.3$ for our
quasi-circular orbit initial data parameters. This value was chosen to
ensure that model $r0$ would have enough evolution time for an
accurate kick measurement, while at the same time model $r8$ would not
require too much evolution time.  In order to check the influence of
the evolution time before plunge on the kick measurements of the $r0$
model, we also calculated initial data for a $r0$ configuration at
larger initial separation $r0$l and at smaller initial separation
$r0$s. The parameters for all the initial data sets are shown in
Table~\ref{tbl:parameters}.

Note that the physical mass $M_i$ of a
single puncture black hole increases when the spin parameter is
increased if the mass parameter $m_i$ is kept constant. For that
reason obtaining $M_1=M_2$ in general requires that $m_1\ne m_2$. Even
in the case where the spins have the same magnitude but different
directions, the two black holes will have different spin-orbit
interactions leading to slightly different physical masses if $m_1 =
m_2$. For this reason, the initial data for $r0$ in
Table~\ref{tbl:parameters} has slightly different puncture mass
parameters $m_1 \ne m_2$.
In contrast, in model $r8$ the black holes have identical
spin parameters and thus also the same spin-orbit interaction,
resulting in identical mass parameters $m_1=m_2$.

\subsection{Convergence tests}
\label{sec:ct}

As described in Section~\ref{sec:NumericalSpecifications}, the finite
difference error of the derivative stencils used in the numerical
algorithm is $O(h^4)$, while the error in the time-interpolation
stencils used for mesh refinement boundary points is $O(\Delta t^3)$.
Thus the expected theoretical convergence rate is three. However, it
is only time-related operations which are at third order, and since
the time step which we use is
smaller than the grid spacing and
much smaller than the dynamical
timescales, we can expect that the error coefficient of the leading
order term is quite small. Third order convergence is expected during
time-periods when the system goes through rapid dynamical changes,
such as the plunge or merger.

The proper convergence of the code was established using the binary
system $r0$, for which we have carried out evolutions using 8 levels
of mesh refinement with fine grid-spacings of $h/M=0.024$, $0.018$,
and $0.012$ (\textit{i.e.}, resolutions ``medium'', ``high'', and
``very-high'', respectively, where ``low'' refers to $h=0.030$ which was
deemed to be of insufficient accuracy for the results of this
paper). Other refinement levels have resolutions that are half of the
next finest grid. The refinement levels on the initial slice are set
up to be identical for the three resolutions and their locations and
sizes evolve according to the same algorithm in each case.

We focus on the convergence of a number of different aspects
of the code. The first of these is the degree of satisfaction of the
Einstein equations, which can be partially determined by examining the
Hamiltonian and momentum
constraints~(\ref{eq:einstein_ham_constraint})--(\ref{eq:einstein_mom_constraints}).
A more stringent requirement is to evaluate how well the Einstein
tensor satisfies the vacuum condition, $G_{\alpha\beta}=0$. For this
we define the positive definite quantity
\begin{equation}
 {\mbf G} \equiv \left\{
\begin{array}{ll}
\sqrt{ G_{00}^2 + G_{01}^2 + \cdots + G_{33}^2 } &
\mbox{outside appar.\ horizons} \\
0 & \mbox{inside appar.\ horizons\ .}
\label{eq:EinsteinNorm}
\end{array}
\right.
\end{equation}
In computing norms over the entire grid, we find it useful to mask out
the interiors of the horizons, where the error at the puncture
locations -- which is not expected to converge -- can
dominate over more relevant errors in the physically observable
domain. In order to compute $G_{\alpha\beta}$ we compute the
$4$-derivatives of the ADM metric, lapse and shift, then construct the
$4$-derivatives of the $4$-metric from which we can compute the
Riemann tensor and then finally obtain
$G_{\alpha\beta}$. Time-derivatives are taken using three time-levels,
centered around the past time-level. Spatial derivatives are taken
using fourth-order accurate centered stencils. Thus the
finite-difference error in computing $G_{\alpha\beta}$ is ${\cal
O}(\Delta t^2)$ in time and ${\cal O}(h^4)$ in the space dimensions.
Effectively we see a minimum of third order accuracy for this
quantity, indicating that the coefficient of the ${\cal O}(\Delta
t^2)$ error term is small compared to the higher-order terms.

Since the metric gradients and hence the truncation errors are the
largest near the black-holes, through the $L_{\infty}$ norm
of~\eqref{eq:EinsteinNorm} we effectively monitor that the Einstein
tensor converges near the horizons for the duration of the
evolution. We regard this as a rather stringent test in comparison
with the common use of the $L_2$ norm, as the latter tends to
dilute errors in small regions or 2D surfaces such as grid boundaries,
as they are normalized over the entire grid volume. By contrast,
the $L_{\infty}$ norm measures the worst error on the grid, which
by propagation of error will also suffer if there are any
non-convergent regions on the grid.

\begin{figure}[t]
\begin{center}
  \centerline{
   \resizebox{8.5cm}{!}{\includegraphics[angle=-0]{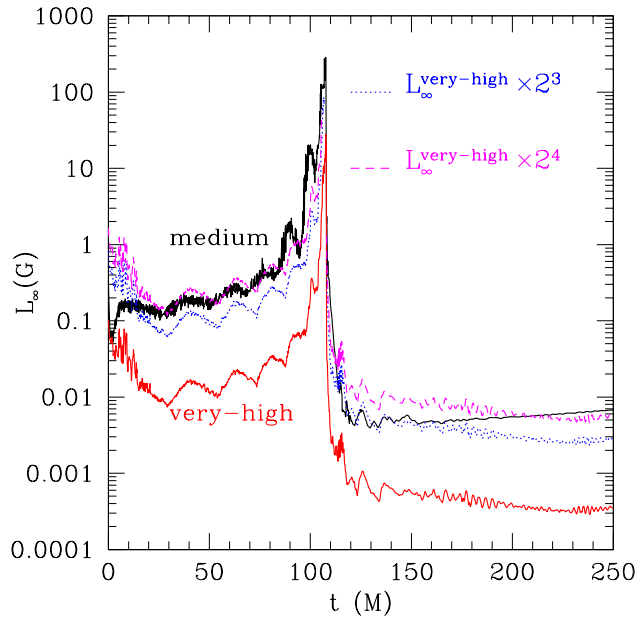}
  }}
  \vskip 0.5cm
  \caption{The $L_\infty$ norm of the Einstein tensor
           Eq.~\eqref{eq:EinsteinNorm} as a function of time. During
           the periods of strong dynamics (\textit{i.e.}, when the
           time derivatives of the evolution variables are large) the
           convergence order is dominated by the accuracy of the
           time-interpolation algorithm used at mesh refinement
           boundaries, thus yielding third-order accuracy. At the
           times when these time-derivatives are small, the
           fourth-order finite-differencing algorithm becomes the
           dominant source of the error. Note that the very large
           violations (of ${\cal O}(300)$ at the medium resolution)
           are confined to a \textit{single} grid point on the trailing
           edge of the apparent horizon and are produced by the very
           steep gradients in the shift. As discussed later, this does
           not affect the fourth-order convergence of the waveforms. At
           the time of the merger a common apparent horizon forms and its
           excision from the calculation of the $L_\infty$ norm is
           responsible for the drop in the violation.}
  \label{fig:ConvOfEinsteinNorm}
\end{center}
\end{figure}

This convergence of ${\mbf G}$ is summarized in
Fig.~\ref{fig:ConvOfEinsteinNorm}, which reports the time evolution of
the $L_{\infty}$ norm of~\eqref{eq:EinsteinNorm} at the medium and
very-high resolutions. Also indicated with dashed and dotted lines are
the expression for the $L_{\infty}$ norm of~\eqref{eq:EinsteinNorm} at
the very-high resolution when rescaled for third (dotted line) and
fourth-order convergence (dashed line).

There is a period at the beginning of the evolutions where the initial
data construction prevents fourth-order convergence. This is due to
the fact that the initial data is computed by an interpolation of the
results of a spectral solver onto the finite difference grid which is
used for evolution. An error is introduced because we keep fixed the
number of spectral coefficients and because the Cartesian grid points
do not coincide with the spectral collocation points of the Chebyshev
polynomials, resulting in a certain amount of high-frequency noise
that spoils the convergence for some time at the beginning of the
simulation. Numerical dissipation and the constraint damping built
into the evolution system implies that the evolution quickly adjusts
itself to actually solving the Einstein equations to a good
accuracy. The effects of these initial transient modes can last for
different amounts of time for the different resolutions,
\textit{e.g.}, $\sim 10\,M$ for the medium resolution and $\sim 30\,M$
for the very-high resolution.

Soon after this transient has disappeared, the code shows the expected
fourth-order convergence, with the largest values of the violation
found in the vicinity of the apparent horizons, where the gradients in
the metric are the steepest. The violations grow rapidly with time as
the binary inspirals and the largest values of the violation of the
Einstein tensor are seen at the time of the merger, $t\approx109\,M$,
with values as large as ${\cal O}(300)$. Such violations are
essentially confined to a \textit{single} grid point on the trailing
edge of the apparent horizon and are produced by the very steep
gradients in the shift. Clearly, violations of this magnitude would
not be revealed when looking at the $L_2$ norms and are a source of
concern. However, as we will show later, such violations do not
propagate away from the horizon to affect the fourth-order convergence
of the waveforms.

At the time of the merger the excision of a common apparent horizon from
the calculation of the $L_\infty$ norm is responsible for the decrease
by about four orders of the violation. After this, the $L_{\infty}$ do
not grow further in time for the very-high resolution simulation,
while a modest increase is seen in the simulation run at medium
resolution.  During this time the code shows a convergence which is
between third-order (right after the merger) and fourth-order (during
the ringdown).

In addition to convergence in the Einstein tensor, we also validate
the correctness of the physically relevant information contained in
the waveforms. We do this by computing convergence rate of the
waveforms $Q^+_{22}$, $Q^+_{33}$, and $Q^{\times}_{21}$ using the ratio
of the integrated differences between the medium and high resolutions,
and the high and very-high resolutions
\begin{equation}
\rho(Q) \equiv
\frac{
\sqrt{\int_{u_1}^{u_2} | Q_{0.024} -  Q_{0.018} |^2 du }
}{
\sqrt{\int_{u_1}^{u_2} | Q_{0.018} -  Q_{0.012} |^2 du }
}\,,
\label{eq:rhodef}
\end{equation}
where $u \equiv t-r_{_{\rm E}}$ is the retarded time at a given
detector, $Q$ stands for either $Q^+_{22}$, $Q^+_{33}$ or
$Q^{\times}_{21}$ and refers to either its amplitude or the phase. As
indicated in Eq.~\eqref{eq:rhodef}, the integrals are evaluated over
the retarded interval $[u_1, u_2]$ which does not include the initial
spurious burst of radiation (which we do not expect to converge) but
contains otherwise the complete waveform including the ringdown.

Assuming a truncation error ${\cal O}(h^p)$ and that the coefficient
of this error does not depend on resolution, the function $\rho$
becomes to leading order
\begin{equation}
\rho = \frac{(h_{0.024})^p - (h_{0.018})^p}{(h_{0.018})^p
 - (h_{0.012})^p}\,,
\label{eq:rhoeq}
\end{equation}
where $h_{0.024} = 0.024\,M$ and we we underline the importance of
having used a full doubling of the resolution between the smallest and
largest resolution to improve the accuracy of this estimate over more
narrowly spaced resolution steps. In practice, we measure $\rho$ and
then solve for the \textit{``effective''} convergence order $p$ using
Eq.~\eqref{eq:rhoeq}. A discussion of the details in this procedure
are presented in Appendix~\ref{appendix_conv} alongside with the
computed convergence rates for the amplitudes and phases of $Q$ which
are found to between two ($\ell=3$) and four ($\ell=2$) (\textit{cf.},
Table~\ref{tbl:Qconv}).

It should be noted that the above definition of convergence rate
naturally results in non-integer values for the exponent $\rho$, even
though our methods are explicitly polynomial.  This is because the
derivation of~(\ref{eq:rhoeq}) assumes a coefficient of one in the
leading order error term that extrapolates between the resolutions. If
the coefficient is in practice different for a given set of
resolutions, then a non-integer value results which is larger if the
coefficient is smaller. As such, values obtained in this way should
not be considered literal polynomial extrapolation orders. By
``convergence order 3.8'' we rather mean that our results are
consistent with third-order finite differencing where the leading
third-order error coefficient is quite small so that at the given
resolutions the convergence appears to be closer to a fourth-order
approximation. Very high convergence exponents are a likely indication
that the lowest resolution is not in the convergent regime for the
measured quantity. Non-integer convergence orders obtained in this way
are resolution dependent, and should themselves converge to the lowest
order finite difference approximation used in the code in the limit of
infinite resolution.

\begin{figure}[t]
\begin{center}
  \centerline{
    \resizebox{8.5cm}{!}{\includegraphics[angle=-0]{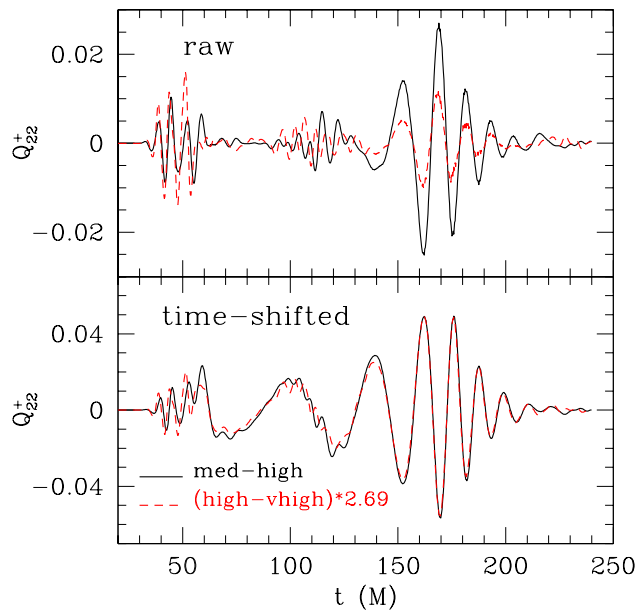}
  }}
  \vskip 0.5cm
  \caption{Convergence of the fiducial waveform $Q^+_{22}$ for the
    binary system $r0$ before and after the time-shift defined in
    Eqs.~\eqref{eq:timeshiftingdef}--\eqref{eq:timeshiftingval}. In
    the upper graph we show the difference between $Q^{+}_{22}$ when
    computed at different resolutions, scaled for fourth-order
    convergence and using raw data (\textit{i.e.}, without
    time-shifting). The overlap between the curves is rather poor
    indicating an over-convergence (\textit{i.e.}, the truncation
    error appears to be smaller than expected). In the lower panel we
    show the same data but after time-shifting. The very good overlap
    of the scaled curves on the indicates that the time-shifting is
    essential for obtaining properly scaling differences between runs
    of various resolutions.}
  \label{fig:convReQ22}
\end{center}
\end{figure}
\begin{figure}[t]
\begin{center}
  \centerline{
    \resizebox{8.5cm}{!}{\includegraphics[angle=-0]{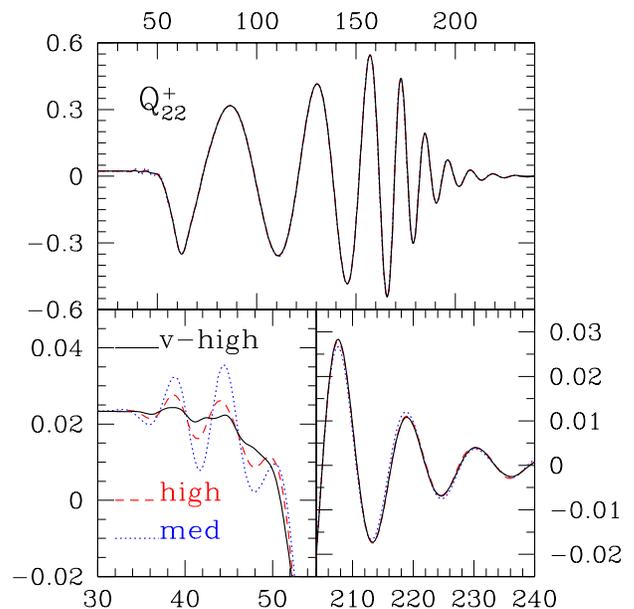}
  }}
  \vskip 0.5cm
  \caption{Accuracy of the fiducial waveform $Q^+_{22}$ for the binary
    system $r0$. In the upper graph we show the waveforms at the three
    different resolutions: very-high (continuous line), high (dashed
    line), medium (dotted line). The accuracy is very good already
    with the lowest resolution and the curves cannot be
    distinguished. The lower panels show magnifications of some
    relevant portions of the waveform, with the lower-left panel
    concentrating on the initial transient radiation produced by the
    truncation error. The lower-right panel, on the other hand, refers
    to the quasi-normal ringing and shows that it is well-captured at
    all resolutions.}
  \label{fig:accuracyReQ22}
\end{center}
\end{figure}

An important property of the waveforms which has emerged when
performing these convergence tests is that the dominant source of
error is a de-phasing which causes the lower resolution evolutions to
``lag'' behind the higher resolution. This delay is usually rather
small and between $0.1\,M$ and $0.5\,M$, but it is clearly visible
when comparing the total amplitude of $Q$ as a function of time. The
most important consequence of this error is that it can spoil the
convergence tests if not properly taken into account: the residuals
errors seem, in fact, to indicate over-convergence. This is shown in
the upper panel of Fig.~\ref{fig:convReQ22}, which reports the
differences between $Q^{+}_{22}$ when computed at different
resolutions scaled for fourth-order convergence. Clearly the overlap
is rather poor and even indicating that the truncation error is
smaller than expected. This is obviously an artifact of the near
cancellation of the lowest-order terms in the truncation error and
induced by the small time-differences at different resolutions.

We remove this effect by shifting the time coordinate of the medium
and high resolution runs by the time interval needed to produce an
alignment of the maxima of the emitted radiation. Details on how to
do this are discussed in Appendix~\ref{appendix_conv}, and we report
in the lower panel of Fig.~\ref{fig:convReQ22} the same data shown in
the upper panel, but after the time-shifting. Clearly, the overlap is
now extremely good suggesting that the time-shifting is essential for
obtaining the expected fourth-order convergence in the waveforms.
In accord with the convergence in the waveforms we
also see  fourth order convergence in the final kick value.   

As a final note we remark that besides validating a proper convergence
of the code, it is also important to assess the accuracy of any
measurable quantity at the relevant resolutions considered here. As a
representative and physically meaningful quantity we have considered
the accuracy of the fiducial waveform $Q^+_{22}$ for the binary
system $r0$. This is shown in Fig.~\ref{fig:accuracyReQ22}, where in
the upper graph we report the waveforms at the three different
resolutions: very-high (continuous line), high (dashed line) and medium (dotted
line). Already with the lowest of these resolutions the accuracy is sufficiently high
so that the curves are essentially
indistinguishable from each other by eye. The lower panels show
magnifications of the relevant portions of the waveform, with the
lower-left panel concentrating on the initial transient radiation
produced by the truncation error. The latter clearly is rather large
at the medium resolution, but it nicely converges away when 
the grid spacing is decreased. The lower-right panel, on the other hand, refers to
the quasi-normal ringing and shows that it is well-captured at all
resolutions.

\section{Linear momentum of black hole spacetimes}
\label{sec:kick_calculation}

In radiating spacetimes where the radiation is emitted asymmetrically,
there will be a net linear momentum imparted to the system. In particular,
in the case of a binary black hole merger, the final black hole
receives a ``kick'' which causes it to move off at a given velocity.
This velocity can be determined by an analysis of the emitted
radiation. In ADM-type numerical simulations, this is typically
done by evaluating some scalar quantity which can be associated
with the wave energy at some large radius within
the computational domain. The chosen radius needs to be large
enough that it is in the ``wave zone'', where non-linear
self-interaction of the gravitational field is negligible and
the waves can be picked out as perturbations of a background.

Two methods have become commonplace to determine the emitted wave
energy. The first uses the Newman-Penrose curvature scalar $\Psi_4$,
which can be identified with the gravitational radiation if a suitable
frame is chosen at the extraction radius. An alternative method
measures the metric of the numerically generated spacetime against a
fixed background at the extraction radius, and determines the
Zerilli-Moncrief perturbation modes. Both methods yield
 data for the gravitational wave energy which can be
integrated to determine a net linear momentum, as described in more
detail in the following sections.

\subsection{Kick measurements via $\Psi_4$}

\begin{figure}[t]
\begin{center}
  \centerline{
    \resizebox{8.5cm}{!}{\includegraphics[angle=-0]{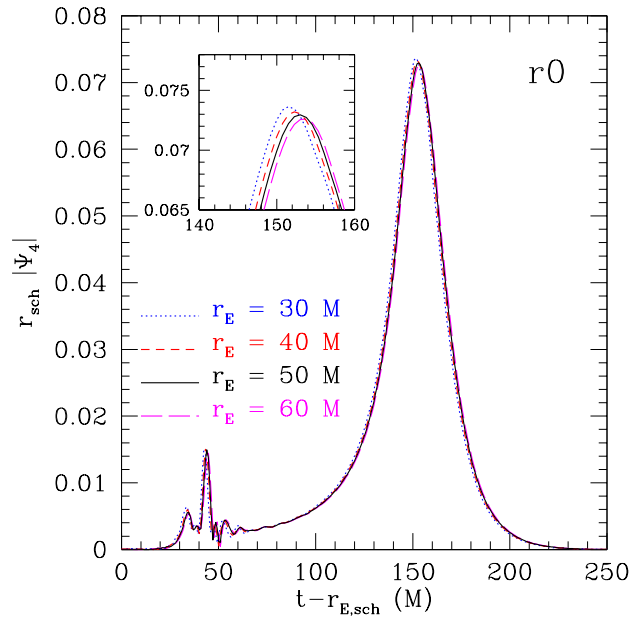}
  }}
  \vskip 0.5cm
  \caption{Amplitude of $r_{_{\rm E, sch}} |\Psi_4|$ for extraction spheres
    at $r_{_{\rm E}}=30\,M$, $40\,M$, $50\,M$ and $60\,M$, demonstrating that
    $\Psi_4$ does indeed fall off as required by the peeling
    property. There is a slight decrease in amplitude with larger
    radius, suggesting that dissipative effects may become important
    at larger radii.  Results in this paper use waveforms from the
    $r_{_{\rm E}}=50\,M$ extraction sphere, unless indicated otherwise.}
  \label{fig:y0_peeling_psi4}
\end{center}
\end{figure}

The Newman-Penrose formalism provides a convenient representation
for a number of radiation related quantities as spin-weighted scalars.
In particular, the curvature component
\begin{equation}
  \Psi_4 \equiv -C_{\alpha\beta\gamma\delta}
    n^\alpha \bar{m}^\beta n^\gamma \bar{m}^\delta,
  \label{eq:psi4def}
\end{equation}
is defined as a particular component of the Weyl curvature,
$C_{\alpha\beta\gamma\delta}$, projected onto a given null frame,
$\{\boldsymbol{l}, \boldsymbol{n}, \boldsymbol{m},
\bar{\boldsymbol{m}}\}$. In practice, we define an orthonormal basis
in the three space $(\hat{\boldsymbol{r}}, \hat{\boldsymbol{\theta}},
\hat{\boldsymbol{\phi}})$, centered on the Cartesian grid center and
oriented with poles along $\hat{\boldsymbol{z}}$. The normal to the
slice defines a time-like vector $\hat{\boldsymbol{t}}$, from which we
construct the null frame
\begin{equation}
   \boldsymbol{l} = \frac{1}{\sqrt{2}}(\hat{\boldsymbol{t}} - \hat{\boldsymbol{r}}),\quad
   \boldsymbol{n} = \frac{1}{\sqrt{2}}(\hat{\boldsymbol{t}} + \hat{\boldsymbol{r}}),\quad
   \boldsymbol{m} = \frac{1}{\sqrt{2}}(\hat{\boldsymbol{\theta}} - 
     {\mathrm i}\hat{\boldsymbol{\phi}}) \ .
\end{equation}
We then calculate $\Psi_4$ via a reformulation of (\ref{eq:psi4def}) 
in terms of ADM variables on the slice~\cite{Shinkai94},
\begin{equation}
  \Psi_4 = C_{ij} \bar{m}^i \bar{m}^j,  \label{eq:psi4_adm}
\end{equation}
where
\begin{equation}
  C_{ij} \equiv R_{ij} - K K_{ij} + K_i{}^k K_{kj} 
    - {\rm i}\epsilon_i{}^{kl} \nabla_l K_{jk}.
\end{equation}
The identification of the Newman-Penrose $\Psi_4$ with the
gravitational radiation content of the spacetime is a result of
the peeling theorem, which states that in an appropriate frame
the $\Psi_4$ component of the curvature has the slowest falloff
with radius, $\mathcal{O}(1/r)$. The conditions of this theorem
are not satisfied exactly at a small radius and in the chosen
frame. While there are proposals for how this situation can
be improved~\cite{Lehner:2007ip}, we find that beyond $r_{_{\rm E}} \geq 30\,M$
in fact our measure of $\Psi_4$ scales extremely well with the different
extraction radii $r_{_{\rm E}}$, suggesting that the peeling property
is satisfied to a reasonable approximation (see
Fig.~\ref{fig:y0_peeling_psi4}).

The gravitational wave polarization amplitudes $h_+$ and $h_\times$
are related to $\Psi_4$ by~\cite{Teukolsky73}
\begin{equation}
\ddot{h}_+ - {\rm i}\ddot{h}_{\times}=\Psi_4 \ ,
\label{eq:psi4_h}
\end{equation}
where the double over-dot stands for second-order time derivative.
The flux of linear momentum emitted in gravitational waves in the
$i$-direction can be computed from the Isaacson's energy-momentum 
tensor and can be written in terms of the two polarization amplitudes
as~\cite{Favata:2004wz}
\begin{equation}
  \label{eq:linmom_flux}
  {\cal F}_i\equiv \dot{P_i}=\dfrac{r^2}{16\pi}\int d\Omega\;
   n_i\left(\dot{h}_+^2+\dot{h}_{\times}^2\right) \ ,
\end{equation}
where $n_i=x_i/r$ is the unit radial vector that points from the
source to the observer and $d\Omega=\sin\theta d\phi d\theta$ is the
line element of our extraction 2-sphere $S^2$. Using
Eq.~(\ref{eq:psi4_h}), this leads to an expression for the momentum
flux in terms of $\Psi_4$ as it is commonly used in recent numerical
relativity calculations~\cite{Campanelli99, Baker:2006nr,
Gonzales06tr, Herrmann:2007ac, Campanelli:2007cg, Gonzalez:2007hi,
Koppitz:2007ev}:
\beq
\label{eq:Pdot}
{\cal F}_i=\lim_{r_{\rm sch} \to \infty} \left\{ \frac{r_{\rm sch}^2}{16\pi}\int
d\Omega\; n_i \left| \int_{-\infty}^t dt \Psi_4  \right|^2
\right\} \;.
\eeq
The Schwarzschild radius, $r_{\rm sch}$, is derived from the
coordinate (isotropic) radius via the standard formula 
\begin{equation}
  r_{\rm sch} = r_{\rm iso} \left( 1 - \frac{M}{2r_{\rm iso}} \right)^2.
\end{equation}
assuming a constant ADM mass $M=M_{_{\mathrm{ADM}}}$ throughout the
simulation. With this choice of radial coordinate,
expression~(\ref{eq:Pdot}) has been shown to provide recoil velocities
which are in better agreement with those obtained through
gauge-invariant perturbations than with the alternative coordinate
radius, (\textit{cf.}  Sect.~\ref{kick_via_gips}) and reported in the
literature (Additional details on the numerical measurement of
$\Psi_4$ are presented in Appendix~\ref{appendix_a}.)

\subsection{Kick measurements via gauge-invariant perturbations}
\label{kick_via_gips}

An independent method to compute the linear momentum carried away by
gravitational radiation is based on the measurements of the
non-spherical gauge-invariant perturbations of a Schwarzschild black
hole (see Refs.~\cite{Allen98a1,Rupright98,Camarda97c} for applications
to Cartesian coordinates grids). In practice, a set of ``observers''
is placed on 2-spheres of fixed coordinate radius $r_{_{\rm E}}$, where
they extract the gauge-invariant, odd-parity (or {\it axial}) current
multipoles $Q_{\ell m}^\times$ and even-parity (or {\it polar}) mass
multipoles $Q_{\ell m}^+$ of the metric
perturbation~\cite{Moncrief74}. The numerical implementations of the
gauge-invariant variables is done by following the multipolar analysis
outlined by Abrahams and Price~\cite{Abrahams95b}.  The $Q^+_{\ell m}$
and $Q^\times_{\ell m}$ variables are related to $h_+$ and $h_\times$
as~\cite{Nagar05}
\begin{eqnarray}
\label{eq:wave_gi}
&&\hskip -0.5cm 
h_+-{\rm i}h_{\times} =
  \dfrac{1}{\sqrt{2}r}\sum_{\ell=2}^{\infty}\sum_{m=-\ell}^{\ell}
  \Biggl( Q_{\ell m}^+ \nonumber \\
&& \hskip 2.5cm 
        -{\rm i}\int_{-\infty}^t Q^\times_{\ell
          m}(t')dt' \Biggr)\,_{-2}Y^{\ell m}\ .
\end{eqnarray}
Here $_{-2}Y^{\ell m}$ are the $s=-2$ spin-weighted 
spherical harmonics and $(\ell, m)$  are the indices 
of the angular decomposition. Validations of this approach in
3D vacuum spacetimes can be found in 
Refs.~\cite{Camarda97c,Rezzolla99a,Baker99a}, while its use with
matter sources has first been reported in~\cite{Font01b}.

We note that the notation introduced in Eq.~(\ref{eq:wave_gi}) could
be misleading as it seems to suggest that $h_\times$ is always of
odd-parity and $h_+$ is always of even-parity. Indeed this is not true
in general and in the absence of axisymmetry, i.e., when
$m\neq0$, both $h_{\times}$ and $h_+$ are a superposition of odd and
even parity modes. It is only for axisymmetric systems, for which only
$m=0$ modes are present, that $Q^\times_{\ell m}$ and $Q^+_{\ell m}$
are {\it real} numbers, that $h_+$ is {\it only} even-parity and
$h_\times$ is {\it only} odd-parity. Despite this possible confusion,
we here prefer to maintain the notation of Eq.~(\ref{eq:wave_gi})
which is the most common in the literature~\cite{Nagar05}.

The flux of linear momentum emitted in gravitational waves in terms of
$Q^+_{\ell m}$ and $Q^\times_{\ell m}$ can be computed by inserting
Eq.~(\ref{eq:wave_gi}) in Eq.~(\ref{eq:linmom_flux}), then decomposing
$n_i$ in spherical harmonics and performing the angular integral. 
This procedure goes along the lines discussed by Thorne in
Ref.~\cite{Thorne80b}, where all the relevant formulae are 
essentially available [\textit{cf.} Eq.~(4.20) there. 
See also Ref.~\cite{Sopuerta:2006wj}], so that we only need to adapt 
them to our notation. In Ref.~\cite{Thorne80b} 
the even-parity (or {\it electric}) multipoles are indicated with
$I_{\ell m}$ and the odd-parity (or {\it magnetic}) ones with  
$S_{\ell m}$. They are related to our notation by 
\begin{align}
^{(\ell)}I_{\lm}   &= Q^+_{\ell m} \ , \\
^{(\ell+1)}S_{\lm} &= Q_{\ell m}^\times\;,
\end{align}
where $^{(\ell)}f_{\lm}\equiv d^{\ell} f_{\ell m}/dt^{\ell}$.  From the
well known property $(Q^{+,\times}_{\lm})^* =
(-1)^mQ_{\ell\,-m}^{+,\times}$, where the asterisk indicates complex
conjugation, one can rewrite Eq.~(4.20) of Ref.~\cite{Thorne80b} in a
more compact form.  Following Ref.~\cite{Damour-Gopakumar-2006} where
the lowest multipolar contribution was explicitly computed in this
way, it is convenient to combine the components of the linear momentum
flux in the equatorial plane in a complex number as ${\cal F}_x+{\rm
i}{\cal F}_y$.  The multipolar expansion of the flux vector can be
written as
\begin{align}
\label{eq:recoil}
{\cal F}_x+{\rm i}{\cal F}_y &=
\sum_{\ell=2}^{\infty}\sum_{m=0}^{\ell}\delta_m\left({\cal
  F}_{x}^{\ell m} + {\rm i}{\cal F}_y^{\ell m}\right) \ , \\
{\cal F}_z &= \sum_{\ell=2}^{\infty}\sum_{m=0}^{\ell}\delta_m {\cal
  F}_{z}^{\ell m} \ ,
\end{align}
where $\delta_m=1$ if $m\neq0$ and $\delta_m=1/2$ if $m=0$.
Each multipole reads
\begin{widetext}
  \begin{align}
    \label{eq:gi}
&{\cal F}_x^{\ell m}+{\rm i}{\cal F}_y^{\ell m} \equiv
  \dfrac{(-1)^m}{16\pi\ell(\ell +1)}\Bigg\{ -2{\rm i}\bigg[a_{\ell
  m}^+ \dot{Q}^+_{\ell-m}Q_{\ell\,m-1}^\times +a_{\ell
  m}^-\dot{Q}^+_{\ell m} Q^\times_{\ell\;-(m+1)}\bigg] \nonumber \\
 & +\sqrt{\dfrac{\ell^2(\ell-1)(\ell+3)}{(2\ell+1)(2\ell+3)}}
   \bigg[ 
    b_{\ell m}^- \left(\dot{Q}^+_{\ell\;-m}\dot{Q}^+_{\ell+1\;m-1} +
    Q_{\ell\;-m}^\times\dot{Q}_{\ell +1\;m-1}^\times \right)
    +b_{\ell m}^+\left(\dot{Q}^+_{\ell m}\dot{Q}^+_{\ell
        +1\;-(m+1)}+Q_{\ell
        m}^\times\dot{Q}^\times_{\ell+1\;-(m+1)}\right)\bigg]
  \Bigg\} \ , \\
\label{eq:kick_z}
&{\cal F}^{\ell m}_z 
 \equiv\dfrac{(-1)^m}{8\pi\ell(\ell+1)}
\bigg\{2m\;{\rm Im}\left[\dot{Q}_{\ell\,-m}^+Q_{\ell m}^{\times}\right]
+c_{\ell m}\sqrt{\dfrac{\ell^2(\ell-1)(\ell+3)}{(2\ell+1)(2\ell+3)}}
{\rm Re}\left[\dot{Q}_{\ell\,-m}^+Q^+_{\ell+1\,m}+Q^\times_{\ell\,-m}\dot{Q}^\times_{\ell+1\,m}\right]\bigg\} \ ,
\end{align}
\end{widetext}
and 
\begin{eqnarray}
&& a_{\ell m}^{\pm}\equiv\sqrt{(\ell \pm m)(\ell \mp m+1)} \ , \\
&&b_{\ell m}^{\pm}\equiv\sqrt{(\ell \pm m+1)(\ell \pm  m+2}) \ , \\
&&c_{\ell m}\equiv\sqrt{(\ell - m+1)(\ell - m+1}) \ .
\end{eqnarray}
Note that here both ${\cal F}_x^{\ell m}$ and ${\cal F}_y^{\ell m}$
are {\it real} numbers and are obtained as the real and imaginary part
of the right-hand-side of Eq.~(\ref{eq:gi}).  For a general system
without symmetries one is expecting ${\cal F}_z^{\ell m}$ to be
nonzero. However, our initial data set-up, an inspiraling binary with
spins anti-aligned and parallel to the orbital angular momentum,
implies that the linear momentum flux vector is completely contained
in the equatorial plane of the system and so that ${\cal F}_z^{\ell
m}=0$ by construction. Since we are imposing equatorial symmetry
(\textit{i.e.}, invariance for $\theta\rightarrow\pi-\theta$) we have
that multipoles with $\ell+m= even$ are purely even-parity
(\textit{i.e.}, $Q^+_{\ell m}\neq0$ and $Q^\times_{\ell m}=0$ ) and
those with $\ell+m=odd$ are purely odd-parity (\textit{i.e.},
$Q^+_{\ell m}=0$ and $Q^\times_{\ell m}\neq0$). As a final remark, we
note that for $\ell=m=2$, our Eq.~(\ref{eq:gi}) reduces to Eq.~(9) of
Ref.~\cite{Damour-Gopakumar-2006}.

\section{Results}
\label{sec:results}

This section collects the results of our analysis of the recoil
velocity of spin-aligned binaries and discusses the different aspects
of the study which combined provide a consistent and accurate picture
of this process. We will first concentrate on the systematic error
introduced by the use of initial data with zero linear momentum and on
the techniques we have developed to remove it. We will then discuss
the actual computation of the recoil velocities and their dependence
on the spin ratio, highlighting the modes of the radiation
which are largely responsible for the asymmetric emission. Finally, we
will discuss the accuracy of our measurements and our ability to
preserve mass and angular momentum to below $1\%$.

\begin{figure}[t]
\begin{center}
  \centerline{
    \resizebox{8.5cm}{!}{\includegraphics[angle=-0]{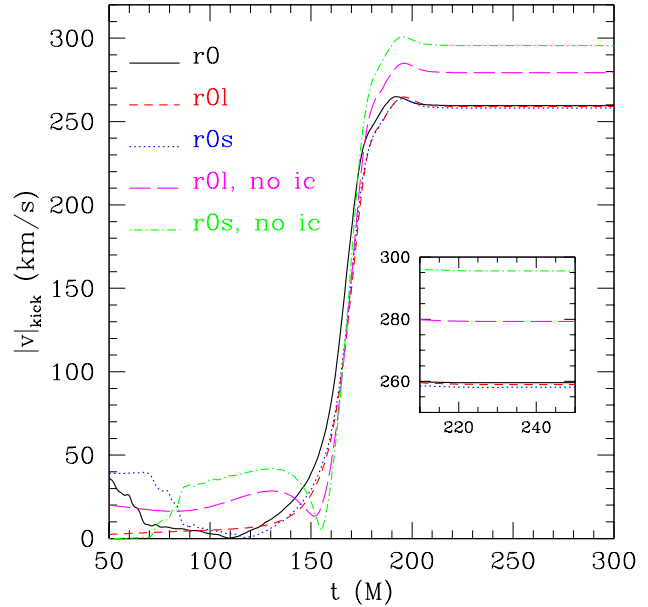}
  }}
  \vskip 0.5cm
  \caption{The recoil velocity of the binary $r0$ is compared to those
    of the same system but with either a larger or a smaller initial
    separation (\textit{i.e.}, $r0l$ and $r0s$, respectively). Note
    the same recoil velocity is obtained when the integration constant
    is properly taken into account, while an error as large as $\sim
    13\%$ is made otherwise.}
  \label{fig:y0_vs_y0x}
\end{center}
\end{figure}

\subsection{Initial transients in the  waveforms}
\label{sbsc:integr_const}

Both Eqs.~(\ref{eq:Pdot}) and~(\ref{eq:gi}) provide an expression of
the recoil velocity in terms of the radiated (linear) momentum per
(infinitesimal) time interval. A time-integration of those equations
is needed in order to compute the recoil and this obviously opens the
question of determining an integration constant which is in practice a
vector. Fortunately, this integration constant has here a clear
physical meaning and it is therefore easy to compute. In essence it
reflects the fact that at the time the simulation is started, the
binary system has already accumulated a non-vanishing net momentum as
a result of the slow inspiral from an infinite separation.

Since the initial data is constructed so as to have a vanishing linear
momentum, there will be a inconsistency between this assumption and
the actual evolution of the initial data. Stated differently, the
numerical evolution of the Einstein equations will soon tend to a
spacetime which is different from the initial one and indeed
corresponding to one with a net linear momentum. This momentum is the
one that the binary has gained when inspiralling from $t=-\infty$ till
$t=0$. Calculating the integration constant amounts therefore to
computing the vector accounting for this mismatch and is essential for
a correct measurement of the recoil velocity. The error made when
neglecting this constant, as routinely done in numerical-relativity
calculations, inevitably produces a systematic deviation from the
correct answer and, as we will show in the next section, it can
altogether prevent from having even the qualitative behavior right.

\begin{figure*}
\begin{center}
  \centerline{
    \resizebox{8.5cm}{!}{\includegraphics[angle=-0]{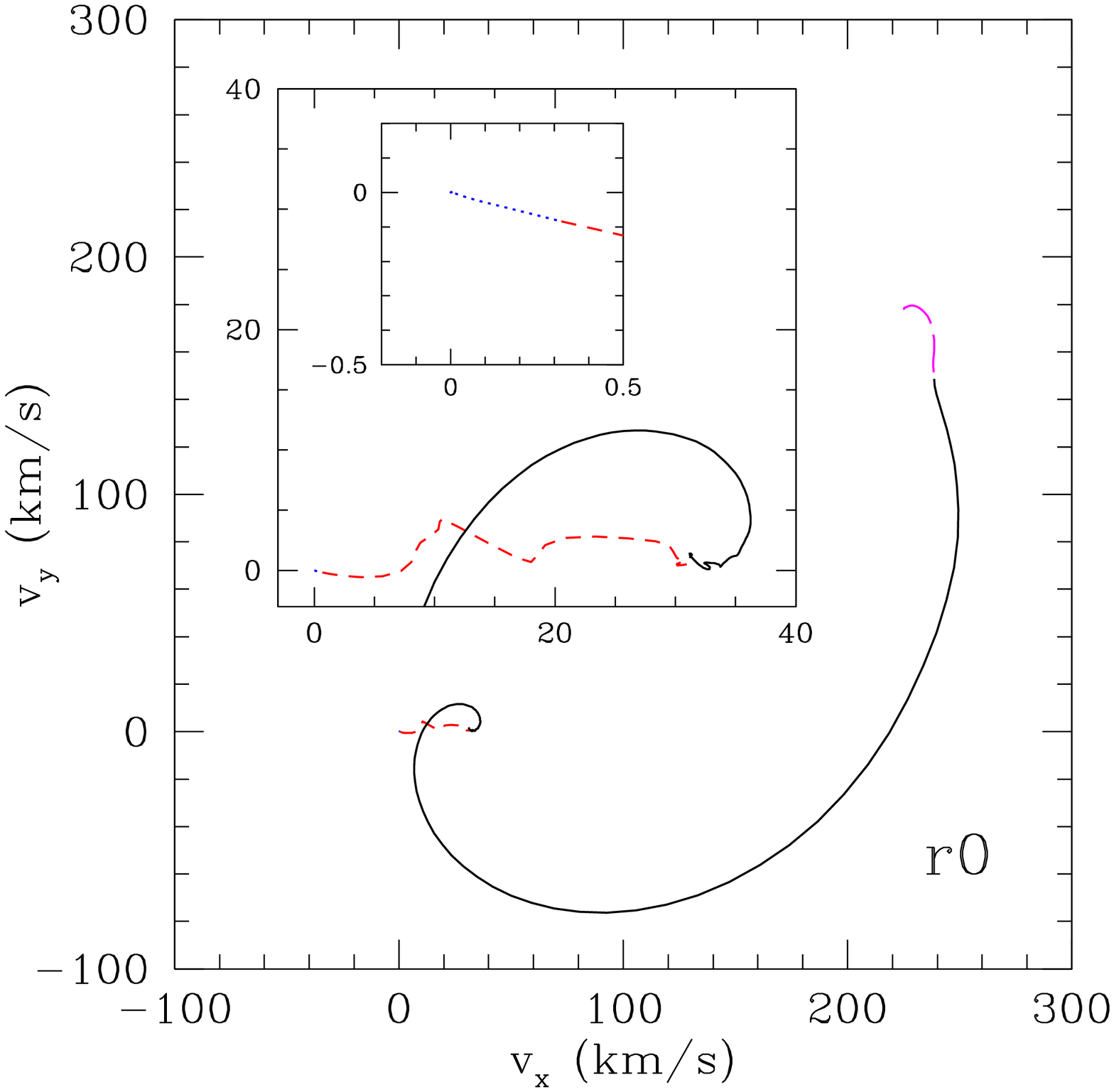}}
    \hskip 0.5cm
    \resizebox{8.5cm}{!}{\includegraphics[angle=-0]{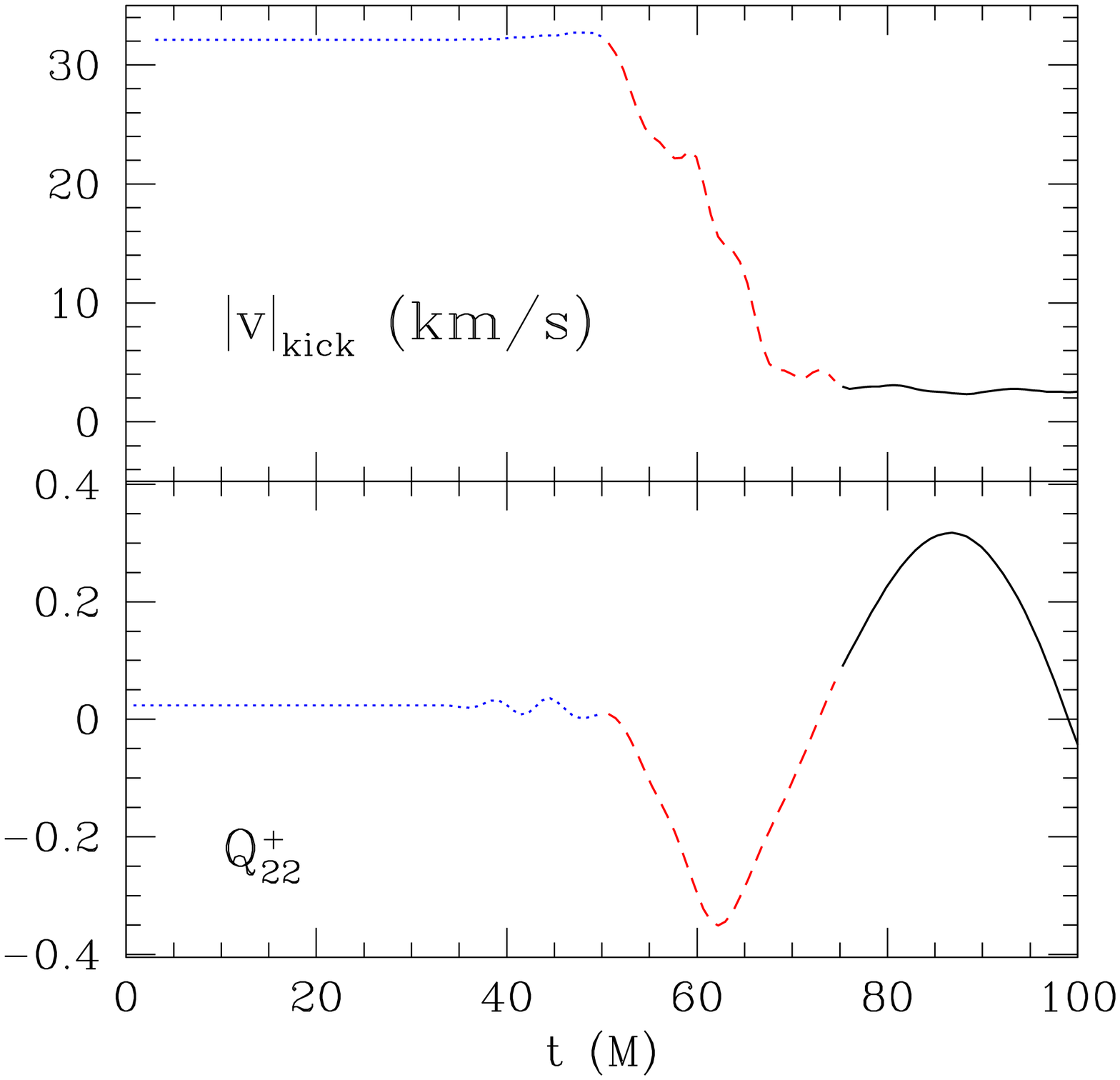}}
  }
  \vskip 0.5cm
  \caption{\textit{Left panel:} Evolution in velocity space of the
    recoil-velocity vector. Very little variation is recorded before
    the radiation reaches the observer at $r_{_{\rm E}} = 50\,M$
    (dotted lines in the two insets). The absence of the proper linear
    momentum in the initial data triggers a rapid and an almost
    straight-line motion (dashed line) of the center of the spiral
    away from the origin of coordinates during the initial stages of
    the evolution. After this transient motion, the evolution is
    slower, with the spiral progressively opening up (solid line). The
    vector to the center of the spiral corresponds to the initial
    linear momentum of the spacetime and is used as integration
    constant for Eqs.~(\ref{eq:Pdot}) and~(\ref{eq:gi}). The final
    part of the evolution is characterized by a change in the spiral
    pattern (long-dashed line) as a result of the interaction of
    different modes in the ringdown of the final black hole. Note that
    the figure has been rotated clockwise of about $30^{\circ}$ to
    allow for the two insets. \textit{Right panel:} Initial behavior
    of the recoil velocity (upper graph) and of the waveform
    ($Q^{+}_{22}$) for model $r0$ (lower graph). This figure should be
    compared with the initial vector evolution of the recoil velocity
    shown in the left panel where the same types of lines have been
    used for the different stages of the evolution.}
  \label{fig:recoil}
\end{center}
\vskip 1.0cm
\begin{center}
  \centerline{
    \resizebox{8.5cm}{!}{\includegraphics[angle=-0]{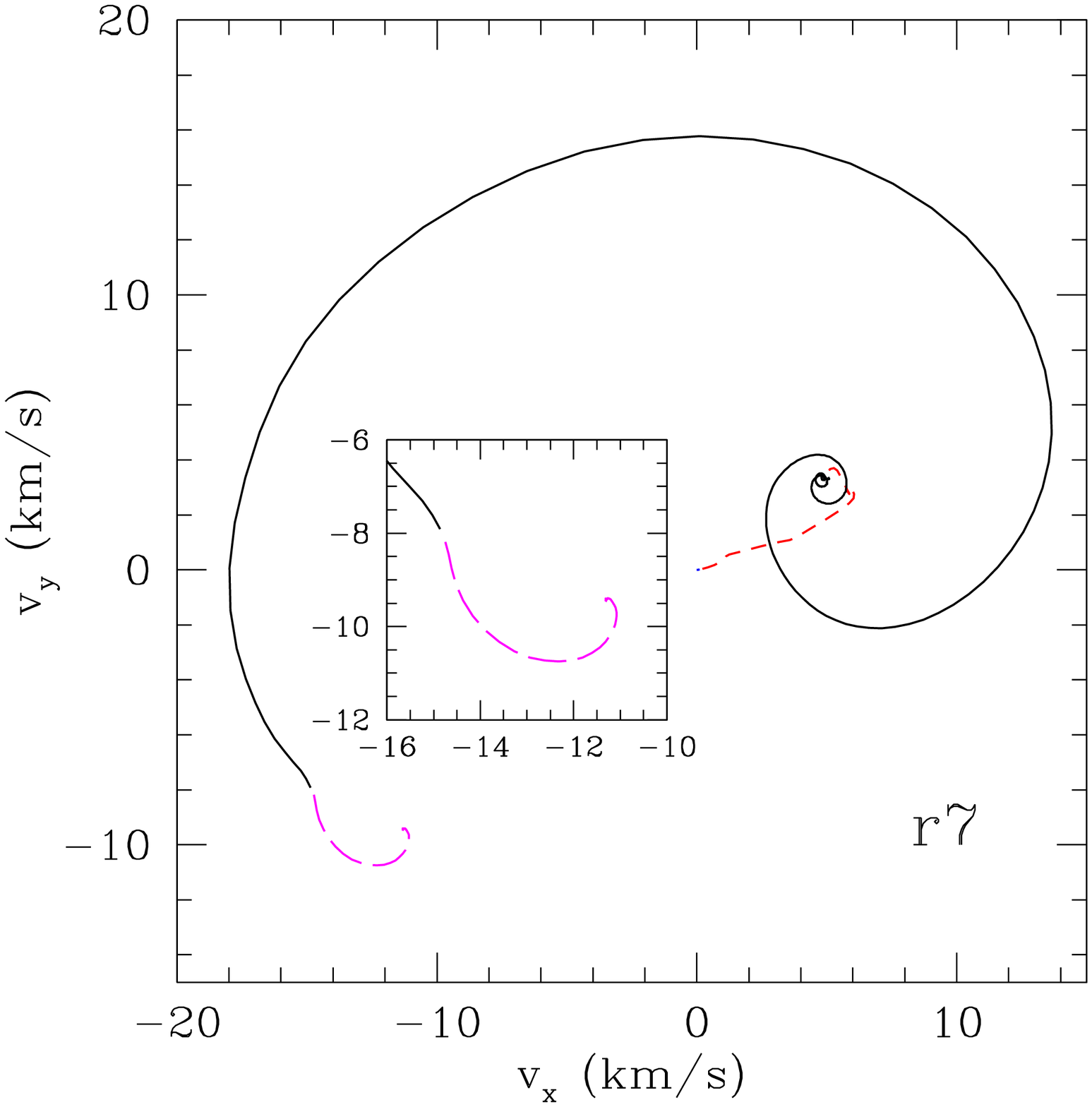}}
      \hskip 0.5cm
    \resizebox{8.5cm}{!}{\includegraphics[angle=-0]{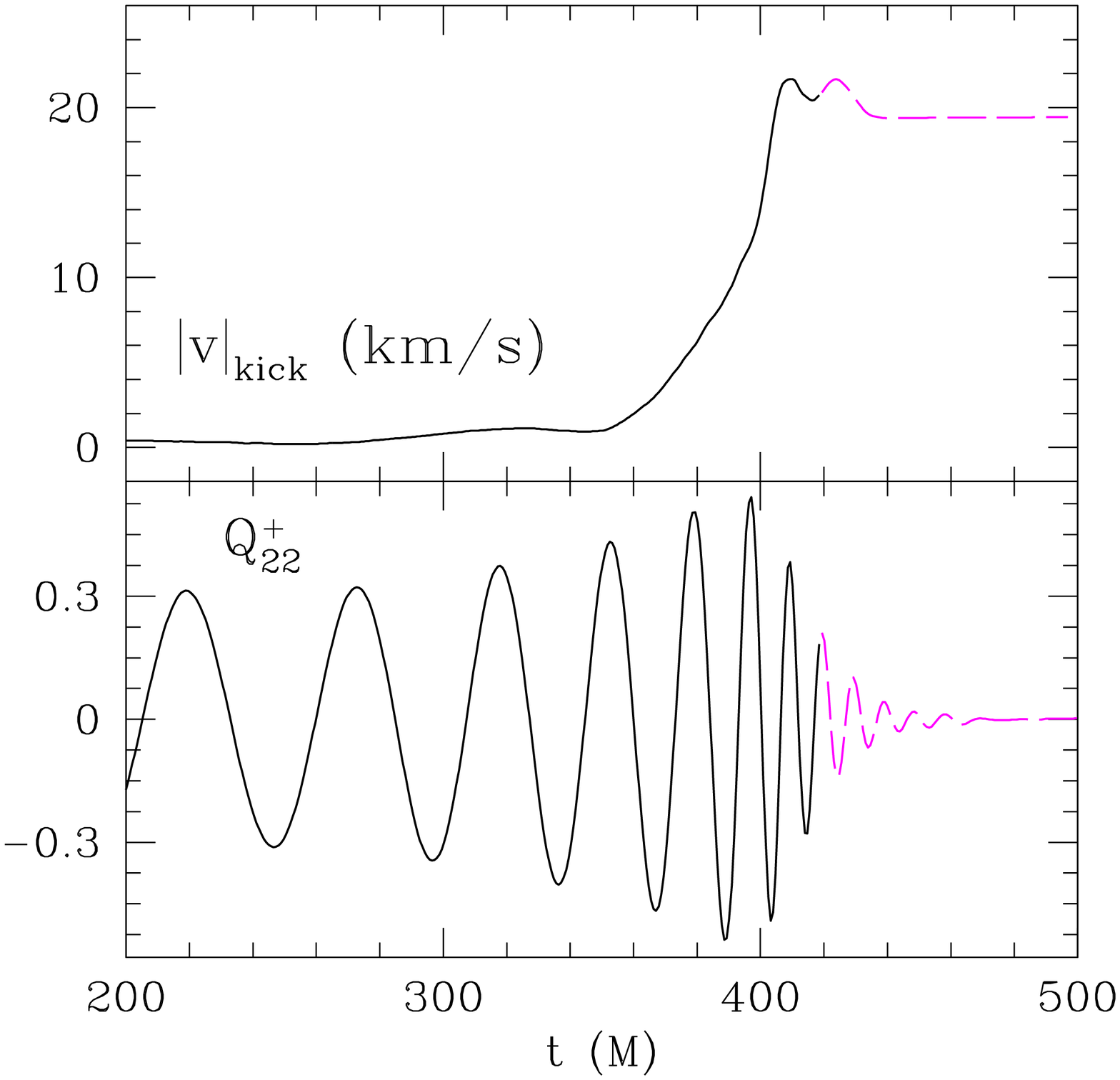}}
}
  \caption{\textit{Left panel:} The same as in the left panel of
    Fig.~\ref{fig:recoil} but for system $r7$. Shown in the inset is
    the sudden re-orientation of the recoil velocity vector during
    ringdown and corresponding to a new spiral with different aperture
    (long-dashed line). Although more pronounced in $r7$, the
    appearance of this ``hook'' at ringdown is seen all the members of
    the sequence. \textit{Right panel:} The same as in the left panel
    of Fig.~\ref{fig:recoil} but for system $r7$. The upper graph
    concentrates on the final stages of the evolution in of the recoil
    velocity and on the appearance of a second peak during ringdown
    (long-dashed curve). The lower graph shows the same but in terms
    of the $Q^{+}_{22}$ waveform. A discussion of these final stages
    of the evolution is made in Sect.~\ref{sec:ModeContribution}.}
  \label{fig:recoil_r7}
\end{center}
\end{figure*}

The relevance of this integration constant depends on the initial
separation and it is more important for binaries that start their
evolution already quite close. This is rather obvious: the tighter the
binary is, the larger the emitted momentum per unit time and the more
important is to evaluate the initial
mismatch. Fig.~\ref{fig:y0_vs_y0x} helps to illustrate this point and
can be discussed before entering into the details of how we actually
compute the integration constant. The figure shows
the time evolution of the recoil velocity $|v|_{\rm kick}\equiv \sqrt
{v^2_x + v^2_y}$ for the same binary system having spin ratio $a_1/a_2
= -1$ but with increasing initial separation. More precisely, we
consider systems $r0l$, $r0$ and $r0s$ which differ only in the
initial separation, which is about $8.4,\ 6.0$ and $5.6\,M$,
respectively. The data Fig.~\ref{fig:y0_vs_y0x} is properly shifted in
time so as to have the curves overlap and shows that \textit{only}
when the integration constant is properly taken into account, do the
three simulations yield the same recoil velocity (\textit{cf.}, solid,
dashed, and dotted lines). On the other hand, when the integration
constant is not included in the calculation, different evolutions will
yield different estimates, with a systematic error that can be as
large as $13\%$ (\textit{cf.}, long-dashed and dot-dashed lines) and
is clearly unacceptable given that the overall precision of the
simulations is below $1\%$ (\textit{cf.},
Figs.~\ref{fig:AngularMomentum}--\ref{fig:Masses} and the discussion
in Sect.~\ref{sec:AngularMomentum}).

Besides providing the right answer, the calculation of the integration
constant also results in a considerable saving in computational
costs. The complete dynamics of the binary $r0l$ including the merger
and ringdown, in fact, requires simulations for about $600\,M$; the
same answer in terms of recoil velocity can be obtained with the
system $r0s$, whose dynamics is fully accounted for with a simulation
lasting only for $340\,M$.

Having stressed the importance of including the integration constant
in the measurement of the recoil velocity, we next illustrate how to
actually compute it. In essence, it is sufficient to look carefully at
the evolution in the velocity-space of the two components $v_x$ and
$v_y$ of the recoil velocity (because of the symmetry the
$z$-component is zero but the method described here can be easily
extended to the case in which $v^z \neq 0$). This is shown in the left
panel of Fig.~\ref{fig:recoil}, which reports the track of the
``center of mass'' for system $r0$ in such a space. Different types of
line refer to different intervals in time during the evolution and,
for an observer at $r_{_{\rm E}} = 50\,M$, the dotted one refers to $t
\lesssim 50\,M$, the dashed one to $50\,M \lesssim t \lesssim 75\,M$,
the continuous one to $75\,M \lesssim t \lesssim 183\,M$, and finally
the long-dashed one to $t \gtrsim 183\,M$.

Clearly, for $t \lesssim 50\,M$ the system undergoes very little
evolution in velocity-space (\textit{cf.}, dotted line in the inset
within the inset of the left panel) but a rapid change, lasting for
about $25\,M$, takes place as the radiation reaches the observer. The
radiation received has information about the ``correct'' linear
momentum of the spacetime which is solution of the Einstein equations
for system $r0$ as if it had inspiraled from infinity, and thus
rapidly moves the center of mass to a net nonzero recoil velocity
(\textit{cf.}, almost-straight dashed line in the inset in the left
panel). Once the system has adjusted for the proper linear momentum,
the evolution proceeds as expected, with the recoil velocity vector
slowly tracking a spiral in velocity space. This is an important point
which we prefer to underline: the rate of change of linear momentum is
very large only initially and this is because as the binary migrates
from the initial non-radiating state (the data is conformally flat) to
the consistent radiating state, it will emit the amount of linear
momentum it would have emitted when inspiralling from infinite
separation. After this burst of linear momentum, the evolution of the
recoil velocity is minute, essentially until it grows very rapidly
during the last orbit.

Computing the integration constant consists then in calculating the
position of the center of the spiral and this can be done either by a
simple inspection of a graph in the velocity-space, from which compute
the center of the spiral or, equivalently, by searching for the
initial vector that would lead to an essentially \textit{monotonic} in
time growth of the recoil velocity\footnote{The presence of a small
eccentricity prevents from a strict monotonicity of the recoil
velocity for binaries starting from a large separation. In this case,
very small oscillations appear over the orbital timescale.}. The
latter procedure does not require a human judgment but we have found
it to yield the same answer (to less than $1\,\mathrm{km/s}$) as the
one guessed by looking at the velocity space.

The right panel of Fig.~\ref{fig:recoil} shows the same evolution as
the left one, but through different quantities. The upper panel, in
particular, shows the time evolution of the recoil velocity and the
rapid changes it undergoes initially when the radiation first invests
the observer. The lower panel, on the other hand, shows the
$Q^{+}_{22}$ amplitude and highlights that, while the initial burst
of radiation stops after $t\sim 50\,M$ (\textit{cf.}, dotted line),
the waveform is still not fully consistent until $t\sim 75\,M$
(\textit{cf.}, dashed line).

The procedure discussed so far for the calculation of the integration
constant relative to the binary system $r0$ applies qualitatively to
all the other members of the sequence, with differences that are due
essentially to the times at which the various stages take place.

It is worth remarking that the evolution of the recoil vector in the
velocity-space has another interesting feature during the final stages
of the evolution and when the final black hole is ringing down. This
is marked as a long-dashed line in the left panel of of
Fig.~\ref{fig:recoil} and shows a break in the building of the spiral
and the appearance of a new spiral with a different aperture (we refer
to this feature as ``the hook''). This is more evident in the left
panel of Fig.~\ref{fig:recoil_r7}, which shows the evolution of the
recoil vector for the binary system $r7$ and offers a magnification of
the hook in the inset. A more detailed description of this feature is
beyond the scope of this paper and will be presented in a future work,
but we can here point out that the hook accounts for a rapid change in
the recoil velocity and it is due to the interplay of different modes
during the ringdown. This is clearly illustrated in the right panel of
Fig.~\ref{fig:recoil_r7} which similarly reports the time evolution of
the recoil velocity and the final stages of the $Q^+_{22}$ waveform.

\subsection{Recoil velocities}

The recoil velocity has been calculated for the sequence of models
listed in Table~\ref{tbl:parameters}. As mentioned in
Sect.~\ref{sec:initial_data}, this sequence corresponds to equal-mass
black holes, whose initial spins are unequal, though always aligned
with the $z$-axis. The $r0$ model has equal but opposite spins, while
the $r8$ model has equal and aligned spins on the black holes, with
other models corresponding to intermediate values, as outlined in
Section~\ref{sec:initial_data}.  Since the total initial orbital
angular momentum $\boldsymbol{L}$ of the system is chosen to be
constant over the sequence, the initial separations of the black holes
increases in the sequence, as well as the time to merger due to
spin-spin effects which contribute to an orbital ``hang-up'' in the
aligned case.

We extract gravitational waves by both the gauge-invariant and the
$\Psi_4$ methods described in the previous section and by
interpolating the radiation-related quantities onto 2-spheres
at coordinate radii $r_{_{\rm E}} =30\,M$, $40\,M$, $50\,M$, and
$60\,M$. The use of multiple extraction radii is made to check the
consistency of the measurement and the precise value of the extraction
radius has little influence on the actual kick calculation. In the
case of the binary system $r0$ we have verified that the recoil
velocity yields the same value with differences that are smaller than
$2\,\mathrm{km/s}$ for extraction 2-spheres at distances larger than
$30\,M$. As a result, we have used $r_{_{\rm E}} =50\,M$ as the
fiducial distance for an observer in the wave-zone and all of the
results presented hereafter will be made at this extraction 2-sphere.
A validation that the gauge-invariant quantities have the proper
scaling with radius is presented in Appendix~\ref{appendix_b}.

\begin{figure*}[t]
\begin{center}
 \centerline{
 \resizebox{8.5cm}{!}{\includegraphics[angle=-0]{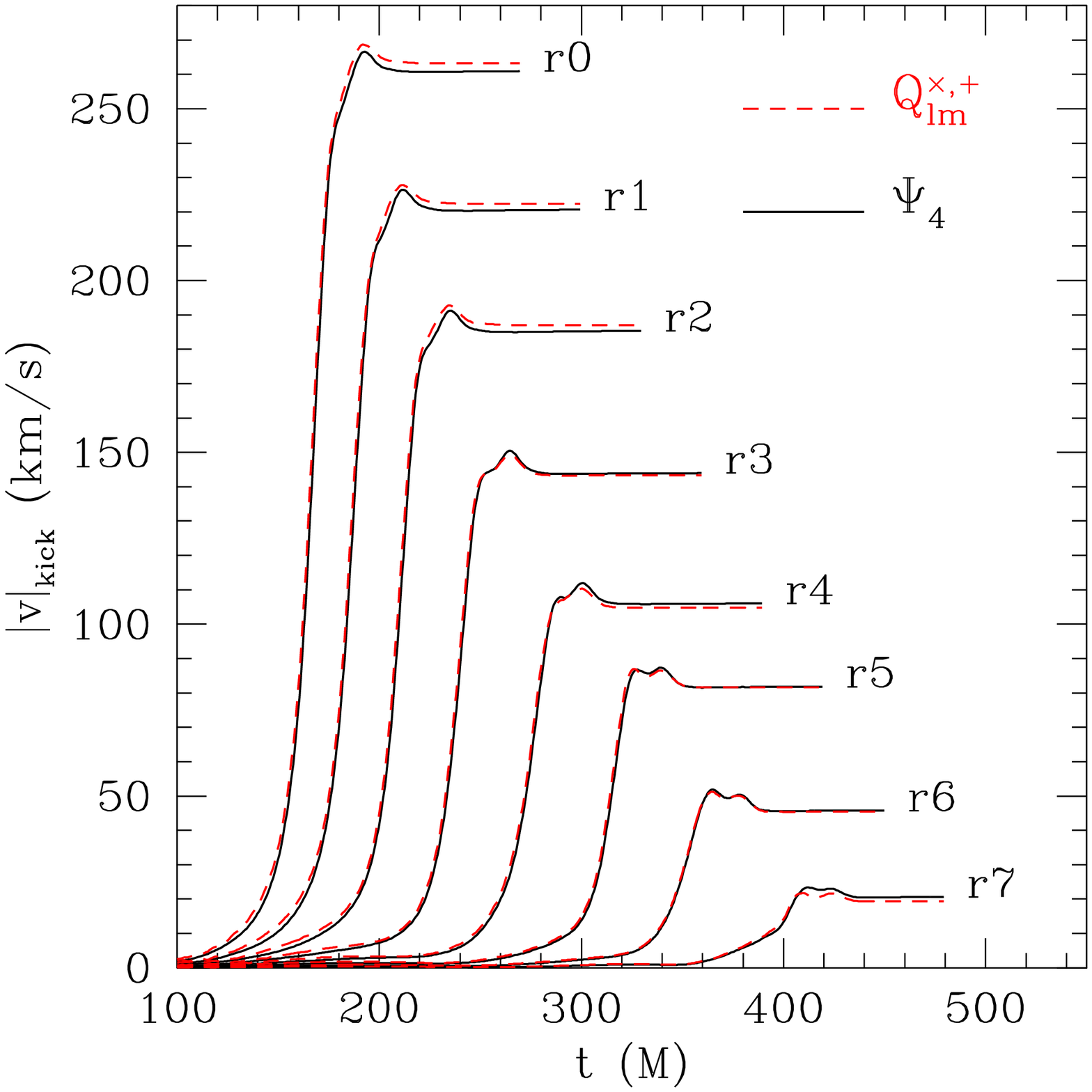}}
  \hskip 0.5cm
 \resizebox{8.5cm}{!}{\includegraphics[angle=-0]{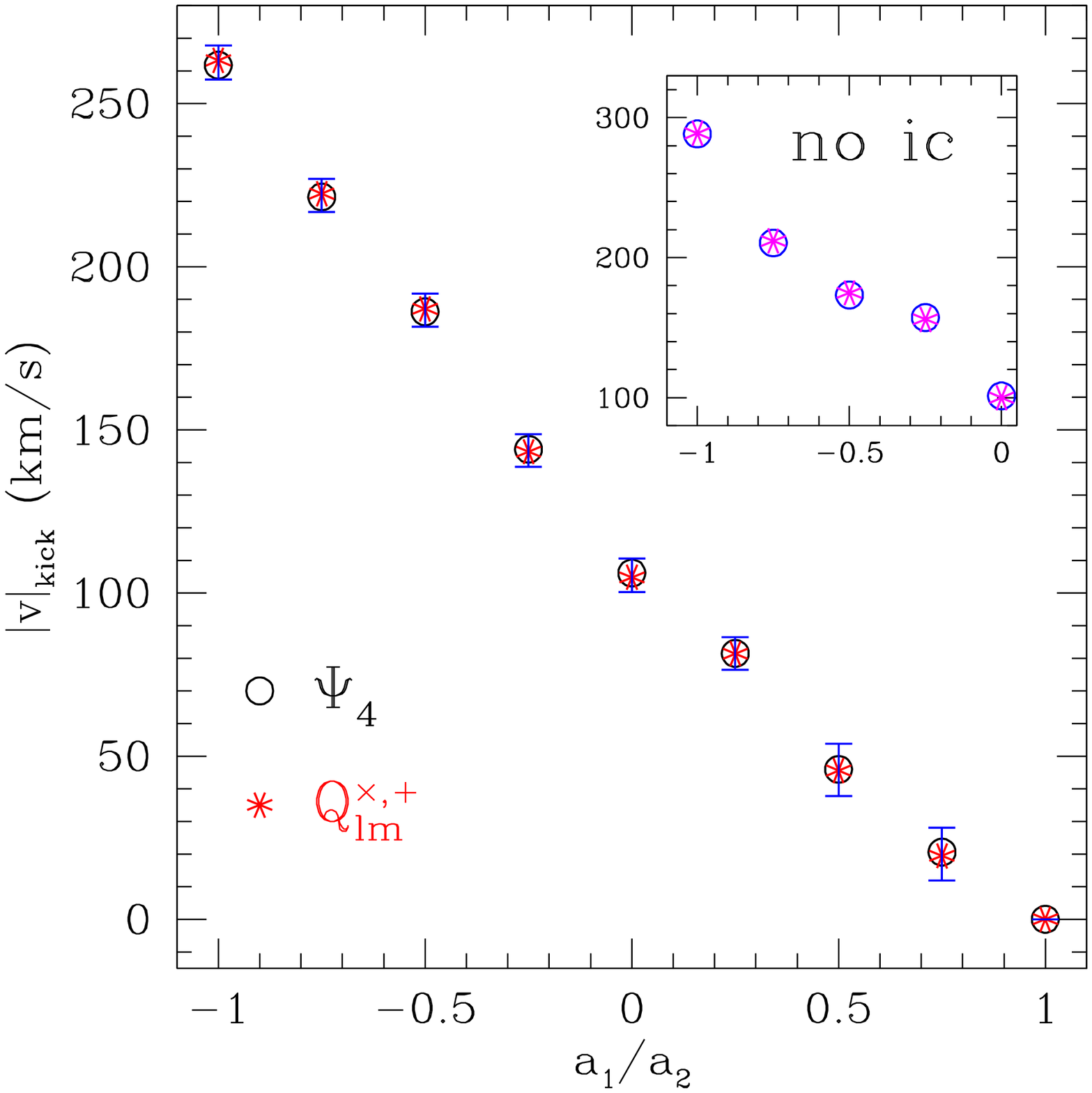}}
  }
  \caption{\textit{Left panel:} Recoil velocity as a function of the
    spin asymmetry parameter $a_1/a_2$ for the models listed in
    Table~\ref{tbl:parameters}. Indicated with a continuous lines are
    the results obtained via $\Psi_4$, while a dashed line is used for
    the gauge-invariant quantities $Q^{+,\times}_{\ell
    m}$. \textit{Right panel:} Final recoil velocity calculated with
    both the use $\Psi_4$ (empty circles) and the gauge-invariant
    quantities (stars). Shown in the inset is the incorrect scaling
    obtained when the correction for the integration constant is not
    made.}
  \label{fig:y_profiles}
\end{center}
\end{figure*}

The evolution of the recoil velocity for the entire sequence listed in
Table~\ref{tbl:parameters} is displayed in the left panel of
Fig.~\ref{fig:y_profiles}. It is apparent that the suitable choice of
the integration constant discussed in the previous section yields
early evolutions that are always monotonic in time and that, as
expected, the largest recoil velocity is generated for the case in
which the asymmetry is the largest, namely for the binary $r0$. The
left panel Fig.~\ref{fig:y_profiles} also shows that the profile for
each case is rather similar, with the largest contribution to the kick
velocity being generated in a period of about $80\,M$, corresponding
roughly to the timescale of the last orbit and merger. Furthermore, it
is notable that $95\%$ of the acceleration occurs $\sim 30\,M$ after
the appearance of the first common apparent horizon, indicating that
the kick is generated not only by the final stages of the inspiral
(\textit{i.e.}, by the ``plunge'') but also and more significantly by
the ringdown of the final black hole. This fact helps to explain why
accurate recoil velocities can be obtained by evolutions involving
very few cycles only, provided the integration constant is properly
taken into account.

It is worth noting that during the final stages of the evolution, the
recoil velocity is not monotonic but shows at least two peaks, whose
relative amplitude depends on the spin ratio. For spin ratios $\sim
-1$ the first peak is hardly visible, while the second one is the most
pronounced one. As the spin ratio increases, however, the first peak
becomes more prominent and for spin ratios $\sim 1$ it becomes
comparable with the second one or even larger for binaries $r6$ and
$r7$. As mentioned in the previous Section and further discussed in
the following one, the appearance of these peaks is related to the
interplay of different mode-contributions during the ringdown. The
second peak, in particular, can be associated to a rapid change in the
recoil-velocity vector and is behind the characteristic ``hook''
discussed in the left panels of Figs.~\ref{fig:recoil} and
\ref{fig:recoil_r7}. While additional work is needed, especially in
thorough perturbative investigations, to fully account for the rich,
post-merger properties of the recoil velocities, we believe the
double-peak evolution to be physically genuine since it is seen in all
binaries and is supported by the highly accurate and convergent
simulations. As a representative measure of the accuracy in
determining these recoil velocities, we mention that we have carried
out simulations also for the binary system $r8$, in which the black
holes have identical spin and thus from which no kick should
result. The computed recoil velocity has been found to be
$10^{-9}\,\mathrm{km/s}$, clearly indicating that our evolutions do an
excellent job in preserving the orbital symmetry of these binaries. 

\begin{table}[t]
  \caption{Final kick velocities in units of $\mathrm{km/s}$ for the
    models listed in Tab.~\ref{tbl:parameters}. Columns two and three
    show the values obtained using the gauge-invariant quantities
    $Q_{\ell m}^{\times, +}$ and $\Psi_4$ respectively and taking into
    account the integration constant. Columns four and five, on the
    other hand, show the results obtained when ignoring the
    integration constant. The same
    data are shown in the right panel of Fig.~\ref{fig:y_profiles}. }
\begin{ruledtabular}
\begin{tabular}{c|dddd}
Model
& \multicolumn{1}{c}{$Q_{\ell m}^{\times, +}$}
& \multicolumn{1}{c}{$\Psi_4$}
& \multicolumn{1}{c}{$Q_{\ell m}^{\times, +}, \textrm{no ic}$}
& \multicolumn{1}{c}{$\Psi_4, \textrm{no ic}$} \\
\hline
$r0$ & 263.2  & 261.8 & 288.9 & 288.4  \\
$r1$ & 222.4  & 221.4 & 211.9 & 210.6  \\
$r2$ & 187.1  & 186.2 & 174.8 & 173.3  \\
$r3$ & 143.3  & 144.0 & 155.9 & 157.3  \\
$r4$ & 104.8  & 106.1 & 100.0 & 101.3  \\
$r5$ &  81.4  &  81.5 &  76.9 &  77.0  \\
$r6$ &  45.6  &  45.9 &  55.4 &  56.2  \\
$r7$ &  19.4  &  20.6 &  13.8 &  14.8  \\
$r8$ &   0.0  &   0.0 &   0.0 &   0.0  \\
\end{tabular}
\end{ruledtabular}
\label{tab:Kicks}
\end{table}

We have found that the evolution of the recoil velocity generated by
spin asymmetries appears to be rather different from the one generated
by mass asymmetries~\cite{Gonzalez:2006md, Baker:2006nr,
Koppitz-etal-2007aa} and which shows much larger variations between
the maximum attained value and the final one. Once again, this
different behavior is related to the different interplay of the
ringdown modes in the case of mass asymmetries and will be presented
in a separate work.

The recoil velocities attained by the final black holes and
shown for in the left panel of Fig~\ref{fig:y_profiles} can be studied
in terms of their dependence on the spin ratio $a_1/a_2$, which can
also be regarded as the ``asymmetry'' parameter of the system, being
the largest for $a_1/a_2=-1$ and zero for $a_1/a_2=1$. These
velocities are collected in Table~\ref{tab:Kicks} and are shown as a
function of $a_1/a_2$ in the right panel of Fig~\ref{fig:y_profiles},
where we have indicated with open circles the values obtained using
$\Psi_4$ and with stars those obtained using the gauge-invariant
perturbations. 

The data in the right panel of Fig~\ref{fig:y_profiles} is shown
together with its error-bars, which include errors from the
determination of the integration constants, from the truncation error
and from the amount of ellipticity contained in the initial data. We
have estimated these errors to be of $5\ {\rm km/s}$ for binaries
$r0$--$r5$ and of $8\ {\rm km/s}$ for binaries $r6$ and $r7$. Shown
also in the inset is the recoil data obtained when ignoring
the integration constant. It is remarkable that when
the proper evaluation of the initial transient is not made, the data
does not show the remarkable correlation with the spin ratio which is
instead shown by the corrected data.  Quite surprisingly, however, the
correlation found the one predicted by PN studies. We recall, in fact,
that using PN theory at the 2.5 order, Kidder~\cite{Kidder:1995zr} has
concluded that in the case of a circular, non-precessing orbit, the
total kick for a binary system of arbitrary mass and spin ratio can be
expressed as~\cite{Favata:2004wz}
\begin{eqnarray}
\label{eq:fhh} 
|v|_{\rm kick}&=&c_1 \frac{q^2(1-q)}{(1+q)^5} + 
        c_2 \frac{a_2 q^2(1 - q  a_1/a_2)}{(1+q)^5}  
\nonumber \\
	&=& {\tilde c}_2 a_2 \left(1 -  \frac{a_1}{a_2}\right)\,, 
\end{eqnarray}
where $q\equiv M_1/M_2$ is the mass ratio and is equal to one for the
binaries considered here, thus leading to the second form of
Eq.~\eqref{eq:fhh}. The coefficients $c_{1}$ and ${\tilde c}_{2}
\equiv c_2/32$ depend on the total mass of the system and on the
orbital separation at which the system stops radiating, which is
intrinsically difficult to determine with precision since it lies in a
region where the PN approximation is not very accurate. Indeed, we
find that the coefficient $c_2$ is not really a constant in the case
of equal-mass binaries but, rather, it can be seen to depend at least
linearly on the spin ratio.

\begin{figure}[t]
\begin{center}
 \centerline{
 \resizebox{8.5cm}{!}{\includegraphics[angle=-0]{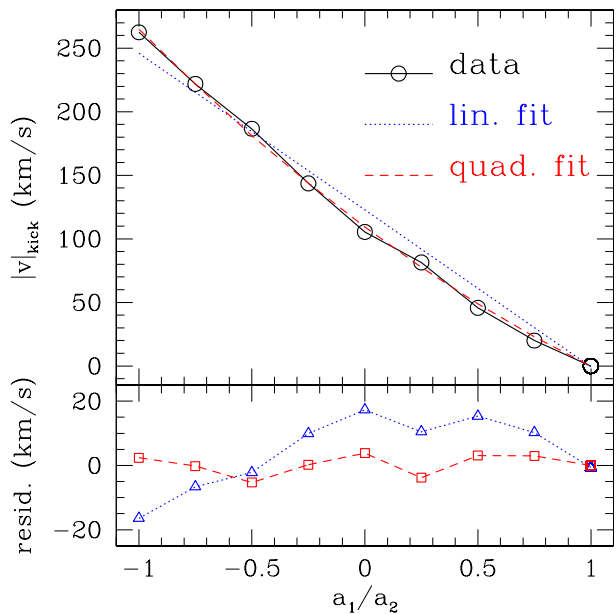}}
  }
  \caption{\textit{Upper panel:} Comparison of the computed data for
    the recoil velocity (open circles) with the least-squares fits
    using either a linear (dotted line) or a quadratic dependence
    (dashed line). \textit{Lower panel:} Point-wise residuals computed
    with the linear (dotted line) or a quadratic fit (dashed line).}
  \label{fig:vkick_fit}
\end{center}
\end{figure}

This is shown in Fig.~\ref{fig:vkick_fit}, whose upper panel offers a
comparison among the computed data for the recoil velocity (open
circles) with the least-squares fits using either a linear (dotted
line) or a quadratic dependence (dashed line). It is quite apparent
that a linear dependence on $a_1/a_2$, such as the one expected in
Eq.~\eqref{eq:fhh} for $c_2={\rm const.}$ does not reproduce well the
numerical data and yields point-wise residuals of the order of
$20\,{\rm km/s}$. These are shown with a dotted line in the lower
panel of Fig.~\ref{fig:vkick_fit}. A quadratic dependence on
$a_1/a_2$, on the other hand, reproduces the numerical data very
nicely, with residuals that are of the order of $5\,{\rm km/s}$, as
shown with a dashed line in the lower panel of the same figure, and
thus compatible with the reported error-bars.


\begin{figure*}[t]
\begin{center}
  \centerline{
    \resizebox{8.5cm}{!}{\includegraphics[angle=-0]{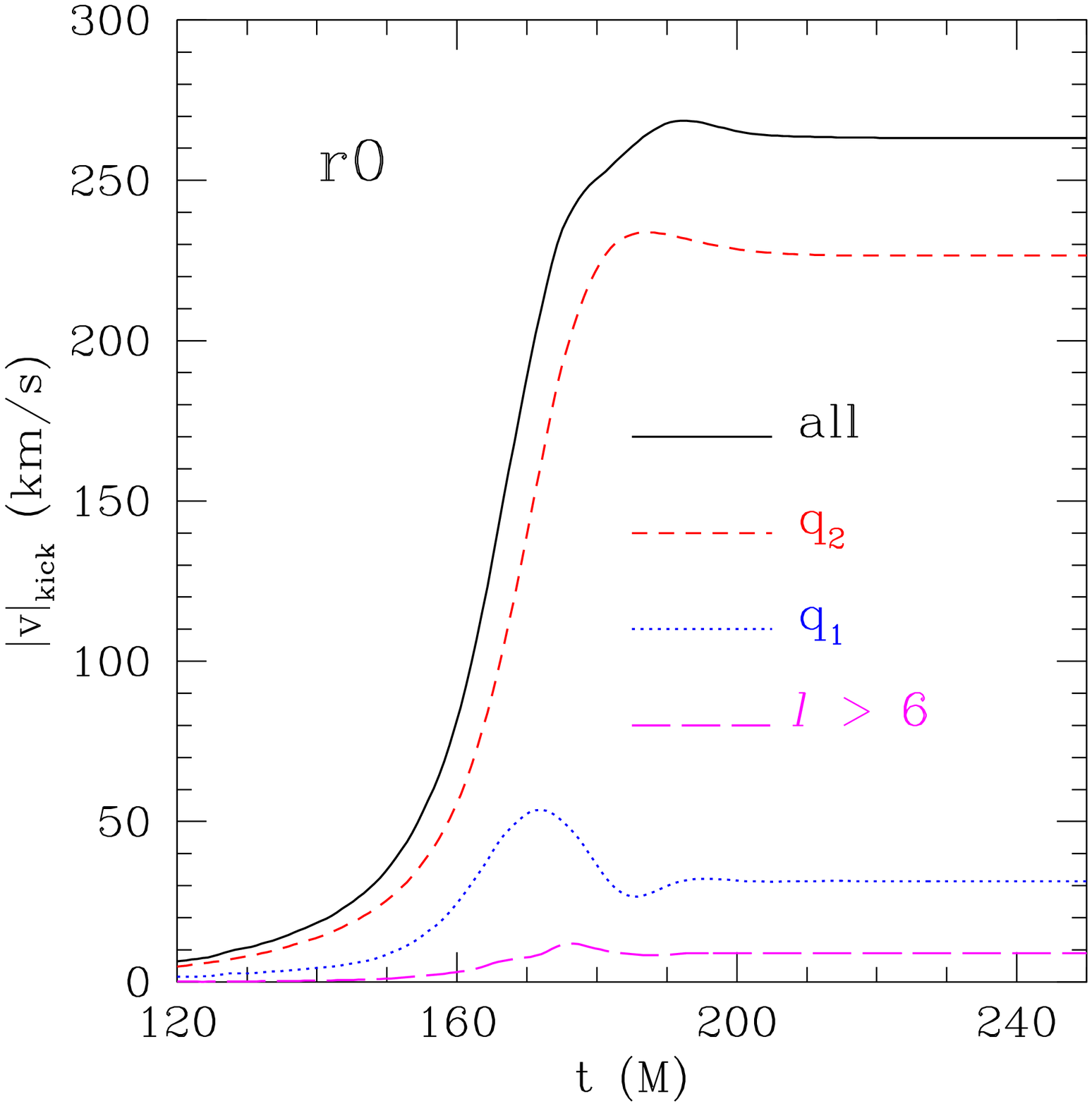}}
      \hskip 0.5cm
    \resizebox{8.5cm}{!}{\includegraphics[angle=-0]{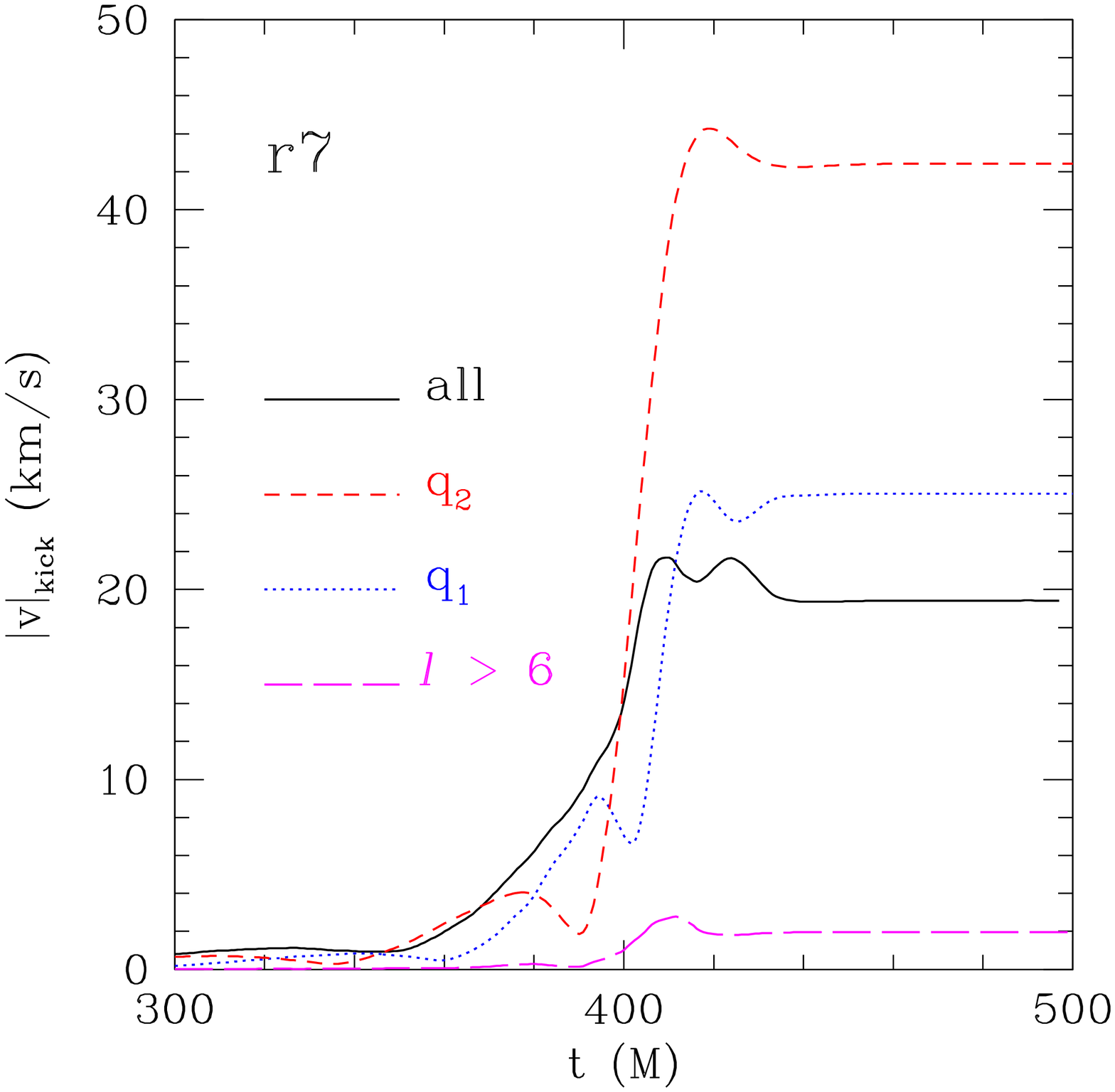}}
  }
  \vskip 0.5cm
  \caption{The total kick calculated via Eq.~(\ref{eq:gi}) up to
    $\ell=7$ is compared to the contributions of individual terms
    $q_1$ and $q_2$, as well as the sum of term excluding these. In
    the case of the $r0$ system (left panel) the spins are
    anti-aligned and the $q_2$ term is dominant and the $q_1$ term
    does not provide a significant contribution. In the case of the
    $r7$ system (right panel), on the other hand, the spins are
    essentially aligned and the while the $q_2$ term is still
    dominant, the $q_1$ term also makes a significant
    contribution. }
  \label{fig:kick_mode_contributions}
\end{center}
\end{figure*}

We can re-express Eq.~\eqref{eq:fhh} in the more generic form
\begin{equation}
\label{eq:n_fhh}
|v|_{\rm kick} \left(a_2,\frac{a_1}{a_2}\right) = 
	|a_2| f\left(\frac{a_1}{a_2}\right)
\end{equation}
where $a_2$ plays here the role of a ``scale-factor''. The function
$f(a_1/a_2)$ with $a_1/a_2 \in [-1,\,1]$ and maximum at $a_1/a_2=-1$
can then be seen as to be determined from numerical-relativity
calculations (or higher-order PN approximations) and our least-squares
fit suggests the expression
\begin{eqnarray}
&&f_{\rm quad.} = 109.3 - 
	132.5\left(\frac{a_1}{a_2}\right)  + 
	23.1 \left(\frac{a_1}{a_2}\right)^2 \ {\rm km/s}\,. \nonumber \\
\end{eqnarray}
The maximum kick velocity for a given $a_2$ is then readily calculated
even without a detailed knowledge of the function $f(a_1/a_2)$ as
\begin{equation}
(|v|_{\rm kick})^{\rm max} ( a_2 ) = |a_2| f(-1) \,.
\end{equation}
Using the data reported in Table~\ref{tab:Kicks} for $a_2=-0.584$ we
obtain for $|a_2|=1$ that the maximum recoil velocity attainable
from a binary system of equal-mass black holes with spins aligned to
the orbital angular momentum is $448\pm 5\,\mathrm{km/s}$. This
is in very good agreement with our previous estimate made in
Ref.~\cite{Koppitz-etal-2007aa} with a smaller sequence and in equally
good agreement with the results reported in
Ref.~\cite{Herrmann:2007ac}.

\subsection{Mode contributions to the recoil velocity}
\label{sec:ModeContribution}

For the models studied in the previous section we have evaluated
Eq.~(\ref{eq:gi}) including modes up to $\ell=7$. In practice,
however, we find that the recoil is strongly determined by the
lower-mode contributions. In particular, the two terms
\begin{align}
q_1 & \equiv \frac{1}{48\pi}\sqrt{\frac{30}{7}}
        \dot{Q}^+_{22}\; Q^+_{3-3}, \\
q_2 & \equiv -\frac{{\rm i}}{48\pi} 
        \dot{Q}^+_{2-2}\; Q^\times_{2\;1}
\end{align}
are the dominant ones. This can be seen in
Fig.~\ref{fig:kick_mode_contributions}, where the time evolutions of
the terms $q_1$ and $q_2$ are plotted (dotted and dashed lines,
respectively) together with the total kick calculated via
Eq.~(\ref{eq:gi}) (solid line), and with the contributions from all
other terms up to $\ell=7$ excluding $q_1$ and $q_2$ (long-dashed
line). A rapid inspection of the figure reveals that the kick is
dominated in particular by the $q_2$ term, whereas the $q_1$ term has
a magnitude of the order of all the other modes combined. A similar
result holds for each member of the sequence, so that the two
contributions determine the final kick to more that $95\%$. It should
be noted that the mode contributions are vector quantities, just as
the kick velocity itself, and are not always aligned or even maintain
the same angle to each other during the duration of the recoil.

This coupling also goes some way to explain some features of the
recoil velocity profiles displayed in Fig.~\ref{fig:y_profiles}. As
mentioned in the previous section, in fact, the binaries $r4$ to $r8$
show a clear double peak in the evolution of the kick velocity before
it settles down to the final value. The same feature can also be seen
in the more asymmetric $r0$ to $r3$ binaries, where it appears as a
flattening of the slope near the maximum. Since the two peaks are
shown both by the gauge-invariant and by the $\Psi_4$-based techniques
(which are rather different in both the assumptions they rely on and
in the practical implementation) we do not believe them to be a simple
numerical artifact. Overall, the properties of the recoil velocity
near its maximum, and before it settles to the final value, are
determined by the relative phases of the two contributions identified
above. An analysis of the terms $q_1$ and $q_2$ in vector-space, and
which will be presented in a subsequent work, reveals that when they
are relatively aligned at the peak of the acceleration, there is a
clear single peak in the evolution. For the more symmetric models, on
the other hand, the two contributions are more anti-aligned and a
double peak results.

These considerations in the vector evolution of the two contributions
$q_1$ and $q_2$ need also to be linked with the evolution in vector
space of the recoil velocity. As stressed in
Sect.~\ref{sbsc:integr_const}, in fact, there is a distinct kink in
the evolution of the velocity vector towards the final stages of the
merger (this feature is indicated with a long-dashed line in the $v_x$
vs.\ $v_y$ plots of Figs.~\ref{fig:recoil}
and~\ref{fig:recoil_r7}). The presence of the kink corresponds to a
local decrease of the recoil velocity and hence to the minimum between
the two peaks. Because this decrease is more pronounced for the
lower-kick binaries $r4$ to $r8$, the first peak becomes more evident
there.

\subsection{Angular Momentum and Mass Conservation}
\label{sec:AngularMomentum}

In this section we discuss the radiated angular momentum and energy
during the evolution of the different initial-data sets. We compute
the radiated angular momentum and mass by calculating their difference
between the initial data and that of the final black hole, and then
compare these quantities with the corresponding ones measured in terms
of the emitted gravitational radiation. The differences in the two
independent estimates serve therefore as stringent indicators of the
conservative properties of our code.

The radiated angular momentum can be simply written as the difference
between the initial and final values
\begin{equation}
  \mbf{J}_{\mathrm{rad}} = \mbf{J}_{\mathrm{fin}}
    - \mbf{J}_{\mathrm{ini}}\,,
\label{eq:FinalMinusInitial}
\end{equation}
where, as a result of the conformal flatness of the initial-data
slice, $\mbf{J}_{\mathrm{ini}}$ is given by the simple expression (see
for example~\cite{Cook94, Cook00a, Pfeiffer:2000um} and discussion in
Sect.~\ref{sec:NumericalMethods})
\begin{equation}
  \mbf{J}_{\mathrm{ini}} \equiv \mbf{J}_{_{\mathrm{ADM}}}
    = \mbf{C}_1 \times \mbf{p}_1 +
    \mbf{C}_2 \times \mbf{p}_2 + \mbf{S}_1 + \mbf{S}_2 \ .
  \label{eq:InitialSpin}
\end{equation}
Here $\mbf{C}_i$, $\mbf{p}_i$ and $\mbf{S}_i$ are the position, the
linear momentum and the spin of the $i$-th black hole.  The final
angular momentum $\boldsymbol{J}_{\mathrm{fin}}$, on the other hand,
is set to be equal to the spin of the final black hole after all the
radiation has left the computational domain.
Two different methods are used to obtain this measure, both of which
are based on properties of the apparent horizon of the
final hole.

\begin{figure*}
\begin{center}
\centerline{
  \resizebox{8.5cm}{!}{\includegraphics[]{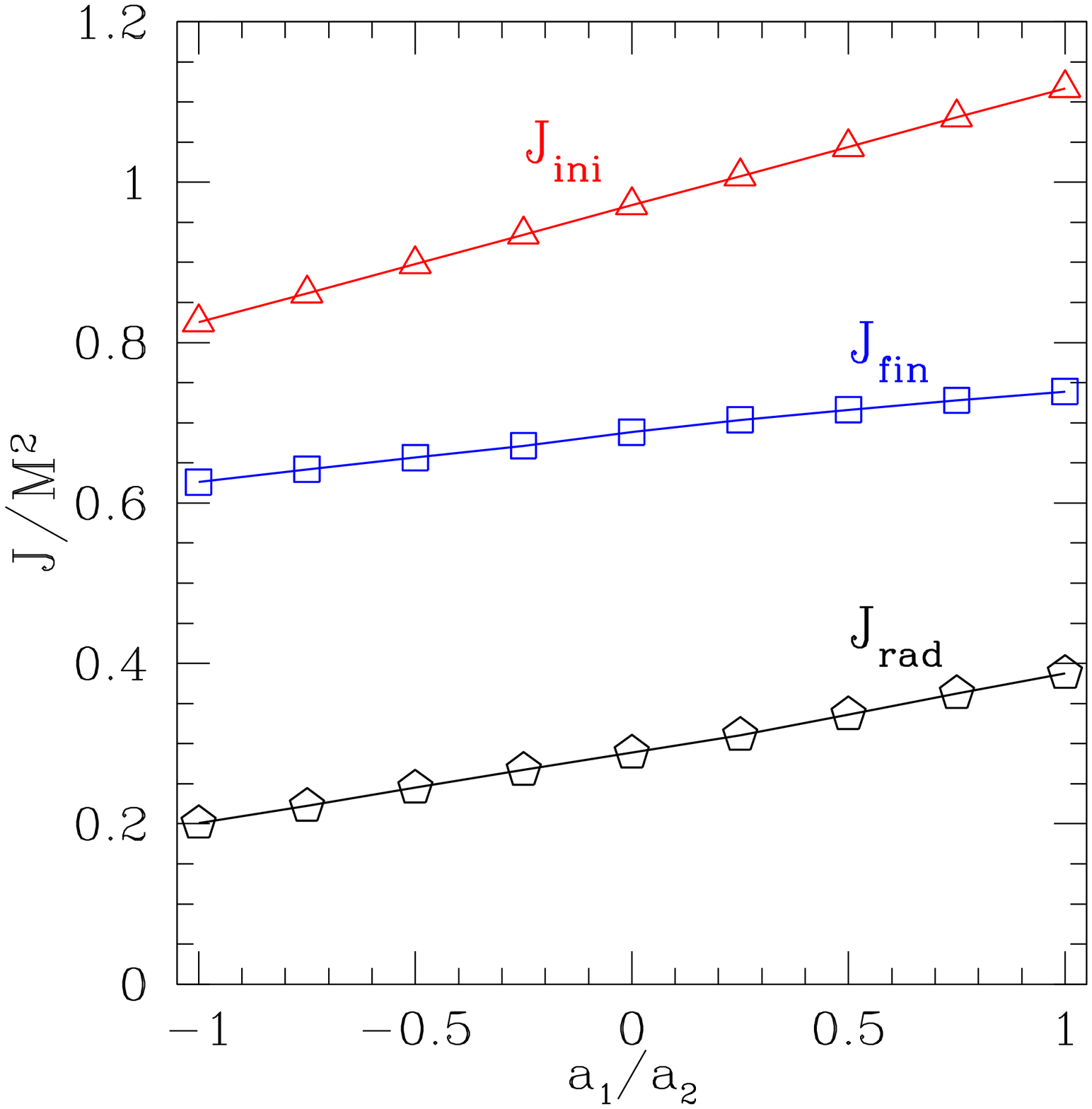}}
    \hskip 0.5cm
  \resizebox{8.5cm}{!}{\includegraphics[]{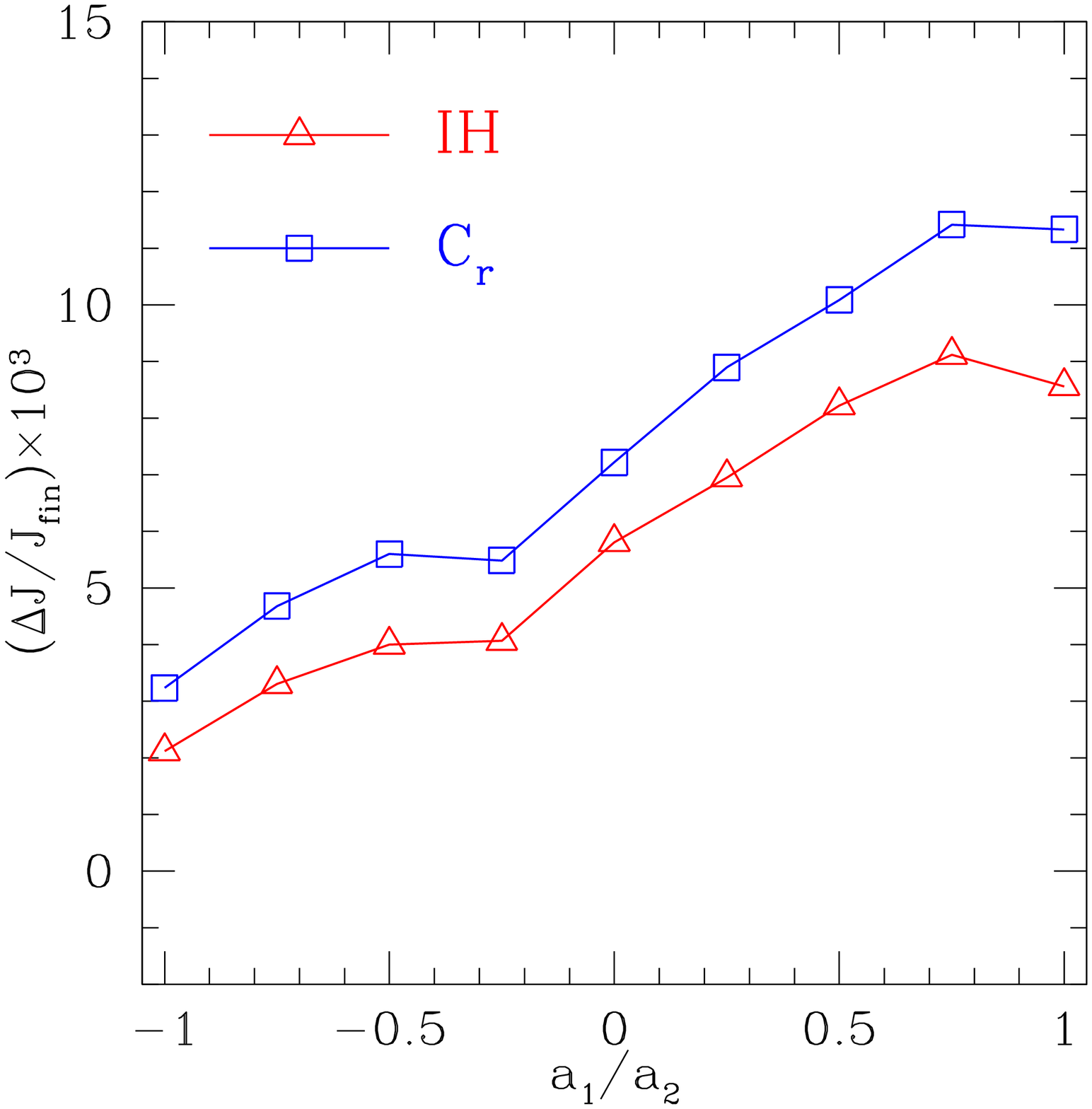}}
}
\caption{\textit{Left panel:} Dependence on the spin ratio of the
  initial total angular momentum $J_{\mathrm{ini}}$ [as computed from
  Eq.~(\ref{eq:InitialSpin})], of the radiated angular momentum
  $J_{\mathrm{rad}}$ [as computed through the gauge-invariant
  waveforms], and of the final spin of the black hole
  $J_{\mathrm{fin}}$. All quantities show a linear behavior, whose
  coefficient are collected in Table~\ref{tab:Fit}. \textit{Right
  panel:} Relative error $\Delta J /J_{\mathrm{ini}}$ in the
  conservation of the angular momentum [\textit{cf.},
  Eq.~(\ref{eq:DeltaJOverJ})]. Different curves refer to whether the
  final spin of the black hole is computed using the
  isolated/dynamical horizon formalism (triangles) or the distortion
  of the apparent horizon (squares). In both cases the error is of
  about $1\%$ at most for simulations at the medium resolution.}
\label{fig:AngularMomentum}
\end{center}
\vskip 1.0cm
\begin{center}
\centerline{
  \resizebox{8.5cm}{!}{\includegraphics[]{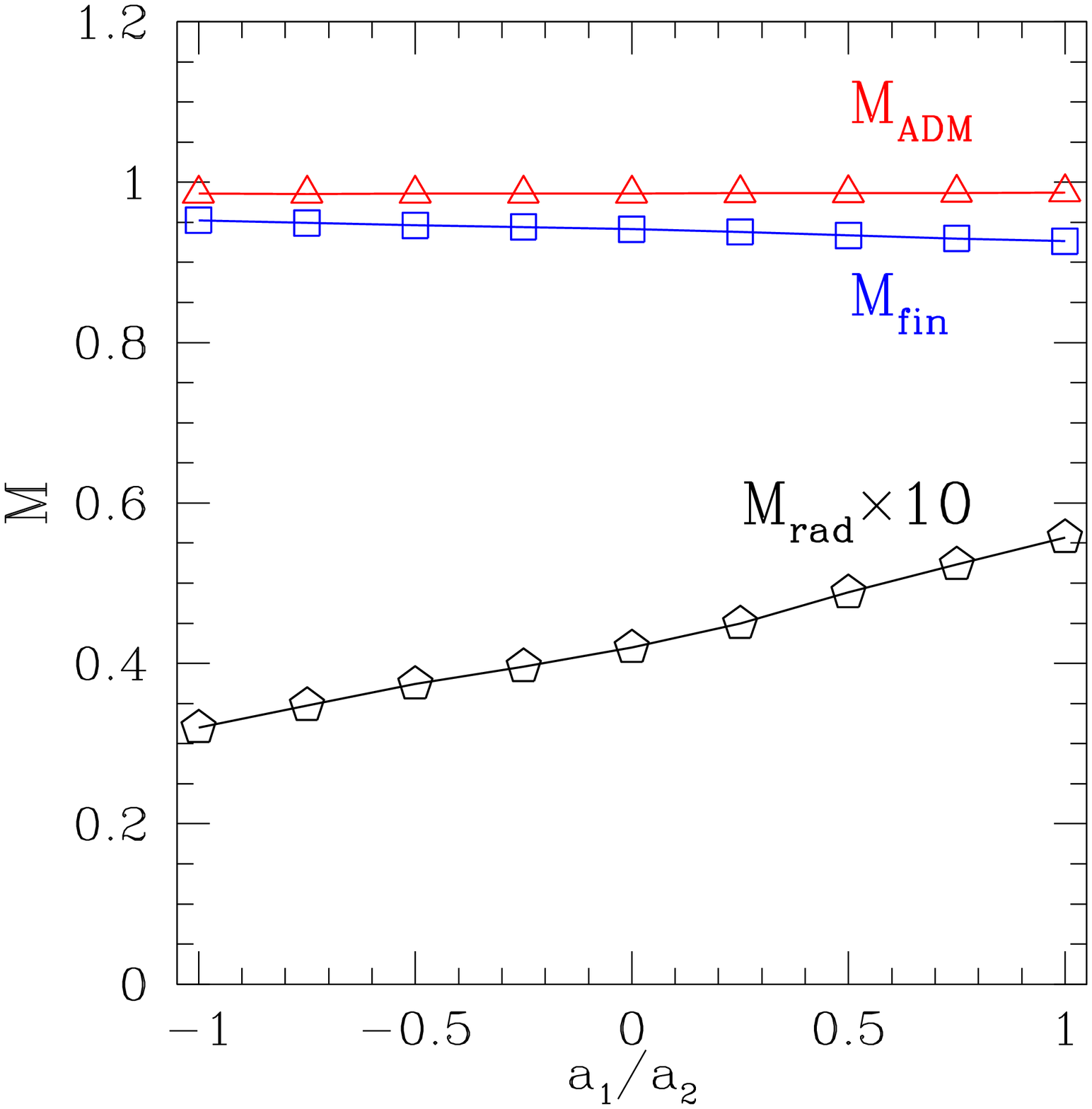}}
    \hskip 0.5cm
  \resizebox{8.5cm}{!}{\includegraphics[]{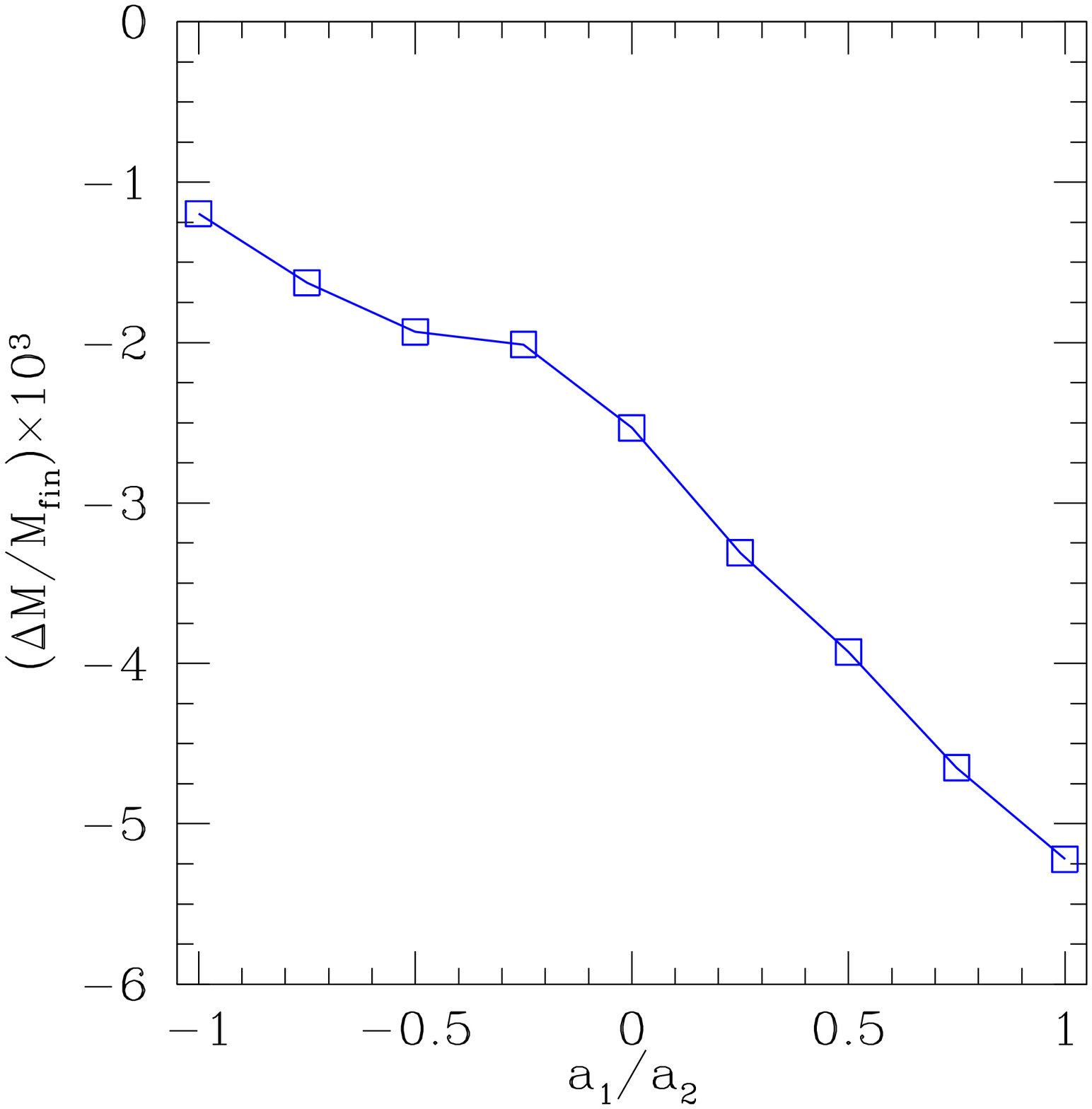}}
}
\caption{\textit{Left panel:} Dependence on the spin ratio of the
  ADM mass $M_{_{\mathrm{ADM}}}$, of the scaled radiated energy
  $M_{\mathrm{rad}}$ [as computed through the gauge-invariant
  waveforms and scaled by a factor of $10$ to make it visible], and of
  the final mass of the black hole $M_{\mathrm{fin}}$. All quantities
  show linear behaviors, whose coefficients are collected in
  Table~\ref{tab:Fit}. \textit{Right panel:} Relative error $\Delta M
  /M_{\mathrm{ini}}$ in the conservation of the energy [\textit{cf.},
  Eq.~(\ref{DeltaMoM})]. Note that the error is of about $0.5\%$ at
  most for simulations at the medium resolution.}
\label{fig:Masses}
\end{center}
\end{figure*}

The first method employs the isolated/dynamical horizon formalism
and searches for a rotational Killing vector $\phi^a$ on the final
apparent horizon so as to measure the spin of the final black hole
as~\cite{Dreyer02a, ashtekar03a, Schnetter-Krishnan-Beyer-2006}
\begin{equation}
  J = -\frac{1}{8\pi} \oint_S K_{ab} \phi^a \hat{r}^b d^2 V\,.
  \label{eq:ih_J}
\end{equation}
We note that this expression~(\ref{eq:ih_J}) is valid on any sphere
where a Killing vector $\phi^a$ can be found, and is therefore a
quasi-local measure of the angular momentum.  In particular, at large
distances where the spacetime is close to axisymmetric, there is a
good approximation to an angular Killing vector, and we can apply this
expression to determine the angular momentum of the spacetime. Note
also that Eq.~(\ref{eq:ih_J}) is identical to the ADM angular momentum
when evaluated at spacelike infinity. (Refs.~\cite{ashtekar03a,
Schnetter-Krishnan-Beyer-2006} also give a quasi-local formula for the
angular momentum flux due to gravitational radiation.)

The second method instead, assumes that the final black hole has
settled to a Kerr one and uses the the rotational-induced distortion
of the apparent horizon of the final black hole to estimate its spin.
Defining $C_p$ and $C_e$ to be respectively the apparent horizon's
polar and equatorial proper circumferences, their ratio $C_r\equiv
C_p/C_e$ will undergo damped oscillations as the perturbed black hole
settles to a Kerr state through the quasi-normal ringing. The final
value of $C_r$ can be expressed as a nonlinear function of the
dimensionless spin parameter $a=J/M^2$
as~\cite{Alcubierre2003:pre-ISCO-coalescence-times, Alcubierre:2004hr,
Seidel99b, Brandt94c}
\begin{equation}
C_r(a)={1+\sqrt{1-a^2} \over \pi} E\left(- {a^2 \over
  (1+\sqrt{1-a^2})^2} \right)\,,
\label{eq:CrOfJ}
\end{equation}
where $E(k)$ is the complete elliptic integral of the second kind
\begin{equation}
E(k)=\int_0^{\pi/2} \sqrt{1-k \sin^2 \theta} d \theta \,.
\end{equation}
By inverting numerically Eq.~(\ref{eq:CrOfJ}) we obtain $a$ from the
late time $C_r$ that is measured from the apparent horizon shape.
Note that for computing $J$ we need to multiply $a$ by the square of
the final mass, which we take to be $M_{_{\mathrm{ADM}}} -
M_{\mathrm{rad}}$. An alternative choice involving the total mass
Eq.~(\ref{eq:AHMass}) as measured from the apparent horizon would lead
to essentially the same results.

As mentioned at the beginning of this section, the determination of
the radiated angular momentum can also be done using directly the
asymptotic waveform amplitudes $h_{+}$ and $h_{\times}$
as~\cite{Poisson04b, Martel:2005, Nagar05}
\begin{equation}
  \frac{d^2 J}{dt\, d\Omega} = 
    -\frac{r^2}{16\pi} \left( \partial_t h_+ \partial_\phi h^*_+
    + \partial_t h_\times \partial_\phi h^*_\times \right) \,,
\label{eq:JFromQ}
\end{equation}
where the amplitude $h_+$ and $h_\times$ themselves can be expressed
either in terms of the Zerilli-Moncrief gauge-invariant variables
$Q^+_{\ell m}$, $Q^\times_{\ell m}$ or, alternatively, in terms of the
Newman-Penrose scalar $\Psi_4$. A comparison between the two
approaches is presented in Appendix~\ref{appendix_b}, where it is
shown that the differences are minute. Because of this, hereafter we
will refer to asymptotic amplitudes measured in terms of the
gauge-invariant variables only. Additional details on the resolution
of the extraction 2-sphere are also presented in
Appendix~\ref{appendix_a}.

The left panel of Fig.~\ref{fig:AngularMomentum} summarizes this
comparison by showing, as functions of the spin ratio $a_1/a_2$,
$J_{\mathrm{fin}}$ from Eq.~(\ref{eq:ih_J}), $J_{\mathrm{rad}}$ from
Eq.~(\ref{eq:JFromQ}) both adding nicely to yield
$J_{\mathrm{ini}}$. Note that $J_{\mathrm{ini}}$ is growing linearly
as it is obvious from Eq.~(\ref{eq:InitialSpin}), but also that that a
similar behavior is shown by the radiated angular momentum (and hence
by the final spin of the black hole). Using a linear fitting we
can derive phenomenological expressions for the relative losses of angular 
momentum 
\begin{equation}
\frac{J_{\rm rad}}{J_{\rm ini}} = \xi^{J}_{\rm rad}
	\left(\frac{a_1}{a_2}\right) + \chi^{J}_{\rm rad}\,,
\label{jrad_phenom}
\end{equation}
and the relative spin-up of the final black hole
\begin{equation}
\frac{J_{\rm fin}}{J_{\rm ini}} = \xi^{J}_{\rm fin}
	\left(\frac{a_1}{a_2}\right) + \chi^{J}_{\rm fin}\,.
\label{jfin_phenom}
\end{equation}
The fitted values for $\xi^{J}_{\rm rad,\,fin}$ and $\chi^{J}_{\rm
rad,\,fin}$ are presented in Table~\ref{tab:Fit} and readily indicate
that the system looses $24\%$ of its initial orbital angular momentum
in the case of anti-aligned spins and up to $34\%$ for aligned spins.

To the best of our knowledge expressions~(\ref{jrad_phenom})
and~(\ref{jfin_phenom}) do not have a PN counterpart and yet, since
they depend only on the spin-ratio, they represent simple and powerful
ways of estimating both the efficiency in the extraction of angular
momentum and the spin of the final black in a binary merger when the
spins are orthogonal to the orbital plane. This information could be
easily injected in those $N$-body simulations in which the interaction
of binary black holes is taken into account~\cite{kupi06} and thus
yield accurate estimates on final distribution of black-hole spins.

Since we have two independent and different ways of computing
$J_{\mathrm{rad}}$ [\textit{i.e.}, either from Eq.~(\ref{eq:JFromQ})
or from Eq.~(\ref{eq:FinalMinusInitial})] we can quantify our ability
to conserve angular momentum by measuring the normalized residual
\begin{equation}
{\Delta J \over J_{\mathrm{ini}}}\equiv
 {J_{\mathrm{fin}}+J_{\mathrm{rad}}-J_{\mathrm{ini}} \over
 J_{\mathrm{ini}}}
\,.
 \label{eq:DeltaJOverJ}
\end{equation}
This is shown in the right panel of Fig.~\ref{fig:AngularMomentum} and
the two different lines refer to the two measures of the final spin of
the black hole, \textit{i.e.}, either via the isolated-horizon
formalism (triangles) or via the distortion of the apparent horizon
(squares). In both cases the error is extremely small, ranging between
$1.1\%$ and $0.2\%$ for simulations at the medium resolution, and thus
providing convincing evidence of our accuracy in the preservation of
angular momentum. It should be noted that while there seems to be a
small advantage in using the isolated horizon measure, the differences
are too small to be significant. Indeed, a small change in the
procedure, such as the use of the mass measured via the apparent
horizon via Eq.~(\ref{eq:CrOfJ}) in place of
$M_{\mathrm{ini}}-M_{\mathrm{fin}}$ (as we are doing in this figure),
would revert the advantage.

\begin{table}[t]
\caption{Coefficients for the phenomenological expressions
(\ref{jrad_phenom}) and~(\ref{jfin_phenom}) (and the corresponding
coefficients for $\Delta M_{\rm rad,\,fin}/M$) by means of which it is
possible to compute the relative losses of energy and angular
momentum, as well as the final mass and spin of the black hole in
binary mergers in which the spins are orthogonal to the orbital
plane.}
\begin{tabular}{lr|lr}
\hline
\hline
        $\xi^J_\mathrm{rad}$  & 0.0513 & $\xi^M_\mathrm{rad}$  & 0.0118 \\ 
        $\chi^J_\mathrm{rad}$ & 0.2967 & $\chi^M_\mathrm{rad}$ & 0.0437 \\
\hline
        $\xi^J_\mathrm{fin}$  &-0.0513 & $\xi^M_\mathrm{fin}$  &-0.0118 \\
        $\chi^J_\mathrm{fin}$ & 0.7033 & $\chi^M_\mathrm{fin}$ & 0.9563 \\
\hline
\hline
\end{tabular}
\label{tab:Fit}
\end{table}

\begin{table*}[t]
\caption{Final and radiated angular momenta and masses, computed from
the gauge-invariant waveforms. Shown is also the
radiated spin and mass relative to their initial values, which are
listed in Tab.~\ref{tbl:parameters}.}
\begin{ruledtabular}
\begin{tabular}{crcccccc}
           &  \multicolumn{1}{c}{$a_1/a_2$} &  $J_{\mathrm{fin}}$ &  $J_{\mathrm{rad}}$ & $J_{\mathrm{rad}} / J_{_{\mathrm{ADM}}}$ &   $M_{\mathrm{fin}}$ &      $M_{\mathrm{rad}}$ &   $M_{\mathrm{rad}}/M_{_{\rm ADM}}$ \\ \hline
      $r0$ &      -1.00 &     0.6244 &     0.2008 &     0.2434 &     0.9536 &     0.0320 &     0.0325 \\

      $r1$ &      -0.75 &     0.6391 &     0.2222 &     0.2580 &     0.9507 &     0.0348 &     0.0353 \\

      $r2$ &      -0.50 &     0.6530 &     0.2449 &     0.2727 &     0.9482 &     0.0374 &     0.0380 \\

      $r3$ &      -0.25 &     0.6676 &     0.2670 &     0.2857 &     0.9461 &     0.0396 &     0.0402 \\

      $r4$ &       0.00 &     0.6827 &     0.2886 &     0.2971 &     0.9439 &     0.0420 &     0.0426 \\

      $r5$ &       0.25 &     0.6966 &     0.3106 &     0.3084 &     0.9412 &     0.0450 &     0.0456 \\

      $r6$ &       0.50 &     0.7075 &     0.3363 &     0.3222 &     0.9376 &     0.0488 &     0.0495 \\

      $r7$ &       0.75 &     0.7181 &     0.3626 &     0.3355 &     0.9344 &     0.0523 &     0.0530 \\

      $r8$ &       1.00 &     0.7292 &     0.3878 &     0.3471 &     0.9315 &     0.0557 &     0.0564 \\
\end{tabular}
\end{ruledtabular}
\label{tab:Radiation}
\end{table*}

We proceed next to a similar analysis for the conservation of the
mass-energy of the system by considering the difference between the
the initial mass and final plus the radiated masses. As for the
initial mass we obviously consider the ADM mass of the system
$M_{_{\rm ADM}}$, while the radiated energy $M_{\mathrm{rad}}$ is
computed through the gravitational waveforms~\cite{Nagar05,
Cunningham78, Cunningham79}
\begin{equation}
{d^2 E \over dt d\Omega} = {r^2 \over 16 \pi} \left( \left |
\dot{h}_+\right|^2 + \left | \dot{h}_\times \right|^2 \right)\,.
\label{eq:RadEnergy}
\end{equation}
As for the angular momenta, we have chosen to express the right hand
side of Eq.~(\ref{eq:RadEnergy}) in terms of the Zerilli-Moncrief
functions and to use as final mass of the black hole
$M_{\mathrm{fin}}$, the one given by Eq.~(\ref{eq:AHMass}) and
measured via the apparent horizon.

The left panel of Fig.~\ref{fig:Masses} shows $M_{_{\rm ADM}}$,
$M_{\mathrm{fin}}$ and $M_{\mathrm{rad}}$, with the latter rescaled
the radiated by a factor of ten to make it more visible. Also in this
case there is a clear linear behavior of both the radiated energy and
of the final mass of the black hole in terms of the spin ratio. As a
result, phenomenological expressions of the type~(\ref{jrad_phenom})
and~(\ref{jfin_phenom}) are possible also for $M_{\mathrm{fin}}$ and
$M_{\mathrm{rad}}$. The corresponding values of the coefficients
$\xi^{M}_{\rm rad,\,fin}$ and $\chi^{M}_{\rm rad,\,fin}$ are also
presented in Table~\ref{tab:Fit}.

Finally, to check the precision at which the energy is conserved, and
in analogy to Eq.~(\ref{eq:DeltaJOverJ}), we have computed the
relative error
\begin{equation}
{\Delta M \over M_{_{\mathrm{ADM}}}} \equiv
 {M_{\mathrm{fin}}+M_{\mathrm{rad}}-M_{_{\mathrm{ADM}}} \over
 M_{_{\mathrm{ADM}}}}\,,
\label{DeltaMoM}
\end{equation}
and plotted this as a function of the spin ratio in the right panel of
Fig.~\ref{fig:Masses}. Clearly, also the energy losses are extremely
small and for all the binaries in the sequence, the error in the
energy balance is below $0.52\%$ at the medium
resolution. Table~\ref{tab:Radiation} summarizes the numerical results
for the radiated energy and angular momentum for the members of the
sequence.

\section{Conclusions}

We have performed a highly-accurate study of recoil velocities in
binary black hole mergers from a sequence of equal-mass black holes
with varying spin configurations. In this sequence, the spins are
aligned with the orbital angular momentum since there are strong
indications that such alignment is preferred in astrophysical
situations. This makes our choice of initial data especially realistic
and our results particularly relevant also within an astrophysical
context.

In practice, the initial configurations are built so that the spin of
one of the black holes is kept at a constant dimensionless value
$a_2=0.584$ while the other varies from $a_1=-a_2$ to $a_1=+a_2$, thus
spanning a range between $-1$ and $1$ in spin ratio.  We have followed
our black hole evolutions for about two to four orbits and then
throughout the plunge, merger, and ringdown phases. This work thus
extends and refines recent results obtained from a reduced but similar
initial-data sequence~\cite{Koppitz-etal-2007aa}.

The main aspects of this work, which revolve around the methods
used, the tests performed and the results obtained, can be summarized
as follows.

\emph{Methods.} To increase the significance of our results and our
confidence in their accuracy, we have implemented two independent
methods for the calculation of the linear momentum from the emitted
gravitational radiation. These are based on either the measure the
Newman-Penrose scalar $\Psi_4$ or on the calculation of the
gauge-invariant perturbations of a Schwarzschild black hole
$Q^{\times, +}_{\ell m}$. Overall, we find that both methods of
calculating the linear momentum loss agree excellently and we are thus
able to obtain accurate recoil measurements with error bars of
$5\,\mathrm{km/s}$ for the anti-aligned spin binaries and of
$8\,\mathrm{km/s}$ in the aligned cases.

Such a good agreement, however, is attainable \textit{only} if the
initial transient in the waveform is properly taken into account. The
transient is produced by the use of initial data not containing the
net linear momentum the system has accumulated since inspiralling from
infinite separation. We discuss the importance of choosing the correct
vector integration constant when calculating the radiated linear
momentum and describe an unambiguous method for doing so.

We remark that a proper choice of this constant is essential not only
because it influences the final recoil velocity with differences of
$10\%$ and more, but also because it allows for a systematic
interpretation of the results. Without it, in fact, the correct
functional dependence of the final recoil velocity on the spin ratio
is irremediably lost and a comparison with the PN prediction
impossible. Last but not least, a proper integration constant can
result in a significant saving of computational time, allowing
simulations to start at much smaller initial separations without
sacrificing accuracy.

\emph{Tests.} In order to show the accuracy of our results, we
demonstrate that both the Zerilli-Moncrief gauge invariant waveforms
and the Einstein tensor converge with an order between three and four,
which is the expected convergence behavior of our numerical methods.

Furthermore, because the Newman-Penrose scalar $\Psi_4$ serves as a
measure for the radiation content of the spacetime in appropriately
chosen frames and at sufficiently large distances from the source, we
show that the peeling property is indeed well satisfied in our
numerical simulations. In particular, we demonstrate that both the
gravitational wave information $\Psi_4$ and the gauge-wave information
$\Psi_3$ satisfy the expected scaling with radius. Similarly, we also
show that, as expected, the gauge invariant quantity $Q^+_{22}$ does
not vary with radius.

Finally, we investigate those systematic effects that may influence
our gravitational-wave measurements. In particular, we study the
effects that the choice of the extraction radius has on the final kick
velocity and find little influence for $r_{_{\rm E}} \ge 30\,M$.
Based on this, we choose $r_{_{\rm E}}=50\,M$ as the fiducial extraction radius
in this paper. Furthermore, to exclude that the effects of the
eccentricity in our initial data are significant for this paper, we
artificially increase or reduce the eccentricity of the initial data
by comparatively large amounts. Also in this case we find that the
differences in the recoil velocities are below the estimated
error-bars. Altogether, the set of tests carried out gives us
confidence that our waveforms and recoil velocities are both correct
and accurate.

\emph{Results.} Using the mathematical and numerical setup as
described and tested above, we have investigate the dependence of the
recoil velocity on the initial data parameters and most notably on the
spin ratio $a_1/a_2$. As expected, a larger asymmetry in the initial
conditions causes a larger recoil, with a velocity of
about $262\,\mathrm{km/s}$ for a binary of equal and anti-aligned
spins, and a numerically computed recoil of $10^{-9}\,{\rm km/s}$ for
a binary of equal and aligned spins.

Using such accurate measurements, we have then studied the functional
dependence of the recoil velocity on the spin ratio finding that a
quadratic behavior reproduces very well the numerical results and
corrects the post-Newtonian prediction of a linear dependence. We
summarize this behavior in a phenomenological expression that can be
readily employed in astrophysical studies on the evolution of binary
black holes in massive galaxies.

With a straightforward extrapolation of the quadratic dependence to
the maximal spinning case $a_1=-a_2=1$ we obtain $448\pm
5\,\mathrm{km/s}$ as the maximal possible recoil velocity attainable
from a binary system of equal-mass black holes with spins aligned to
the orbital angular momentum. This recoil velocity is in very good
agreement with our previous estimate made in
Ref.~\cite{Koppitz-etal-2007aa} with a smaller sequence and in equally
good agreement with the results reported in
Ref.~\cite{Herrmann:2007ac}.

As mentioned above, the inclusion of the integration constant has been
essential to obtain physically consistent results. At the same time,
its investigation has allowed to highlight some important features of
the evolution of the recoil velocity in vector space. Most
importantly, it has shown that even when all non-spherical modes up to
$\ell=7$ are taken into account, the recoil is dominated by lower mode
contributions, especially $\ell=2, m=-2, 1, 2$ and $\ell=3, m=-3$. The
interplay of these contributions in vector space and during ringdown
is what is responsible for the rich features observed in the final
evolution of the recoil velocity.

Finally, we provide accurate measurements of the radiated energy and
angular momentum. These measurements reveal a clear linear dependence
on the spin ratio $a_2/a_1$, and we derive phenomenological
expressions for the relative losses of angular momentum and the
relative spin-up of the final black hole. These relations can be
easily used in $N$-body simulations if the interaction of binary black
holes is to be taken into account, and when an accurate estimate on
the final distribution of black hole spins is important.

\begin{acknowledgments}

It is a pleasure to thank Thibault Damour for interesting
discussions. AN and ES also acknowledge kind hospitality from the AEI
during the development of part of this work.  The computations were
performed on the clusters Peyote, Belladonna, and Damiana of the Albert
Einstein Institute. This work was supported in part by the DFG grant
SFB/Transregio~7 ``Gravitational Wave Astronomy''.

\end{acknowledgments}

\begin{appendix}

\section{On the convergence tests}
\label{appendix_conv}

The effects of the initial transient modes can last for different
amounts of time for the different resolutions.  A comparison of the
$Q^+_{22}$ waveforms between the three resolutions confirms this
shift in time -- the waveform maxima are seen at slightly different
times for the different resolutions.  We attempt to undo this effect
by manually shifting the time-coordinate of the medium and high
resolution runs
\begin{equation}
t \rightarrow t + \delta t.
\label{eq:timeshiftingdef}
\end{equation}
The value of $\delta t$ is set for the medium and high resolution runs
independently, using the minimization condition
\begin{equation}
\frac{\partial}{\partial(\delta t)}
\int_{150}^{170} | Q(t\rightarrow t+\delta t) -  Q_{\rm vhigh} |^2 dt = 0.
\label{eq:timeshiftingcalc}
\end{equation}
This effectively means aligning in time the peak amplitude of the
three runs, at $t\approx 160\,M$. Solving
Eq.~\eqref{eq:timeshiftingcalc} numerically for the $Q^+_{22}$
waveforms gives
\begin{equation}
\delta t_{0.024} = 0.4756 \quad \mbox{and} \quad
\delta t_{0.018} = 0.1078 .
\label{eq:timeshiftingval}
\end{equation}

Applying the time-shifting condition Eq.~\eqref{eq:timeshiftingdef}
to the coarse and medium resolution data, and inserting the result into
Eqs.~\eqref{eq:rhodef}--\eqref{eq:rhoeq} gives convergence rates
that are consistent with the theoretical expectations.

\begin{figure*}[htbp]
\begin{center}
\resizebox{8.5cm}{!}{\includegraphics[]{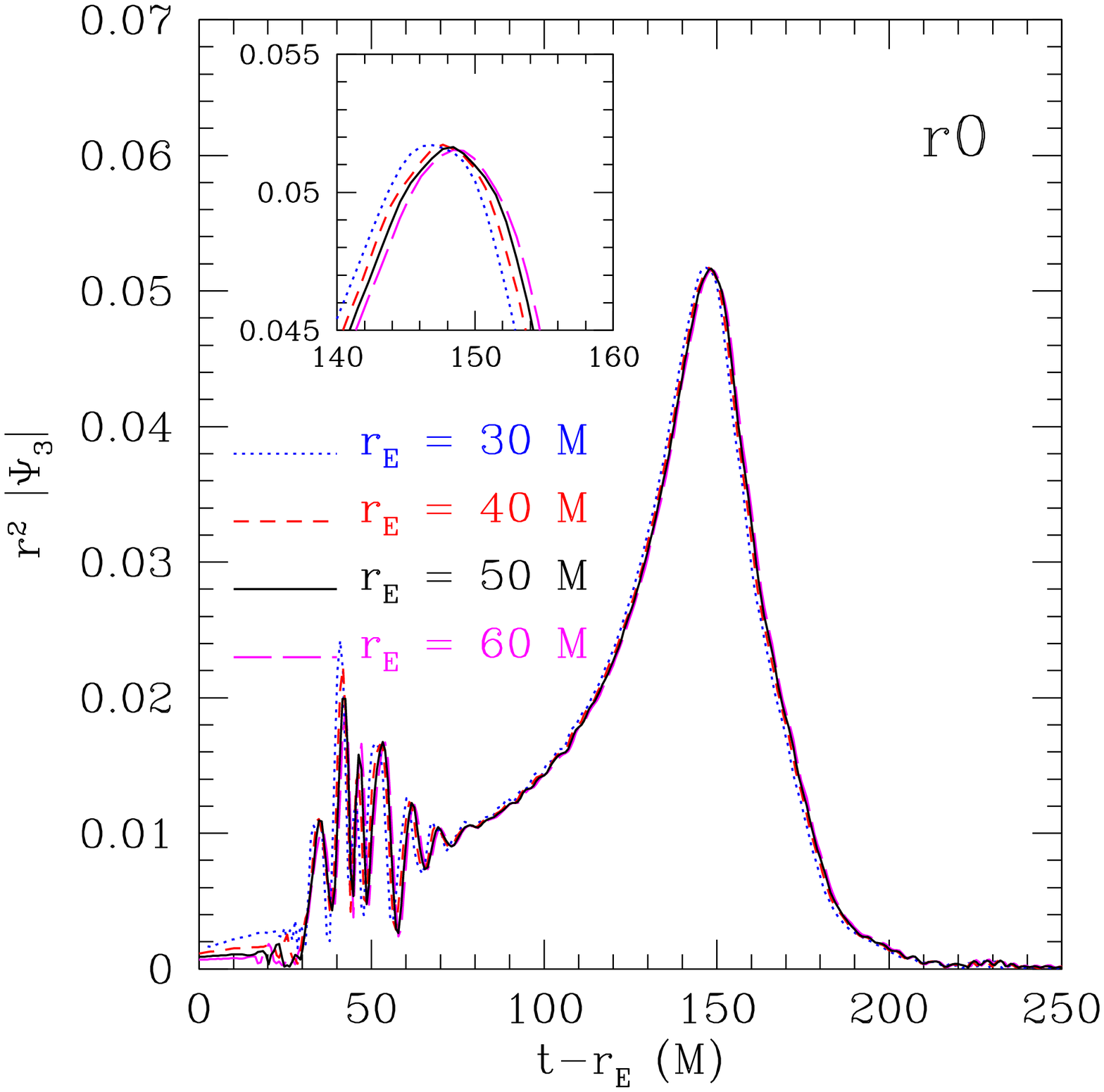}}
	\hskip 0.5cm
\resizebox{8.5cm}{!}{\includegraphics[]{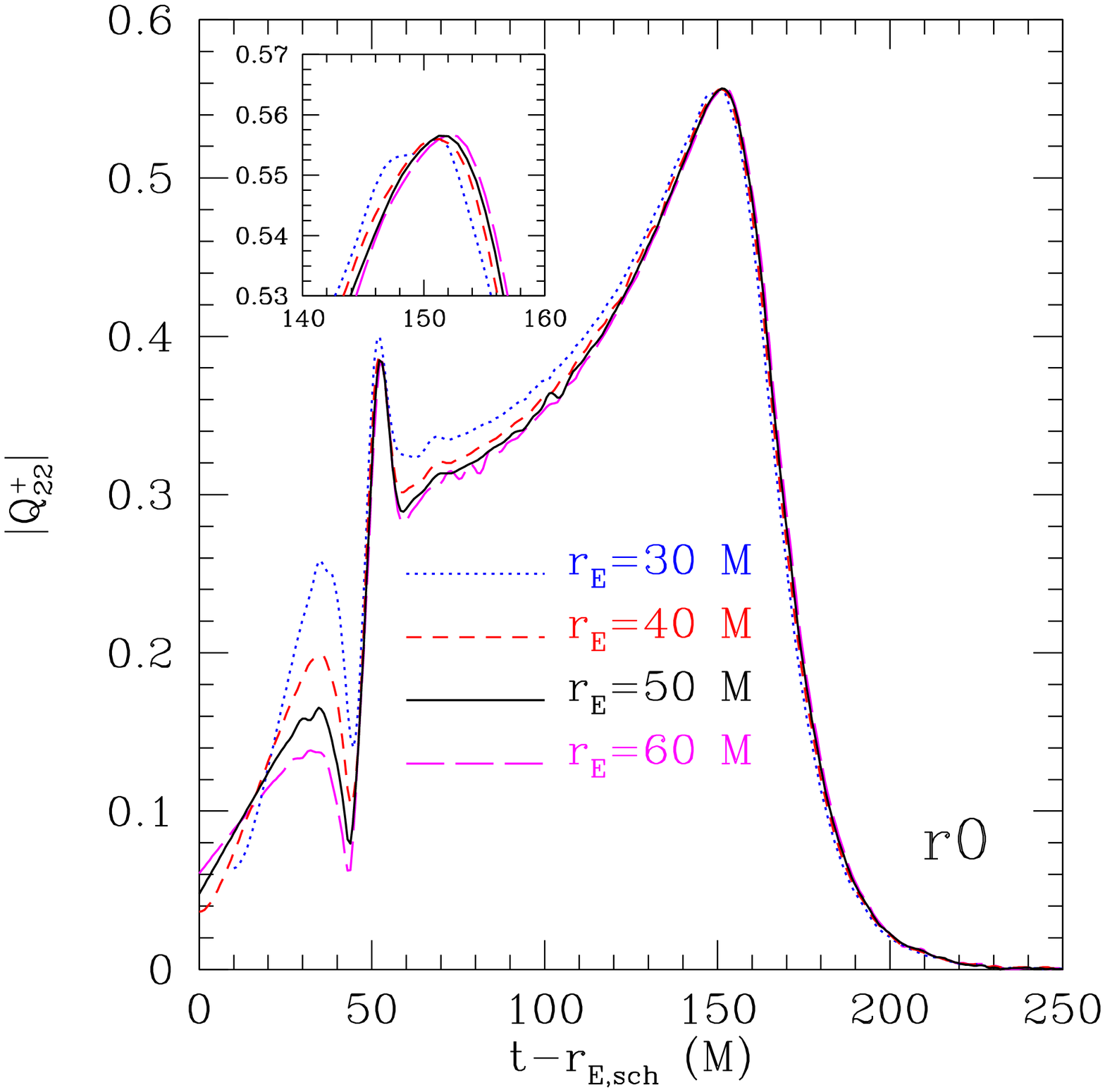}}
\caption{\textit{Left panel:} Evidence that the conditions for the
  Peeling theorem are met also for $\Psi_{3}$, which scales as
  $r^{-2}$ when extracted at isotropic radii $r_{_{\rm E}}=30\,M$,
  $40\,M$, $50\,M$, and $60\,M$. This figure should be compared to the
  corresponding Fig.~\ref{fig:y0_peeling_psi4}. \textit{Right panel:}
  The same as the left panel but for the gauge-invariant quantity
  $Q^+_{22}$, which is shown to be constant when extracted at
  isotropic radii $r_{_{\rm E}}=30\,M$, $40\,M$, $50\,M$, and
  $60\,M$.}
\label{fig:y0_peeling_psi3}
\end{center}
\end{figure*}

In Table~\ref{tbl:Qconv} we report the convergence rates as calculated
from Eq.~\ref{eq:rhodef} for the time interval $0 \leq u \leq 190$
($u$ is the retarded time as defined in Sec.~\ref{sec:ct})
which excludes the initial burst but contains the rest of the
waveform. We see close to fourth-order convergence for the $\ell=2$
modes $Q^+_{22}$ and $Q^\times_{21}$. The $\ell=m=3$ mode $Q^+_{33}$,
on the other hand, shows second order convergence in phase, which is
most likely related to the fact that the magnitude of this mode is the
same size as the finite difference error in $Q^+_{22}$ and is a factor
of $40$ smaller than the magnitude of $Q^+_{22}$ itself.

The final kick-velocity magnitude in units of $\mathrm{km/s}$ is 
\begin{equation}
|v|_{\rm kick} = 263.49, \quad 259.75, \quad \mbox{and} \quad 261.00
\end{equation}
for the medium, high and very-high resolutions.
This gives $\rho(|v|_{\rm kick}) = 2.98$ which can be inserted
into Eq.~\eqref{eq:rhoeq}
to obtain a calculated convergence rate of $4.32$.

\begin{table}[t]
\caption{Integrated convergence rates of the Zerilli-Moncrief
gauge-invariant variables providing the dominant contribution in the
kick-velocity measurements. As the numbers indicate, we achieve at
least third order convergence both in amplitude and phase. A time-shift
as given by
Eqs.~\eqref{eq:timeshiftingdef}--\eqref{eq:timeshiftingval} was made
on the raw data to remove the near cancellation of the lowest-order
error terms.}
\begin{ruledtabular}
\begin{tabular}{ccccccc}
$Q$ &
\multicolumn{2}{c}{$Q^\times_{21} $}&
\multicolumn{2}{c}{$Q^+_{22}$} &
\multicolumn{2}{c}{$Q^+_{33}$} \\
$r_{_{\rm E}}/M$&
amp & phase &
amp & phase &
amp & phase \\
\hline
$30$ & $4.51$ & $3.95$ & $4.65$ & $4.31$ & $4.32$ & $2.13$ \\
$40$ & $4.08$ & $3.70$ & $4.61$ & $4.34$ & $4.26$ & $2.62$ \\
$50$ & $3.83$ & $4.44$ & $4.35$ & $4.76$ & $4.02$ & $2.39$ \\
\end{tabular}
\end{ruledtabular}
\vskip -0.25cm
\label{tbl:Qconv}
\end{table}

\section{Details on the extraction of $\Psi_4$}
\label{appendix_a}

The numerical solution of Eqs.~(\ref{eq:Pdot}) involves first an
interpolation of $\Psi_4$ as calculated according to Eqs.~(\ref{eq:psi4_adm}) 
from its values on the Cartesian grid to
those onto the extraction sphere by using fourth-order Lagrange
interpolants. Because of the symmetry across the $z=0$ plane the
interpolation is effectively done on the upper hemisphere only, thus
using a spherical coordinate system with $\theta, \phi \in[0, \pi/2]
\times [0, 2\pi]$ and applying cell-centered discretization along the
$\theta$-direction to avoid the coordinate singularities at the poles
on the sphere.  

The angular resolution is chosen so that the spacings $\Delta\theta$
and $\Delta\phi$ are equal and of the same order as the corresponding
Cartesian spacings of the refinement level in which the largest
extraction 2-sphere is located. As an example, for the fiducial finest
resolution of $h = 0.024\,M$, the largest extraction radius is at
$r_{_{\rm E}}=60\,M$ and in a region covered by the second refinement
level with spacing $\Delta_{rl=2}^{0.024}=1.536 \, M$. To obtain an
equivalent spacing on the 2-sphere, we solve for $\Delta\theta$ and
$\Delta\phi$ such that
\begin{equation} 
r_{_{\rm E}} \Delta\theta = r_{_{\rm E}} \Delta\phi \approx \Delta_{rl=2}^{0.024}= 1.536 \, M\,.
\end{equation}
The resulting number of grid points is $N_{\theta}=56$ along the
$\theta$-direction and $N_{\phi}=224$ along the $\phi$-direction.

After interpolation onto the extraction sphere, we first calculate the
time integral of $\Psi_4|_{S^2}$ and afterwards, the surface integral
of the absolute square of the former according to
Eqs.~(\ref{eq:Pdot}). These integrals are both computed using
fourth-order schemes. In particular, for the surface integral, we use
Simpson's rule in the form
\begin{eqnarray}
\int_{x_0}^{x_N}dx\, f(x) &\approx& \Delta x
\left[ \frac{17}{48} f_0 + \frac{59}{48} f_1 + \frac{43}{48} f_2 +
\frac{49}{48} f_3 \right. 
\nonumber \\ && 
\left.\qquad + \langle f_k
\rangle \right. 
\nonumber \\ && 
\hskip -1.75cm + \frac{49}{48} f_{N-3} +
\frac{43}{48} f_{N-2} 
\left. +
\frac{59}{48} f_{N-1} + \frac{17}{48} f_N \right]\,,
\label{eq:variant_Simpsons_rule} 
\end{eqnarray}
where $\langle f_k \rangle$ is the sum over all $f_k$ with $3\, <\,
k\, <\, N-3$. The integral over $d\theta d\phi$ is obtained by
computing the tensor product of the RHS of
Eqs.~(\ref{eq:variant_Simpsons_rule}), \textit{i.e.},
\bea
\int_{\theta_0}^{\theta_N}d\theta \int_{\phi_0}^{\phi_N}d\phi\,
f(\theta,\, \phi) \approx
\Delta\theta\Delta\phi\sum_{i=0}^{N_\theta}\sum_{j=0}^{N_\phi} c_i c_j\,
f_{ij} \,, 
\nonumber \\ 
\eea 
where the $c_i, c_j$ are the coefficients in the RHS of
Eqs.~(\ref{eq:variant_Simpsons_rule}).

The time integral of Eqs.~(\ref{eq:Pdot}) is generically calculated by
using the fourth-order Simpson's rule in such a way that the integral for
the time step $k$ uses only past time steps $i$ with $0 \le i \le k$.
Care is required for the very first time steps, for which we have less
than 7 evaluations of the integrand. In this case, we use the
2nd-order accurate trapezoid rule if $N=1, 3$, or $5$
\beq
\int_{x_0}^{x_N}dx\, f(x) \approx
\Delta x \left[
           \frac{1}{2} f_0 
         + \langle f_k \rangle
         + \frac{1}{2} f_N
         \right] \,,
\label{eq:trapezoid_rule}
\eeq
or the fourth-order accurate Simpson's rule 
\bea
\int_{x_0}^{x_N}dx\, f(x) &\approx&
\Delta x \left[
           \frac{1}{3} f_0
         + \frac{4}{3} f_1 \nonumber \right. \\
&& \left.
         + \langle \frac{2}{3} f_{2k} + \frac{4}{3} f_{2k+1} \rangle
         + \frac{1}{3} f_N
         \right]\,,
\eea
if $N=2, 4$ or $6$. For $N \geq 7$ we simply use Simpson's rule in the
form~(\ref{eq:variant_Simpsons_rule}). It should be noted that the use
of a higher-order time integration scheme improves the overall
accuracy in the calculation of the final recoil velocity by more than
a factor of 10.

\section{A comparison of wave-extraction methods}
\label{appendix_b}

In Fig.~\ref{fig:y0_peeling_psi4}, we have shown that $\Psi_4$ as
extracted at different radii correctly scales with the $1/r$ falloff
as predicted by the peeling theorem.  Here, we also check if all other
components of the Weyl tensor exhibit the correct
$r^{5-n}\Psi_{n}={\rm const.}$ scaling.

The left panel of Fig.~\ref{fig:y0_peeling_psi3} indeed shows that
the scaling property of all $\Psi_n$ behave as expected. In the course
of the same analysis, it is also worth looking at the waveforms as
calculated by using the gauge-invariant formalism. In particular, we
focus on the real part of the $\ell=2, m=2$ even parity wave mode
$Q_{22}^{+}$ and check for the correct scaling for the different
extraction radii. The right panel of Fig.~\ref{fig:y0_peeling_psi3}
shows that $Q^+_{22}$ is constant for all extraction radii as
expected.

\begin{figure}[htbp]
\begin{center}
\resizebox{8.5cm}{!}{
\includegraphics[]{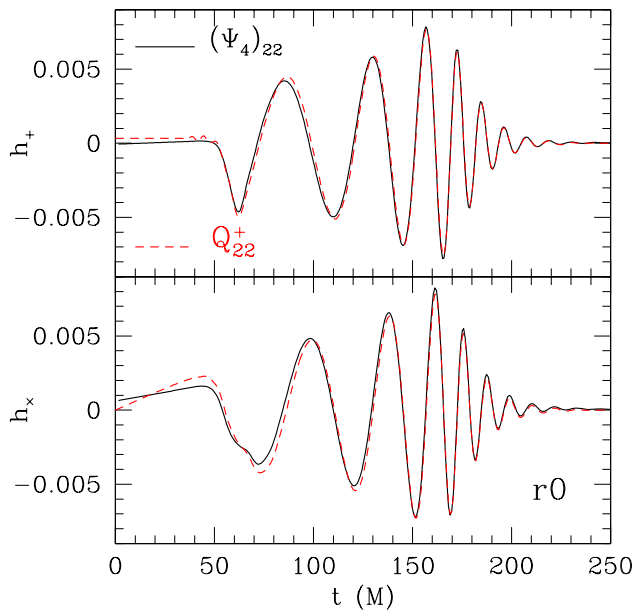}}
\caption{Comparison of the two polarization amplitudes $h_+$ (upper
  graph) and $h_{\times}$ (lower graph) as computed with $\Psi_4$
  (continuous black line) or with the gauge invariant quantities
  $Q^+_{\ell m}$ (dashed red line). Note the two polarizations are
  computed using the lowest (and dominant) multipole $\ell=2,\ m=2$
  and are extracted at $r_{_{\rm E}}=50\,M$.}
\label{fig:hplus_hcross}
\end{center}
\end{figure}

As a final remark, we will also compare the $h_{+}$ and $h_\times$ as
calculated by using the odd and even master functions in the
gauge-invariant formalism according to Eq.~(\ref{eq:wave_gi}) and the
spin-weighted spherical harmonic amplitudes of the Weyl component
$\Psi_4^{\ell m}$ decomposed on the extraction spheres.  Using these
amplitudes, the metric perturbations $h_+, h_\times$ recovered by a
double time integral of Eq.~(\ref{eq:psi4_h})
\beq h_+ -
{\rm i} h_\times = \lim_{r\rightarrow \infty} \sum_{\ell,m} \int_{0}^{t}dt'
\int_{0}^{t'}dt'' \Psi_4^{\ell m}\, _{-2}Y_{\ell m} \ .  \label{eq:h_psi4} 
\eeq 
The numerical integration of Eq.~(\ref{eq:h_psi4}) requires knowledge
of an integration constant for the calculation of the second integral
to eliminate the linear offset. This constant is determined by
searching for minima in the $\Psi_4^{\ell m}$ mode and averaging over them. The
resulting value is used as the integration constant. In both cases, we
only consider the dominant contribution from mode $\ell=2\, , m=2$.

\section{On the influence of orbital eccentricity}

\begin{figure*}[t]
\begin{center}
\centerline{
\resizebox{8.5cm}{!}{\includegraphics[angle=-0]{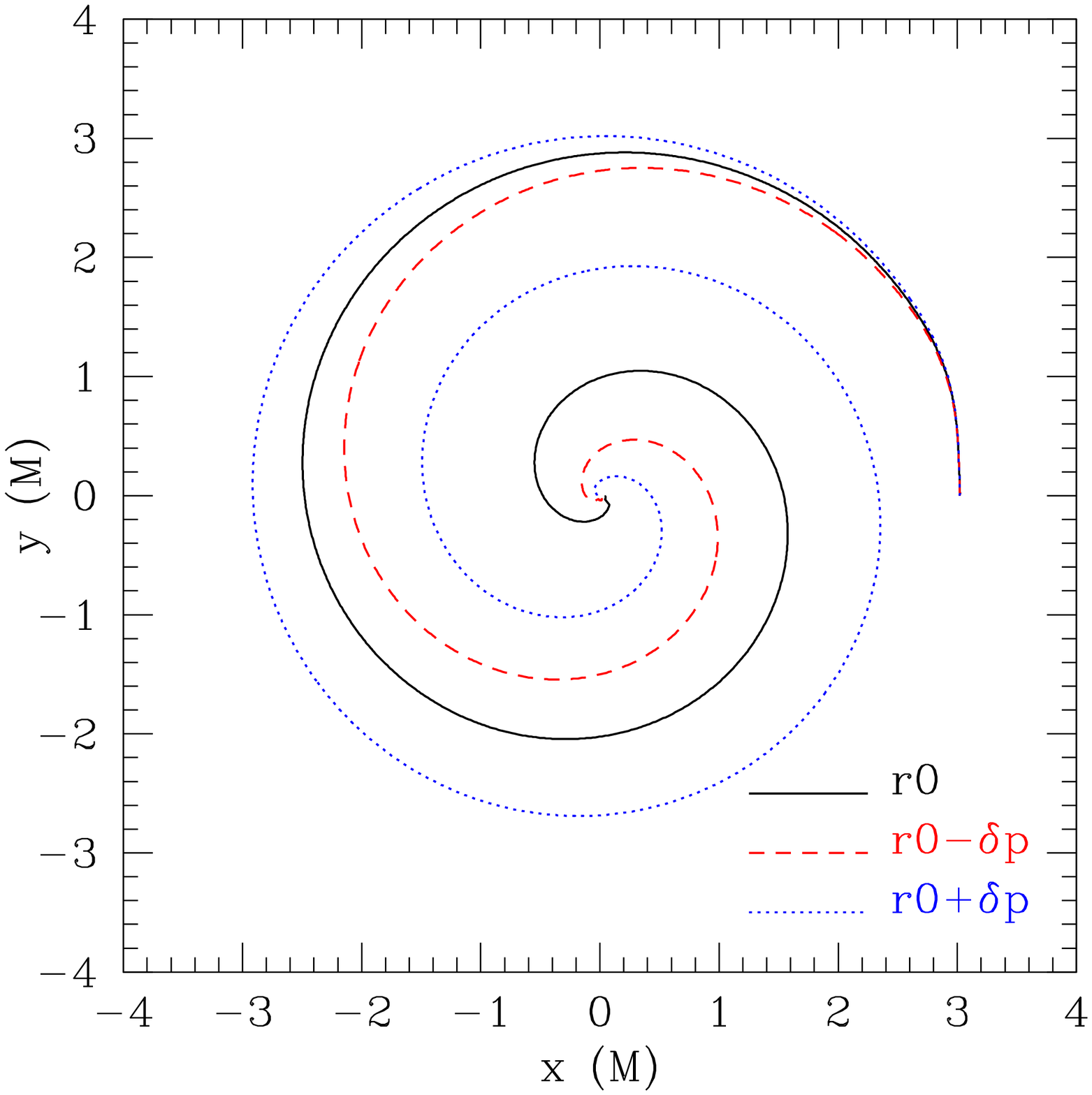}}
 \hskip 0.5cm
\resizebox{8.5cm}{!}{\includegraphics[angle=-0]{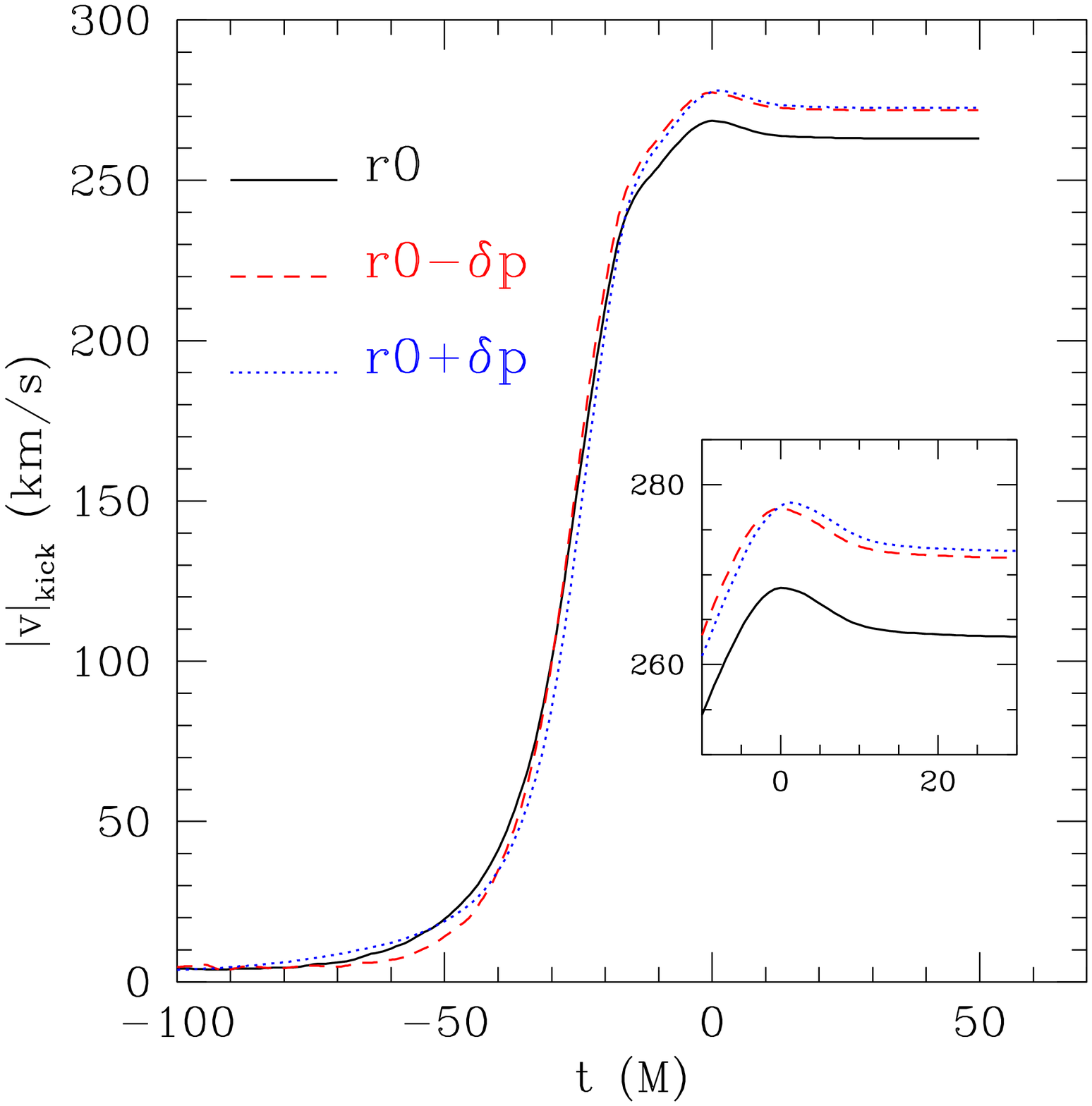}}
}
  \caption{\textit{Left panel:} Coordinate trajectories for one of the
    black holes for the $r0$ compared with similar models where the
    initial linear momenta have been changed by $\pm 3\%$ in order to
    modify the eccentricity of the inspiral. \textit{Right panel:}
    Recoil velocity for the $r0$ case is compared with similar models
    for which the initial eccentricity has been increased by adding
    and subtracting 3\% of the initial linear momentum of the black
    holes relative to the $r0$ values.  The effect of increased
    eccentricity in the final merger is to increase the size of the
    kick, by about 4\% in both cases.}
  \label{fig:trajectory_y0_vs_ecc}
\end{center}
\end{figure*}

Another source of potential error in calculating a ``physical'' kick
comes from the choice of initial data parameters. Our evolutions begin
from fairly close separations, comprising at most the last 2-3
orbits. As such, parameters for quasi-circular orbits determined by
the effective potential method, give only approximations to the true
orbital parameters for black holes that have spiraled in from
infinity, and it is known that the method produces a non-trivial
residual eccentricity for initial data at close separation. This
eccentricity can have significant effects on the orbital trajectories
before merger, and a potential influence on the calculated recoil. To
test this we have evolved two modified $r0$ models, one in which the
initial linear momenta of the black holes is $3\%$ larger than that
specified in Table~\ref{tbl:parameters}, and another in which the
linear momenta are $3\%$ smaller. The modified momenta have the effect
of changing the orbital energy of the bodies from the minima
determined by the effective potential method, introducing an
additional eccentricity to the evolution. The resulting black hole
trajectories and kick determinations are shown respectively in
Fig.~\ref{fig:trajectory_y0_vs_ecc}. We see that although the level of
applied eccentricity is large, and in fact much larger than the expected
eccentricity due to the intrinsic inaccuracy of the effective
potential method, it modifies the recoil by only about
$10\,\mathrm{km/s}$, that is, $~4\%$. Further, in both the high and
low energy cases, the recoil is increased over the fiducial $r0$ case,
suggesting that increased eccentricity generically leads to a slightly
larger recoil.

\end{appendix}

\bibliography{aeireferences}

\begin{thebibliography}{73}
\expandafter\ifx\csname natexlab\endcsname\relax\def\natexlab#1{#1}\fi
\expandafter\ifx\csname bibnamefont\endcsname\relax
  \def\bibnamefont#1{#1}\fi
\expandafter\ifx\csname bibfnamefont\endcsname\relax
  \def\bibfnamefont#1{#1}\fi
\expandafter\ifx\csname citenamefont\endcsname\relax
  \def\citenamefont#1{#1}\fi
\expandafter\ifx\csname url\endcsname\relax
  \def\url#1{\texttt{#1}}\fi
\expandafter\ifx\csname urlprefix\endcsname\relax\def\urlprefix{URL }\fi
\providecommand{\bibinfo}[2]{#2}
\providecommand{\eprint}[2][]{\url{#2}}

\bibitem[{\citenamefont{Peres}(1962)}]{peres:1962}
\bibinfo{author}{\bibfnamefont{A.}~\bibnamefont{Peres}},
  \bibinfo{journal}{Phys. Rev.} \textbf{\bibinfo{volume}{128}},
  \bibinfo{pages}{2471} (\bibinfo{year}{1962}).

\bibitem[{\citenamefont{Bekenstein}(1973)}]{Bekenstein:1973mi}
\bibinfo{author}{\bibfnamefont{J.~D.} \bibnamefont{Bekenstein}},
  \bibinfo{journal}{Phys. Rev.} \textbf{\bibinfo{volume}{D7}},
  \bibinfo{pages}{949} (\bibinfo{year}{1973}).

\bibitem[{\citenamefont{Fitchett}(1983)}]{fitchett:1983}
\bibinfo{author}{\bibfnamefont{M.~J.} \bibnamefont{Fitchett}},
  \bibinfo{journal}{Mon. Not. R. astr. Soc.} \textbf{\bibinfo{volume}{203}},
  \bibinfo{pages}{1049} (\bibinfo{year}{1983}).

\bibitem[{\citenamefont{{Fitchett} and
  {Detweiler}}(1984)}]{1984MNRAS.211..933F}
\bibinfo{author}{\bibfnamefont{M.~J.} \bibnamefont{{Fitchett}}}
  \bibnamefont{and}
  \bibinfo{author}{\bibfnamefont{S.}~\bibnamefont{{Detweiler}}},
  \bibinfo{journal}{Mon. Not. R. astr. Soc.} \textbf{\bibinfo{volume}{211}},
  \bibinfo{pages}{933} (\bibinfo{year}{1984}).

\bibitem[{\citenamefont{Favata et~al.}(2004)\citenamefont{Favata, Hughes, and
  Holz}}]{Favata:2004wz}
\bibinfo{author}{\bibfnamefont{M.}~\bibnamefont{Favata}},
  \bibinfo{author}{\bibfnamefont{S.~A.} \bibnamefont{Hughes}},
  \bibnamefont{and} \bibinfo{author}{\bibfnamefont{D.~E.} \bibnamefont{Holz}},
  \bibinfo{journal}{Astrophys. J.} \textbf{\bibinfo{volume}{607}},
  \bibinfo{pages}{L5} (\bibinfo{year}{2004}), \eprint{astro-ph/0402056}.

\bibitem[{\citenamefont{Herrmann et~al.}(2006)\citenamefont{Herrmann,
  Shoemaker, and Laguna}}]{Herrmann:2006ks}
\bibinfo{author}{\bibfnamefont{F.}~\bibnamefont{Herrmann}},
  \bibinfo{author}{\bibfnamefont{D.}~\bibnamefont{Shoemaker}},
  \bibnamefont{and} \bibinfo{author}{\bibfnamefont{P.}~\bibnamefont{Laguna}}
  (\bibinfo{year}{2006}), \eprint{gr-qc/0601026}.

\bibitem[{\citenamefont{Baker et~al.}(2006{\natexlab{a}})}]{Baker:2006vn}
\bibinfo{author}{\bibfnamefont{J.~G.} \bibnamefont{Baker}}
  \bibnamefont{et~al.}, \bibinfo{journal}{Astrophys. J.}
  \textbf{\bibinfo{volume}{653}}, \bibinfo{pages}{L93}
  (\bibinfo{year}{2006}{\natexlab{a}}), \eprint{astro-ph/0603204}.

\bibitem[{\citenamefont{Gonzalez et~al.}(2007)\citenamefont{Gonzalez, Sperhake,
  Bruegmann, Hannam, and Husa}}]{Gonzalez:2006md}
\bibinfo{author}{\bibfnamefont{J.~A.} \bibnamefont{Gonzalez}},
  \bibinfo{author}{\bibfnamefont{U.}~\bibnamefont{Sperhake}},
  \bibinfo{author}{\bibfnamefont{B.}~\bibnamefont{Bruegmann}},
  \bibinfo{author}{\bibfnamefont{M.}~\bibnamefont{Hannam}}, \bibnamefont{and}
  \bibinfo{author}{\bibfnamefont{S.}~\bibnamefont{Husa}},
  \bibinfo{journal}{Phys. Rev. Lett.} \textbf{\bibinfo{volume}{98}},
  \bibinfo{pages}{091101} (\bibinfo{year}{2007}), \eprint{gr-qc/0610154}.

\bibitem[{\citenamefont{Herrmann et~al.}(0100)\citenamefont{Herrmann, Hinder,
  Shoemaker, Laguna, and Matzner}}]{Herrmann:2007ac}
\bibinfo{author}{\bibfnamefont{F.}~\bibnamefont{Herrmann}},
  \bibinfo{author}{\bibfnamefont{I.}~\bibnamefont{Hinder}},
  \bibinfo{author}{\bibfnamefont{D.}~\bibnamefont{Shoemaker}},
  \bibinfo{author}{\bibfnamefont{P.}~\bibnamefont{Laguna}}, \bibnamefont{and}
  \bibinfo{author}{\bibfnamefont{R.~A.} \bibnamefont{Matzner}}
  (\bibinfo{year}{0100}), \eprint{gr-qc/0701143}.

\bibitem[{\citenamefont{Koppitz
  et~al.}(2007{\natexlab{a}})\citenamefont{Koppitz, Pollney, Reisswig,
  Rezzolla, Thornburg, Diener, and Schnetter}}]{Koppitz-etal-2007aa}
\bibinfo{author}{\bibfnamefont{M.}~\bibnamefont{Koppitz}},
  \bibinfo{author}{\bibfnamefont{D.}~\bibnamefont{Pollney}},
  \bibinfo{author}{\bibfnamefont{C.}~\bibnamefont{Reisswig}},
  \bibinfo{author}{\bibfnamefont{L.}~\bibnamefont{Rezzolla}},
  \bibinfo{author}{\bibfnamefont{J.}~\bibnamefont{Thornburg}},
  \bibinfo{author}{\bibfnamefont{P.}~\bibnamefont{Diener}}, \bibnamefont{and}
  \bibinfo{author}{\bibfnamefont{E.}~\bibnamefont{Schnetter}}
  (\bibinfo{year}{2007}{\natexlab{a}}), \eprint{gr-qc/0701163}.

\bibitem[{\citenamefont{Gonzalez et~al.}(0200)\citenamefont{Gonzalez, Hannam,
  Sperhake, Brugmann, and Husa}}]{Gonzalez:2007hi}
\bibinfo{author}{\bibfnamefont{J.~A.} \bibnamefont{Gonzalez}},
  \bibinfo{author}{\bibfnamefont{M.~D.} \bibnamefont{Hannam}},
  \bibinfo{author}{\bibfnamefont{U.}~\bibnamefont{Sperhake}},
  \bibinfo{author}{\bibfnamefont{B.}~\bibnamefont{Brugmann}}, \bibnamefont{and}
  \bibinfo{author}{\bibfnamefont{S.}~\bibnamefont{Husa}}
  (\bibinfo{year}{0200}), \eprint{gr-qc/0702052}.

\bibitem[{\citenamefont{Campanelli
  et~al.}(2007{\natexlab{a}})\citenamefont{Campanelli, Lousto, Zlochower, and
  Merritt}}]{Campanelli:2007ew}
\bibinfo{author}{\bibfnamefont{M.}~\bibnamefont{Campanelli}},
  \bibinfo{author}{\bibfnamefont{C.~O.} \bibnamefont{Lousto}},
  \bibinfo{author}{\bibfnamefont{Y.}~\bibnamefont{Zlochower}},
  \bibnamefont{and} \bibinfo{author}{\bibfnamefont{D.}~\bibnamefont{Merritt}},
  \bibinfo{journal}{Astrophys. J.} \textbf{\bibinfo{volume}{659}},
  \bibinfo{pages}{L5} (\bibinfo{year}{2007}{\natexlab{a}}),
  \eprint{gr-qc/0701164}.

\bibitem[{\citenamefont{Campanelli
  et~al.}(2007{\natexlab{b}})\citenamefont{Campanelli, Lousto, Zlochower, and
  Merritt}}]{Campanelli:2007cg}
\bibinfo{author}{\bibfnamefont{M.}~\bibnamefont{Campanelli}},
  \bibinfo{author}{\bibfnamefont{C.~O.} \bibnamefont{Lousto}},
  \bibinfo{author}{\bibfnamefont{Y.}~\bibnamefont{Zlochower}},
  \bibnamefont{and} \bibinfo{author}{\bibfnamefont{D.}~\bibnamefont{Merritt}},
  \bibinfo{journal}{Phys. Rev. Lett.} \textbf{\bibinfo{volume}{98}},
  \bibinfo{pages}{231102} (\bibinfo{year}{2007}{\natexlab{b}}),
  \eprint{gr-qc/0702133}.

\bibitem[{\citenamefont{Bruegmann et~al.}(0700)\citenamefont{Bruegmann,
  Gonzalez, Hannam, Husa, and Sperhake}}]{Bruegmann:2007zj}
\bibinfo{author}{\bibfnamefont{B.}~\bibnamefont{Bruegmann}},
  \bibinfo{author}{\bibfnamefont{J.}~\bibnamefont{Gonzalez}},
  \bibinfo{author}{\bibfnamefont{M.}~\bibnamefont{Hannam}},
  \bibinfo{author}{\bibfnamefont{S.}~\bibnamefont{Husa}}, \bibnamefont{and}
  \bibinfo{author}{\bibfnamefont{U.}~\bibnamefont{Sperhake}}
  (\bibinfo{year}{0700}), \eprint{arXiv:0707.0135 [gr-qc]}.

\bibitem[{\citenamefont{Schnittman}(2004)}]{Schnittman:2004vq}
\bibinfo{author}{\bibfnamefont{J.~D.} \bibnamefont{Schnittman}},
  \bibinfo{journal}{Phys. Rev.} \textbf{\bibinfo{volume}{D70}},
  \bibinfo{pages}{124020} (\bibinfo{year}{2004}), \eprint{astro-ph/0409174}.

\bibitem[{\citenamefont{Bogdanovic et~al.}(0300)\citenamefont{Bogdanovic,
  Reynolds, and Miller}}]{Bogdanovic:2007hp}
\bibinfo{author}{\bibfnamefont{T.}~\bibnamefont{Bogdanovic}},
  \bibinfo{author}{\bibfnamefont{C.~S.} \bibnamefont{Reynolds}},
  \bibnamefont{and} \bibinfo{author}{\bibfnamefont{M.~C.} \bibnamefont{Miller}}
  (\bibinfo{year}{0300}), \eprint{astro-ph/0703054}.

\bibitem[{\citenamefont{Goodale et~al.}(2003)\citenamefont{Goodale, Allen,
  Lanfermann, Mass{\'o}, Radke, Seidel, and Shalf}}]{Goodale02a}
\bibinfo{author}{\bibfnamefont{T.}~\bibnamefont{Goodale}},
  \bibinfo{author}{\bibfnamefont{G.}~\bibnamefont{Allen}},
  \bibinfo{author}{\bibfnamefont{G.}~\bibnamefont{Lanfermann}},
  \bibinfo{author}{\bibfnamefont{J.}~\bibnamefont{Mass{\'o}}},
  \bibinfo{author}{\bibfnamefont{T.}~\bibnamefont{Radke}},
  \bibinfo{author}{\bibfnamefont{E.}~\bibnamefont{Seidel}}, \bibnamefont{and}
  \bibinfo{author}{\bibfnamefont{J.}~\bibnamefont{Shalf}}, in
  \emph{\bibinfo{booktitle}{Vector and Parallel Processing -- VECPAR'2002, 5th
  International Conference, Lecture Notes in Computer Science}}
  (\bibinfo{publisher}{Springer}, \bibinfo{address}{Berlin},
  \bibinfo{year}{2003}).

\bibitem[{cac()}]{cactusweb1}
\bibinfo{note}{{Cactus} Computational Toolkit home page},
  \urlprefix\url{http://www.cactuscode.org/}.

\bibitem[{\citenamefont{Alcubierre et~al.}(2000)\citenamefont{Alcubierre,
  Br{\"u}gmann, Dramlitsch, Font, Papadopoulos, Seidel, Stergioulas, and
  Takahashi}}]{Alcubierre99d}
\bibinfo{author}{\bibfnamefont{M.}~\bibnamefont{Alcubierre}},
  \bibinfo{author}{\bibfnamefont{B.}~\bibnamefont{Br{\"u}gmann}},
  \bibinfo{author}{\bibfnamefont{T.}~\bibnamefont{Dramlitsch}},
  \bibinfo{author}{\bibfnamefont{J.~A.} \bibnamefont{Font}},
  \bibinfo{author}{\bibfnamefont{P.}~\bibnamefont{Papadopoulos}},
  \bibinfo{author}{\bibfnamefont{E.}~\bibnamefont{Seidel}},
  \bibinfo{author}{\bibfnamefont{N.}~\bibnamefont{Stergioulas}},
  \bibnamefont{and}
  \bibinfo{author}{\bibfnamefont{R.}~\bibnamefont{Takahashi}},
  \bibinfo{journal}{Phys. Rev. D} \textbf{\bibinfo{volume}{62}},
  \bibinfo{pages}{044034} (\bibinfo{year}{2000}), \eprint{gr-qc/0003071}.

\bibitem[{\citenamefont{Alcubierre et~al.}(2003)\citenamefont{Alcubierre,
  Br{\"u}gmann, Diener, Koppitz, Pollney, Seidel, and
  Takahashi}}]{Alcubierre02a}
\bibinfo{author}{\bibfnamefont{M.}~\bibnamefont{Alcubierre}},
  \bibinfo{author}{\bibfnamefont{B.}~\bibnamefont{Br{\"u}gmann}},
  \bibinfo{author}{\bibfnamefont{P.}~\bibnamefont{Diener}},
  \bibinfo{author}{\bibfnamefont{M.}~\bibnamefont{Koppitz}},
  \bibinfo{author}{\bibfnamefont{D.}~\bibnamefont{Pollney}},
  \bibinfo{author}{\bibfnamefont{E.}~\bibnamefont{Seidel}}, \bibnamefont{and}
  \bibinfo{author}{\bibfnamefont{R.}~\bibnamefont{Takahashi}},
  \bibinfo{journal}{Phys. Rev. D} \textbf{\bibinfo{volume}{67}},
  \bibinfo{pages}{084023} (\bibinfo{year}{2003}), \eprint{gr-qc/0206072}.

\bibitem[{\citenamefont{Alcubierre
  et~al.}(2005{\natexlab{a}})\citenamefont{Alcubierre, Br{\"u}gmann, Diener,
  Guzm{\'a}n, Hawke, Hawley, Herrmann, Koppitz, Pollney, Seidel
  et~al.}}]{Alcubierre2003:pre-ISCO-coalescence-times}
\bibinfo{author}{\bibfnamefont{M.}~\bibnamefont{Alcubierre}},
  \bibinfo{author}{\bibfnamefont{B.}~\bibnamefont{Br{\"u}gmann}},
  \bibinfo{author}{\bibfnamefont{P.}~\bibnamefont{Diener}},
  \bibinfo{author}{\bibfnamefont{F.~S.} \bibnamefont{Guzm{\'a}n}},
  \bibinfo{author}{\bibfnamefont{I.}~\bibnamefont{Hawke}},
  \bibinfo{author}{\bibfnamefont{S.}~\bibnamefont{Hawley}},
  \bibinfo{author}{\bibfnamefont{F.}~\bibnamefont{Herrmann}},
  \bibinfo{author}{\bibfnamefont{M.}~\bibnamefont{Koppitz}},
  \bibinfo{author}{\bibfnamefont{D.}~\bibnamefont{Pollney}},
  \bibinfo{author}{\bibfnamefont{E.}~\bibnamefont{Seidel}},
  \bibnamefont{et~al.}, \bibinfo{journal}{Phys. Rev. D}
  \textbf{\bibinfo{volume}{72}}, \bibinfo{pages}{044004}
  (\bibinfo{year}{2005}{\natexlab{a}}), \eprint{gr-qc/0411149},
  \urlprefix\url{http://link.aps.org/abstract/PRD/v72/e044004}.

\bibitem[{\citenamefont{Baker et~al.}(2006{\natexlab{b}})\citenamefont{Baker,
  Centrella, Choi, Koppitz, and van Meter}}]{Baker:2006yw}
\bibinfo{author}{\bibfnamefont{J.~G.} \bibnamefont{Baker}},
  \bibinfo{author}{\bibfnamefont{J.}~\bibnamefont{Centrella}},
  \bibinfo{author}{\bibfnamefont{D.-I.} \bibnamefont{Choi}},
  \bibinfo{author}{\bibfnamefont{M.}~\bibnamefont{Koppitz}}, \bibnamefont{and}
  \bibinfo{author}{\bibfnamefont{J.}~\bibnamefont{van Meter}},
  \bibinfo{journal}{Phys. Rev. D} \textbf{\bibinfo{volume}{73}},
  \bibinfo{pages}{104002} (\bibinfo{year}{2006}{\natexlab{b}}),
  \eprint{gr-qc/0602026}.

\bibitem[{\citenamefont{Campanelli et~al.}(2006)\citenamefont{Campanelli,
  Lousto, Marronetti, and Zlochower}}]{Campanelli:2005dd}
\bibinfo{author}{\bibfnamefont{M.}~\bibnamefont{Campanelli}},
  \bibinfo{author}{\bibfnamefont{C.~O.} \bibnamefont{Lousto}},
  \bibinfo{author}{\bibfnamefont{P.}~\bibnamefont{Marronetti}},
  \bibnamefont{and}
  \bibinfo{author}{\bibfnamefont{Y.}~\bibnamefont{Zlochower}},
  \bibinfo{journal}{Phys. Rev. Lett.} \textbf{\bibinfo{volume}{96}},
  \bibinfo{pages}{111101} (\bibinfo{year}{2006}), \eprint{gr-qc/0511048}.

\bibitem[{\citenamefont{Nakamura et~al.}(1987)\citenamefont{Nakamura, Oohara,
  and Kojima}}]{Nakamura87}
\bibinfo{author}{\bibfnamefont{T.}~\bibnamefont{Nakamura}},
  \bibinfo{author}{\bibfnamefont{K.}~\bibnamefont{Oohara}}, \bibnamefont{and}
  \bibinfo{author}{\bibfnamefont{Y.}~\bibnamefont{Kojima}},
  \bibinfo{journal}{Prog. Theor. Phys. Suppl.} \textbf{\bibinfo{volume}{90}},
  \bibinfo{pages}{1} (\bibinfo{year}{1987}).

\bibitem[{\citenamefont{Shibata and Nakamura}(1995)}]{Shibata95}
\bibinfo{author}{\bibfnamefont{M.}~\bibnamefont{Shibata}} \bibnamefont{and}
  \bibinfo{author}{\bibfnamefont{T.}~\bibnamefont{Nakamura}},
  \bibinfo{journal}{Phys. Rev. D} \textbf{\bibinfo{volume}{52}},
  \bibinfo{pages}{5428} (\bibinfo{year}{1995}).

\bibitem[{\citenamefont{Baumgarte and Shapiro}(1999)}]{Baumgarte99}
\bibinfo{author}{\bibfnamefont{T.~W.} \bibnamefont{Baumgarte}}
  \bibnamefont{and} \bibinfo{author}{\bibfnamefont{S.~L.}
  \bibnamefont{Shapiro}}, \bibinfo{journal}{Phys. Rev. D}
  \textbf{\bibinfo{volume}{59}}, \bibinfo{pages}{024007}
  (\bibinfo{year}{1999}), \eprint{gr-qc/9810065}.

\bibitem[{\citenamefont{York}(1979)}]{York79}
\bibinfo{author}{\bibfnamefont{J.~W.} \bibnamefont{York}}, in
  \emph{\bibinfo{booktitle}{Sources of gravitational radiation}}, edited by
  \bibinfo{editor}{\bibfnamefont{L.~L.} \bibnamefont{Smarr}}
  (\bibinfo{publisher}{Cambridge University Press},
  \bibinfo{address}{Cambridge, UK}, \bibinfo{year}{1979}), pp.
  \bibinfo{pages}{83--126}, ISBN \bibinfo{isbn}{0-521-22778-X}.

\bibitem[{\citenamefont{Misner et~al.}(1973)\citenamefont{Misner, Thorne, and
  Wheeler}}]{misner73}
\bibinfo{author}{\bibfnamefont{C.~W.} \bibnamefont{Misner}},
  \bibinfo{author}{\bibfnamefont{K.~S.} \bibnamefont{Thorne}},
  \bibnamefont{and} \bibinfo{author}{\bibfnamefont{J.~A.}
  \bibnamefont{Wheeler}}, \emph{\bibinfo{title}{Gravitation}}
  (\bibinfo{publisher}{W. H. Freeman}, \bibinfo{address}{San Francisco},
  \bibinfo{year}{1973}).

\bibitem[{\citenamefont{Baker et~al.}(2006{\natexlab{c}})\citenamefont{Baker,
  Centrella, Choi, Koppitz, and van Meter}}]{Baker05a}
\bibinfo{author}{\bibfnamefont{J.~G.} \bibnamefont{Baker}},
  \bibinfo{author}{\bibfnamefont{J.}~\bibnamefont{Centrella}},
  \bibinfo{author}{\bibfnamefont{D.-I.} \bibnamefont{Choi}},
  \bibinfo{author}{\bibfnamefont{M.}~\bibnamefont{Koppitz}}, \bibnamefont{and}
  \bibinfo{author}{\bibfnamefont{J.}~\bibnamefont{van Meter}},
  \bibinfo{journal}{Phys. Rev. Lett.} \textbf{\bibinfo{volume}{96}},
  \bibinfo{pages}{111102} (\bibinfo{year}{2006}{\natexlab{c}}),
  \eprint{gr-qc/0511103}.

\bibitem[{\citenamefont{van Meter et~al.}(2006)\citenamefont{van Meter, Baker,
  Koppitz, and Choi}}]{Baker:2006mp}
\bibinfo{author}{\bibfnamefont{J.}~\bibnamefont{van Meter}},
  \bibinfo{author}{\bibfnamefont{J.~G.} \bibnamefont{Baker}},
  \bibinfo{author}{\bibfnamefont{M.}~\bibnamefont{Koppitz}}, \bibnamefont{and}
  \bibinfo{author}{\bibfnamefont{D.-I.} \bibnamefont{Choi}}
  (\bibinfo{year}{2006}), \bibinfo{note}{unpublished, gr-qc/0605030},
  \eprint{gr-qc/0605030}.

\bibitem[{\citenamefont{Schnetter et~al.}(2004)\citenamefont{Schnetter, Hawley,
  and Hawke}}]{Schnetter-etal-03b}
\bibinfo{author}{\bibfnamefont{E.}~\bibnamefont{Schnetter}},
  \bibinfo{author}{\bibfnamefont{S.~H.} \bibnamefont{Hawley}},
  \bibnamefont{and} \bibinfo{author}{\bibfnamefont{I.}~\bibnamefont{Hawke}},
  \bibinfo{journal}{Class. Quantum Grav.} \textbf{\bibinfo{volume}{21}},
  \bibinfo{pages}{1465} (\bibinfo{year}{2004}), \eprint{gr-qc/0310042}.

\bibitem[{car()}]{carpetweb}
\bibinfo{note}{Mesh Refinement with {Carpet}},
  \urlprefix\url{http://www.carpetcode.org/}.

\bibitem[{\citenamefont{Thornburg}(1996)}]{Thornburg95}
\bibinfo{author}{\bibfnamefont{J.}~\bibnamefont{Thornburg}},
  \bibinfo{journal}{Phys. Rev. D} \textbf{\bibinfo{volume}{54}},
  \bibinfo{pages}{4899} (\bibinfo{year}{1996}), \eprint{gr-qc/9508014}.

\bibitem[{\citenamefont{Thornburg}(2004{\natexlab{a}})}]{Thornburg2003:AH-find%
ing_nourl}
\bibinfo{author}{\bibfnamefont{J.}~\bibnamefont{Thornburg}},
  \bibinfo{journal}{Class. Quantum Grav.} \textbf{\bibinfo{volume}{21}},
  \bibinfo{pages}{743} (\bibinfo{year}{2004}{\natexlab{a}}),
  \eprint{gr-qc/0306056}.

\bibitem[{\citenamefont{Brandt and Br{\"u}gmann}(1997)}]{Brandt97b}
\bibinfo{author}{\bibfnamefont{S.}~\bibnamefont{Brandt}} \bibnamefont{and}
  \bibinfo{author}{\bibfnamefont{B.}~\bibnamefont{Br{\"u}gmann}},
  \bibinfo{journal}{Phys. Rev. Lett.} \textbf{\bibinfo{volume}{78}},
  \bibinfo{pages}{3606} (\bibinfo{year}{1997}), \eprint{gr-qc/9703066}.

\bibitem[{\citenamefont{Ansorg et~al.}(2004)\citenamefont{Ansorg, Br{\"u}gmann,
  and Tichy}}]{Ansorg:2004ds}
\bibinfo{author}{\bibfnamefont{M.}~\bibnamefont{Ansorg}},
  \bibinfo{author}{\bibfnamefont{B.}~\bibnamefont{Br{\"u}gmann}},
  \bibnamefont{and} \bibinfo{author}{\bibfnamefont{W.}~\bibnamefont{Tichy}},
  \bibinfo{journal}{Phys. Rev. D} \textbf{\bibinfo{volume}{70}},
  \bibinfo{pages}{064011} (\bibinfo{year}{2004}), \eprint{gr-qc/0404056}.

\bibitem[{\citenamefont{Smarr}(1973)}]{Smarr73a}
\bibinfo{author}{\bibfnamefont{L.~L.} \bibnamefont{Smarr}},
  \bibinfo{journal}{Phys. Rev. Lett.} \textbf{\bibinfo{volume}{30}},
  \bibinfo{pages}{71} (\bibinfo{year}{1973}).

\bibitem[{\citenamefont{Christodoulou}(1970)}]{Christodoulou70}
\bibinfo{author}{\bibfnamefont{D.}~\bibnamefont{Christodoulou}},
  \bibinfo{journal}{Phys. Rev. Lett.} \textbf{\bibinfo{volume}{25}},
  \bibinfo{pages}{1596} (\bibinfo{year}{1970}).

\bibitem[{\citenamefont{Cook}(1994)}]{Cook94}
\bibinfo{author}{\bibfnamefont{G.~B.} \bibnamefont{Cook}},
  \bibinfo{journal}{Phys. Rev. D} \textbf{\bibinfo{volume}{50}},
  \bibinfo{pages}{5025} (\bibinfo{year}{1994}).

\bibitem[{\citenamefont{Pfeiffer et~al.}(2000)\citenamefont{Pfeiffer,
  {T}eukolsky, and Cook}}]{Pfeiffer:2000um}
\bibinfo{author}{\bibfnamefont{H.~P.} \bibnamefont{Pfeiffer}},
  \bibinfo{author}{\bibfnamefont{S.~A.} \bibnamefont{{T}eukolsky}},
  \bibnamefont{and} \bibinfo{author}{\bibfnamefont{G.~B.} \bibnamefont{Cook}},
  \bibinfo{journal}{Phys. Rev. D} \textbf{\bibinfo{volume}{62}},
  \bibinfo{pages}{104018} (\bibinfo{year}{2000}), \eprint{gr-qc/0006084}.

\bibitem[{\citenamefont{Cook}(2000)}]{Cook00a}
\bibinfo{author}{\bibfnamefont{G.~B.} \bibnamefont{Cook}},
  \bibinfo{journal}{Living Rev. Relativity} \textbf{\bibinfo{volume}{3}},
  \bibinfo{pages}{5} (\bibinfo{year}{2000}),
  \urlprefix\url{http://www.livingreviews.org/lrr-2000-5}.

\bibitem[{\citenamefont{Thornburg}(2004{\natexlab{b}})}]{Thornburg2003:AH-find%
ing}
\bibinfo{author}{\bibfnamefont{J.}~\bibnamefont{Thornburg}},
  \bibinfo{journal}{Class. Quantum Grav.} \textbf{\bibinfo{volume}{21}},
  \bibinfo{pages}{743} (\bibinfo{year}{2004}{\natexlab{b}}),
  \eprint{gr-qc/0306056},
  \urlprefix\url{http://stacks.iop.org/0264-9381/21/743}.

\bibitem[{\citenamefont{Gunnarsen et~al.}(1995)\citenamefont{Gunnarsen,
  Shinkai, and Maeda}}]{Shinkai94}
\bibinfo{author}{\bibfnamefont{L.}~\bibnamefont{Gunnarsen}},
  \bibinfo{author}{\bibfnamefont{H.}~\bibnamefont{Shinkai}}, \bibnamefont{and}
  \bibinfo{author}{\bibfnamefont{K.}~\bibnamefont{Maeda}},
  \bibinfo{journal}{Class. Quantum Grav.} \textbf{\bibinfo{volume}{12}},
  \bibinfo{pages}{133} (\bibinfo{year}{1995}), \eprint{gr-qc/9406003}.

\bibitem[{\citenamefont{Lehner and Moreschi}(2007)}]{Lehner:2007ip}
\bibinfo{author}{\bibfnamefont{L.}~\bibnamefont{Lehner}} \bibnamefont{and}
  \bibinfo{author}{\bibfnamefont{O.~M.} \bibnamefont{Moreschi}}
  (\bibinfo{year}{2007}), \eprint{arXiv:0706.1319 [gr-qc]}.

\bibitem[{\citenamefont{Teukolsky}(1973)}]{Teukolsky73}
\bibinfo{author}{\bibfnamefont{S.~A.} \bibnamefont{Teukolsky}},
  \bibinfo{journal}{Astrophys. J.} \textbf{\bibinfo{volume}{185}},
  \bibinfo{pages}{635} (\bibinfo{year}{1973}).

\bibitem[{\citenamefont{Campanelli and Lousto}(1999)}]{Campanelli99}
\bibinfo{author}{\bibfnamefont{M.}~\bibnamefont{Campanelli}} \bibnamefont{and}
  \bibinfo{author}{\bibfnamefont{C.~O.} \bibnamefont{Lousto}},
  \bibinfo{journal}{Phys. Rev. D} \textbf{\bibinfo{volume}{59}},
  \bibinfo{pages}{124022} (\bibinfo{year}{1999}), \eprint{gr-qc/9811019}.

\bibitem[{\citenamefont{Baker et~al.}(2006{\natexlab{d}})\citenamefont{Baker,
  Centrella, Choi, Koppitz, van Meter, and Miller}}]{Baker:2006nr}
\bibinfo{author}{\bibfnamefont{J.~G.} \bibnamefont{Baker}},
  \bibinfo{author}{\bibfnamefont{J.}~\bibnamefont{Centrella}},
  \bibinfo{author}{\bibfnamefont{D.-I.} \bibnamefont{Choi}},
  \bibinfo{author}{\bibfnamefont{M.}~\bibnamefont{Koppitz}},
  \bibinfo{author}{\bibfnamefont{J.}~\bibnamefont{van Meter}},
  \bibnamefont{and} \bibinfo{author}{\bibfnamefont{M.~C.}
  \bibnamefont{Miller}}, \bibinfo{journal}{Astrophys. J.}
  \textbf{\bibinfo{volume}{653}}, \bibinfo{pages}{L93}
  (\bibinfo{year}{2006}{\natexlab{d}}), \eprint{astro-ph/0603204}.

\bibitem[{\citenamefont{Gonz{\'a}lez et~al.}(2006)\citenamefont{Gonz{\'a}lez,
  Sperhake, Br{\"u}gmann, Hannam, and Husa}}]{Gonzales06tr}
\bibinfo{author}{\bibfnamefont{J.~A.} \bibnamefont{Gonz{\'a}lez}},
  \bibinfo{author}{\bibfnamefont{U.}~\bibnamefont{Sperhake}},
  \bibinfo{author}{\bibfnamefont{B.}~\bibnamefont{Br{\"u}gmann}},
  \bibinfo{author}{\bibfnamefont{M.}~\bibnamefont{Hannam}}, \bibnamefont{and}
  \bibinfo{author}{\bibfnamefont{S.}~\bibnamefont{Husa}}
  (\bibinfo{year}{2006}), \bibinfo{note}{gr-qc/0610154},
  \eprint{gr-qc/0610154}.

\bibitem[{\citenamefont{Koppitz et~al.}(2007{\natexlab{b}})}]{Koppitz:2007ev}
\bibinfo{author}{\bibfnamefont{M.}~\bibnamefont{Koppitz}} \bibnamefont{et~al.}
  (\bibinfo{year}{2007}{\natexlab{b}}), \eprint{gr-qc/0701163}.

\bibitem[{\citenamefont{Allen et~al.}(1998)\citenamefont{Allen, Camarda, and
  Seidel}}]{Allen98a1}
\bibinfo{author}{\bibfnamefont{G.}~\bibnamefont{Allen}},
  \bibinfo{author}{\bibfnamefont{K.}~\bibnamefont{Camarda}}, \bibnamefont{and}
  \bibinfo{author}{\bibfnamefont{E.}~\bibnamefont{Seidel}}
  (\bibinfo{year}{1998}), \bibinfo{note}{gr-qc/9806036}.

\bibitem[{\citenamefont{Rupright et~al.}(1998)\citenamefont{Rupright, Abrahams,
  and Rezzolla}}]{Rupright98}
\bibinfo{author}{\bibfnamefont{M.~E.} \bibnamefont{Rupright}},
  \bibinfo{author}{\bibfnamefont{A.~M.} \bibnamefont{Abrahams}},
  \bibnamefont{and} \bibinfo{author}{\bibfnamefont{L.}~\bibnamefont{Rezzolla}},
  \bibinfo{journal}{Phys. Rev. D} \textbf{\bibinfo{volume}{58}},
  \bibinfo{pages}{044005} (\bibinfo{year}{1998}).

\bibitem[{\citenamefont{Camarda and Seidel}(1999)}]{Camarda97c}
\bibinfo{author}{\bibfnamefont{K.}~\bibnamefont{Camarda}} \bibnamefont{and}
  \bibinfo{author}{\bibfnamefont{E.}~\bibnamefont{Seidel}},
  \bibinfo{journal}{Phys. Rev. D} \textbf{\bibinfo{volume}{59}},
  \bibinfo{pages}{064019} (\bibinfo{year}{1999}), \eprint{gr-qc/9805099}.

\bibitem[{\citenamefont{Moncrief}(1974)}]{Moncrief74}
\bibinfo{author}{\bibfnamefont{V.}~\bibnamefont{Moncrief}},
  \bibinfo{journal}{Annals of Physics} \textbf{\bibinfo{volume}{88}},
  \bibinfo{pages}{323} (\bibinfo{year}{1974}).

\bibitem[{\citenamefont{Abrahams and Price}(1996)}]{Abrahams95b}
\bibinfo{author}{\bibfnamefont{A.}~\bibnamefont{Abrahams}} \bibnamefont{and}
  \bibinfo{author}{\bibfnamefont{R.}~\bibnamefont{Price}},
  \bibinfo{journal}{Phys. Rev. D} \textbf{\bibinfo{volume}{53}},
  \bibinfo{pages}{163} (\bibinfo{year}{1996}).

\bibitem[{\citenamefont{Nagar and Rezzolla}(2005)}]{Nagar05}
\bibinfo{author}{\bibfnamefont{A.}~\bibnamefont{Nagar}} \bibnamefont{and}
  \bibinfo{author}{\bibfnamefont{L.}~\bibnamefont{Rezzolla}},
  \bibinfo{journal}{Class. Quantum Grav.} \textbf{\bibinfo{volume}{22}},
  \bibinfo{pages}{R167} (\bibinfo{year}{2005}), \bibinfo{note}{erratum-ibid.
  \textbf{23}, 4297, (2006)}.

\bibitem[{\citenamefont{Rezzolla et~al.}(1999)\citenamefont{Rezzolla, Abrahams,
  Matzner, Rupright, and Shapiro}}]{Rezzolla99a}
\bibinfo{author}{\bibfnamefont{L.}~\bibnamefont{Rezzolla}},
  \bibinfo{author}{\bibfnamefont{A.~M.} \bibnamefont{Abrahams}},
  \bibinfo{author}{\bibfnamefont{R.~A.} \bibnamefont{Matzner}},
  \bibinfo{author}{\bibfnamefont{M.~E.} \bibnamefont{Rupright}},
  \bibnamefont{and} \bibinfo{author}{\bibfnamefont{S.~L.}
  \bibnamefont{Shapiro}}, \bibinfo{journal}{Phys. Rev. D}
  \textbf{\bibinfo{volume}{59}}, \bibinfo{pages}{064001}
  (\bibinfo{year}{1999}), \eprint{gr-qc/9807047}.

\bibitem[{\citenamefont{Baker et~al.}(2000)\citenamefont{Baker, Brandt,
  Campanelli, Lousto, Seidel, and Takahashi}}]{Baker99a}
\bibinfo{author}{\bibfnamefont{J.}~\bibnamefont{Baker}},
  \bibinfo{author}{\bibfnamefont{S.~R.} \bibnamefont{Brandt}},
  \bibinfo{author}{\bibfnamefont{M.}~\bibnamefont{Campanelli}},
  \bibinfo{author}{\bibfnamefont{C.~O.} \bibnamefont{Lousto}},
  \bibinfo{author}{\bibfnamefont{E.}~\bibnamefont{Seidel}}, \bibnamefont{and}
  \bibinfo{author}{\bibfnamefont{R.}~\bibnamefont{Takahashi}},
  \bibinfo{journal}{Phys. Rev. D} \textbf{\bibinfo{volume}{62}},
  \bibinfo{pages}{127701} (\bibinfo{year}{2000}),
  \bibinfo{note}{gr-qc/9911017}.

\bibitem[{\citenamefont{Font et~al.}(2002)\citenamefont{Font, Goodale, Iyer,
  Miller, Rezzolla, Seidel, Stergioulas, Suen, and Tobias}}]{Font01b}
\bibinfo{author}{\bibfnamefont{J.~A.} \bibnamefont{Font}},
  \bibinfo{author}{\bibfnamefont{T.}~\bibnamefont{Goodale}},
  \bibinfo{author}{\bibfnamefont{S.}~\bibnamefont{Iyer}},
  \bibinfo{author}{\bibfnamefont{M.}~\bibnamefont{Miller}},
  \bibinfo{author}{\bibfnamefont{L.}~\bibnamefont{Rezzolla}},
  \bibinfo{author}{\bibfnamefont{E.}~\bibnamefont{Seidel}},
  \bibinfo{author}{\bibfnamefont{N.}~\bibnamefont{Stergioulas}},
  \bibinfo{author}{\bibfnamefont{W.-M.} \bibnamefont{Suen}}, \bibnamefont{and}
  \bibinfo{author}{\bibfnamefont{M.}~\bibnamefont{Tobias}},
  \bibinfo{journal}{Phys. Rev. D} \textbf{\bibinfo{volume}{65}},
  \bibinfo{pages}{084024} (\bibinfo{year}{2002}), \eprint{gr-qc/0110047}.

\bibitem[{\citenamefont{Thorne}(1980)}]{Thorne80b}
\bibinfo{author}{\bibfnamefont{K.}~\bibnamefont{Thorne}},
  \bibinfo{journal}{Rev. Mod. Phys.} \textbf{\bibinfo{volume}{52}},
  \bibinfo{pages}{299} (\bibinfo{year}{1980}).

\bibitem[{\citenamefont{C.~F.~Sopuerta and Laguna}(2006)}]{Sopuerta:2006wj}
\bibinfo{author}{\bibfnamefont{N.~Y.} \bibnamefont{C.~F.~Sopuerta}}
  \bibnamefont{and} \bibinfo{author}{\bibfnamefont{P.}~\bibnamefont{Laguna}},
  \bibinfo{journal}{Phys. Rev. D} p. \bibinfo{pages}{124010}
  (\bibinfo{year}{2006}), \bibinfo{note}{erratum-ibid.75, 069903 (2007)},
  \eprint{[arXiv:astro-ph/0608600]}.

\bibitem[{\citenamefont{Damour and Gopakumar}(2006)}]{Damour-Gopakumar-2006}
\bibinfo{author}{\bibfnamefont{T.}~\bibnamefont{Damour}} \bibnamefont{and}
  \bibinfo{author}{\bibfnamefont{A.}~\bibnamefont{Gopakumar}},
  \bibinfo{journal}{Phys. Rev. D} \textbf{\bibinfo{volume}{73}},
  \bibinfo{pages}{124006} (\bibinfo{year}{2006}), \eprint{gr-qc/0602117}.

\bibitem[{\citenamefont{Kidder}(1995)}]{Kidder:1995zr}
\bibinfo{author}{\bibfnamefont{L.~E.} \bibnamefont{Kidder}},
  \bibinfo{journal}{Phys. Rev. D} \textbf{\bibinfo{volume}{52}},
  \bibinfo{pages}{821} (\bibinfo{year}{1995}), \eprint{gr-qc/9506022}.

\bibitem[{\citenamefont{Dreyer et~al.}(2003)\citenamefont{Dreyer, Krishnan,
  Shoemaker, and Schnetter}}]{Dreyer02a}
\bibinfo{author}{\bibfnamefont{O.}~\bibnamefont{Dreyer}},
  \bibinfo{author}{\bibfnamefont{B.}~\bibnamefont{Krishnan}},
  \bibinfo{author}{\bibfnamefont{D.}~\bibnamefont{Shoemaker}},
  \bibnamefont{and}
  \bibinfo{author}{\bibfnamefont{E.}~\bibnamefont{Schnetter}},
  \bibinfo{journal}{Phys. Rev. D} \textbf{\bibinfo{volume}{67}},
  \bibinfo{pages}{024018} (\bibinfo{year}{2003}), \eprint{gr-qc/0206008},
  \urlprefix\url{http://link.aps.org/abstract/PRD/v67/e024018}.

\bibitem[{\citenamefont{Ashtekar and Krishnan}(2003)}]{ashtekar03a}
\bibinfo{author}{\bibfnamefont{A.}~\bibnamefont{Ashtekar}} \bibnamefont{and}
  \bibinfo{author}{\bibfnamefont{B.}~\bibnamefont{Krishnan}},
  \bibinfo{journal}{Phys. Rev. D} \textbf{\bibinfo{volume}{68}},
  \bibinfo{pages}{104030} (\bibinfo{year}{2003}), \eprint{gr-qc/0308033}.

\bibitem[{\citenamefont{Schnetter et~al.}(2006)\citenamefont{Schnetter,
  Krishnan, and Beyer}}]{Schnetter-Krishnan-Beyer-2006}
\bibinfo{author}{\bibfnamefont{E.}~\bibnamefont{Schnetter}},
  \bibinfo{author}{\bibfnamefont{B.}~\bibnamefont{Krishnan}}, \bibnamefont{and}
  \bibinfo{author}{\bibfnamefont{F.}~\bibnamefont{Beyer}},
  \bibinfo{journal}{Phys. Rev. D} \textbf{\bibinfo{volume}{74}},
  \bibinfo{pages}{024028} (\bibinfo{year}{2006}), \eprint{gr-qc/0604015}.

\bibitem[{\citenamefont{Alcubierre
  et~al.}(2005{\natexlab{b}})\citenamefont{Alcubierre, Br{\"u}gmann, Diener,
  Guzm{\'a}n, Hawke, Hawley, Herrmann, Koppitz, Pollney, Seidel
  et~al.}}]{Alcubierre:2004hr}
\bibinfo{author}{\bibfnamefont{M.}~\bibnamefont{Alcubierre}},
  \bibinfo{author}{\bibfnamefont{B.}~\bibnamefont{Br{\"u}gmann}},
  \bibinfo{author}{\bibfnamefont{P.}~\bibnamefont{Diener}},
  \bibinfo{author}{\bibfnamefont{F.~S.} \bibnamefont{Guzm{\'a}n}},
  \bibinfo{author}{\bibfnamefont{I.}~\bibnamefont{Hawke}},
  \bibinfo{author}{\bibfnamefont{S.}~\bibnamefont{Hawley}},
  \bibinfo{author}{\bibfnamefont{F.}~\bibnamefont{Herrmann}},
  \bibinfo{author}{\bibfnamefont{M.}~\bibnamefont{Koppitz}},
  \bibinfo{author}{\bibfnamefont{D.}~\bibnamefont{Pollney}},
  \bibinfo{author}{\bibfnamefont{E.}~\bibnamefont{Seidel}},
  \bibnamefont{et~al.}, \bibinfo{journal}{Phys. Rev. D}
  \textbf{\bibinfo{volume}{72}}, \bibinfo{pages}{044004}
  (\bibinfo{year}{2005}{\natexlab{b}}), \eprint{gr-qc/0411149}.

\bibitem[{\citenamefont{Seidel}(1999)}]{Seidel99b}
\bibinfo{author}{\bibfnamefont{E.}~\bibnamefont{Seidel}},
  \bibinfo{journal}{Prog. Theor. Phys. Suppl.} \textbf{\bibinfo{volume}{136}},
  \bibinfo{pages}{87} (\bibinfo{year}{1999}),
  \urlprefix\url{http://ptp.ipap.jp/link?PTPS/136/87/}.

\bibitem[{\citenamefont{Brandt and Seidel}(1995)}]{Brandt94c}
\bibinfo{author}{\bibfnamefont{S.}~\bibnamefont{Brandt}} \bibnamefont{and}
  \bibinfo{author}{\bibfnamefont{E.}~\bibnamefont{Seidel}},
  \bibinfo{journal}{Phys. Rev. D} \textbf{\bibinfo{volume}{52}},
  \bibinfo{pages}{870} (\bibinfo{year}{1995}),
  \urlprefix\url{http://link.aps.org/abstract/PRD/v52/p870}.

\bibitem[{\citenamefont{Poisson}(2004)}]{Poisson04b}
\bibinfo{author}{\bibfnamefont{E.}~\bibnamefont{Poisson}},
  \bibinfo{journal}{Phys. Rev.} \textbf{\bibinfo{volume}{D70}},
  \bibinfo{pages}{084044} (\bibinfo{year}{2004}), \eprint{gr-qc/0407050}.

\bibitem[{\citenamefont{Martel and Poisson}(2005)}]{Martel:2005}
\bibinfo{author}{\bibfnamefont{K.}~\bibnamefont{Martel}} \bibnamefont{and}
  \bibinfo{author}{\bibfnamefont{E.}~\bibnamefont{Poisson}},
  \bibinfo{journal}{Phys. Rev. D} \textbf{\bibinfo{volume}{71}},
  \bibinfo{pages}{104003} (\bibinfo{year}{2005}).

\bibitem[{\citenamefont{{Kupi} et~al.}(2006)\citenamefont{{Kupi},
  {Amaro-Seoane}, and {Spurzem}}}]{kupi06}
\bibinfo{author}{\bibfnamefont{G.}~\bibnamefont{{Kupi}}},
  \bibinfo{author}{\bibfnamefont{P.}~\bibnamefont{{Amaro-Seoane}}},
  \bibnamefont{and}
  \bibinfo{author}{\bibfnamefont{R.}~\bibnamefont{{Spurzem}}},
  \bibinfo{journal}{MNRAS} \textbf{\bibinfo{volume}{371}}, \bibinfo{pages}{L45}
  (\bibinfo{year}{2006}), \eprint{arXiv:astro-ph/0602125}.

\bibitem[{\citenamefont{Cunningham et~al.}(1978)\citenamefont{Cunningham,
  Price, and Moncrief}}]{Cunningham78}
\bibinfo{author}{\bibfnamefont{C.~T.} \bibnamefont{Cunningham}},
  \bibinfo{author}{\bibfnamefont{R.~H.} \bibnamefont{Price}}, \bibnamefont{and}
  \bibinfo{author}{\bibfnamefont{V.}~\bibnamefont{Moncrief}},
  \bibinfo{journal}{Astrophys. J.} \textbf{\bibinfo{volume}{224}},
  \bibinfo{pages}{643} (\bibinfo{year}{1978}).

\bibitem[{\citenamefont{Cunningham et~al.}(1979)\citenamefont{Cunningham,
  Price, and Moncrief}}]{Cunningham79}
\bibinfo{author}{\bibfnamefont{C.}~\bibnamefont{Cunningham}},
  \bibinfo{author}{\bibfnamefont{R.}~\bibnamefont{Price}}, \bibnamefont{and}
  \bibinfo{author}{\bibfnamefont{V.}~\bibnamefont{Moncrief}},
  \bibinfo{journal}{Astrophys. J.} \textbf{\bibinfo{volume}{230}},
  \bibinfo{pages}{870} (\bibinfo{year}{1979}).

\end{thebibliography}

\end{document}